\documentclass[aps,prb,twocolumn,superscriptaddress]{revtex4-1}

\usepackage[dvips]{graphicx}
\usepackage{braket}
\usepackage{amsmath}%

\begin{document}

\renewcommand{\vec}[1]{\mathbf{#1}}




\title{Many-body Green's function theory of electrons and nuclei beyond the Born-Oppenheimer approximation}


\author{Ville J. H\"{a}rk\"{o}nen}
\email[]{ville.j.harkonen@gmail.com}
\affiliation{Max Planck Institute of Microstructure Physics, Weinberg 2, D-06112 Halle, Germany}
\author{Robert van Leeuwen}
\affiliation{Department of Physics, Nanoscience Center, P.O. Box 35 FI-40014, University of Jyv\"{a}skyl\"{a}, Jyv\"{a}skyl\"{a}, Finland}
\author{E. K. U. Gross}
\affiliation{Max Planck Institute of Microstructure Physics, Weinberg 2, D-06112 Halle, Germany}
\affiliation{Fritz Haber Center for Molecular Dynamics, Institute of Chemistry, The Hebrew University
of Jerusalem, Jerusalem 91904, Israel}



\date{\today}

\begin{abstract}
The method of many-body Green's functions is developed for arbitrary systems of electrons and nuclei starting from the full (beyond Born-Oppenheimer) Hamiltonian of Coulomb interactions and kinetic energies. The theory presented here resolves the problems arising from the translational and rotational invariance of this Hamiltonian that afflict the existing many-body Green's function theories. We derive a coupled set of exact equations for the electronic and nuclear Green's functions and provide a systematic way to approximately compute the properties of arbitrary many-body systems of electrons and nuclei beyond the Born-Oppenheimer approximation. The case of crystalline solids is discussed in detail.
\end{abstract}

\pacs{63.20.kd,71.10.-w,63.20.kg,63.20.D-}
\keywords{Many-body Green's function, Born-Oppenheimer approximation, Lattice dynamics}

\maketitle


\section{Introduction}
\label{cha:Introduction}

The Born-Oppenheimer (BO) approximation \cite{Born-Oppenheimer-Adiabatic-Approx.1927,born-huang-dynamical-1954} is among the most fundamental ingredients of modern condensed-matter theory. Much of our current understanding of molecules and crystalline solids is heavily based on this approximation and its validity. The BO approximation not only makes calculations computationally feasible, the motion of nuclear wave packets in the lowest BO potential energy surface also provides us with an intuitive picture of many chemical reactions.

The numerical efficiency of the BO approximation comes from treating the electronic and nuclear problems separately in such a way that the nuclear coordinates only enter as parameters in the electronic many-body problem while the nuclear Hamiltonian contains, as scalar potential, the total electronic ground-state energy as a function of the nuclear coordinates. Then the total wave function of the system is a single product of a nuclear wave function (typically a vibrational state) and the many-electron wave function which parametrically depends on the nuclear coordinates.

To deal with the electronic-structure problem at fixed nuclear configuration, various approaches are available, such as density functional theory \cite{Hohenberg-DFT-PhysRev.136.B864-1964,KohnSham-DFT-PhysRev.140.A1133-1965,Parr-DFTbook-1989,DreizlerGross-DFTbook-1990} or the method of many-body Green's functions \cite{Martin-SchwingerPhysRev.115.1342-Theory-Many-Part.Syst.-1959,Baym-field-1961,Maksimov-SelfConsistentElectronPhonon-1975,Hedin-NewMethForCalcTheOneParticleGreensFunctWithApplToTheElectronGasProb-PhysRev.139.A796-1965,Mahan-many-particle-1990,Fetter-Walencka-q-theory-of-many-particle-1971,Stefanucci-Leeuwen-many-body-book-2013,Golze-TheGWcompendiumApracticalGuideToTheoreticalPhotoemissionSpectroscopy-2019}. The nuclear many-body Schr\"{o}dinger equation is usually treated differently: One first rewrites the nuclear Hamiltonian in terms of normal coordinates which represent the nuclear vibrational and rotational degrees of freedom. In terms of these coordinates, the BO surface (i.e. the scalar potential in the nuclear Schr\"{o}dinger equation) is expanded to second order around the equilibrium positions, so that the solution can be written analytically as a product of harmonic-oscillator wave functions (phonons). Higher orders of the expansion appear as phonon-phonon interactions in the nuclear Schr\"{o}dinger equation and can be dealt with using bosonic many-body Green's functions \cite{Baym-field-1961,Maradudin-Fein-PhysRev.128.2589-Scat-Neutr.1962,Cowley-lattice-1963,Shukla-Helmholtz-1971,maradudin-dyn-prop-solids-1974,Rickayzen-GreensFunctions-1980}. With this strategy, phonon spectra \cite{Giannozzi-AbInitioCalcOfPhononDispersInSemicond-PhysRevB.43.7231-1991,Baroni-PhononsAndRelatedCrystalPropFromDFTPT-RevModPhys.73.515-2001,Togo-FirstPrincPhononCalcInMaterScience-2015,Ribeiro-StrongAnharmonInThePhononSpectraOfPbTeAndSnTeFromFirstPrinc-PhysRevB.97.014306-2018}, thermal conductivities and other thermal properties \cite{Baroni-DFPTForQuasiHarmonicCalculations-2010,Harkonen-NTE-2014,Mittal-PhononsAndAnomalousThermalExpBhaviourInCrystSolids-2018,Harkonen-Tcond-II-VIII-PhysRevB.93.024307-2016,Linnera-AbInitStudyOfTheLattTCondOfCu2OUsingTheGGAAndHybridDFMethodsPhysRevB.96.014304-2017,Harkonen-AbInitioComputStudyOnTheLattThermalCondOfZintlClathrates-PhysRevB.94.054310-2016,Norouzzadeh-TCondOfTypeITypeIIAndTypeVIIIPristineSiliconClathAFirstPrincStudy-PhysRevB.96.245201-2017,Euchner-UnderstLatTCondInThermoelClathrADFTStudyOnBinarySiBasedTypeVClathratesPhysRevB.97.014304-2018,Tadano-QuarticAnharmOfRattlersAndItsEffOnLatTCondOfClathrFromFirstPrinciplesPhysRevLett.120.105901-2018,Shen-Si2GeANewVIITypeClathrateWithUltralowThermalConductivityAndHighThermoelectricProperty-2020} have been successfully calculated. In most cases, especially in solids, the results of such BO-based calculations are in very good agreement with experiment. Somewhat surprisingly, this is also true for metals. As there is no gap in metals, there is no clear separation of electronic and nuclear energy scales in the low-lying excitations, so the naive expectation would have been that the validity of the BO approximation is questionable. Still, BO-based calculations of phononic properties of metals are usually very successful. There are only a few notable exceptions such as $\text{MgB}_{2}$ \cite{Calandra-AnharmAndNonAdiabaticEffectsInMgB2ImplicForTheIsotopeEffectAndInterpretOfRamanSpectra-2007}, graphene \cite{Pisana-BreakdownOfTheAdiabaticBornOppenhApproximationInGraphene-2007} and, possibly, the recently discovered lanthanum hydride high-temperature superconductors \cite{Somayazulu-EvidForSupercAbove260KInLanthanumSuperhydrAtMegabarPressures-PhysRevLett.122.027001-2019,Flores-APredictionForHotSuperconductivity-2019,Errea-QuantumCrystalStructureInThe250kelvinSuperconductingLanthanumHydride-2020}.

In spite of this tremendous success of the BO approximation, a number of fascinating phenomena in physics and chemistry appear in the so-called non-adiabatic regime where the coupled motion of electrons and nuclei \textit{beyond the Born-Oppenheimer approximation} is essential. Examples are the process of vision \cite{Polli-ConicalIntersectionDynamicsOfThePrimaryPhotoisomerizEventInVision-2010}, exciton dynamics in photovoltaic systems \cite{Tully-PerspectiveNonadiabaticDynamicsTheory-2012,Nelson-NonadiabaticExcitedStateMolecularDynamicsModelingPhotophysicsInOrganicConjugatedMaterials-2014}, the splitting of the nuclear wave packet in the Zewail Nobel prize experiments \cite{Dantus-RealTimeFemtosecondProbingOfTransitionStatesInChemicalReactions-1987,Potter-FemtosecondLaserControlOfaChemicalReaction-1992,Zewail-FemtochemistryAtomicScaleDynamicsOfTheChemicalBond-2000} and the occurrence of local electronic currents associated with nuclear motion \cite{Barth-ConcertedQuantumEffectsOfElectronicAndNuclearFluxesInMolecules-2009,Schild-ElectronicFluxDensityBeyondTheBornOppenheimerApproximation-2016,Scherrer-OnTheMassOfAtomsInMoleculesBeyondTheBornOppenheimerApproximation-PhysRevX.7.031035-2017}.

To go beyond the BO approximation is notoriously difficult because one has to start from scratch, that is from the complete many-body Hamiltonian of interacting electrons and nuclei. In the non-relativistic regime, this Hamiltonian is given by:
\begin{equation} 
H = T_{n} + T_{e} + V_{ee} + V_{en} + V_{nn},
\label{eq:HamiltonianEq_1}
\end{equation}
where, in position representation, the kinetic energies are
\begin{equation} 
T_{n}  = - \sum^{N_{n}}_{k = 1} \frac{ \hbar^{2} }{ 2 M_{k} } \nabla^{2}_{\vec{R}_{k}}, \quad T_{e} = - \frac{ \hbar^{2} }{ 2 m_{e} } \sum^{N_{e}}_{i = 1}  \nabla^{2}_{\vec{r}_{i}},
\label{eq:HamiltonianEq_2}
\end{equation}
while the potential energy contributions are of the form
\begin{eqnarray} 
V_{ee}  &=& \sum^{N_{e}}_{ i,i'= 1 }{}^{'} v_{e}\left( \vec{r}_{i}, \vec{r}_{i'} \right), \quad V_{en} = \sum^{N_{e}}_{ i= 1 }\sum^{N_{n}}_{ k= 1 } v_{en}\left( \vec{r}_{i}, \vec{R}_{k} \right), \nonumber \\
V_{nn}  &=& \sum^{N_{n}}_{ k,k'= 1 }{}^{'} v_{n}\left( \vec{R}_{k}, \vec{R}_{k'} \right).
\label{eq:HamiltonianEq_2_1}
\end{eqnarray}
Here the primed sums are over the values $i \neq i'$, $k \neq k'$ and $v_{en}$, $v_{e}$ and $v_{n}$ are the bare Coulomb interactions
\begin{eqnarray} 
v_{en}\left( \vec{r}_{i},\vec{R}_{k} \right)   &=& \frac{ - Z_{k} \varsigma }{ \left| \vec{r}_{i} - \vec{R}_{k} \right| }, \nonumber \\
v_{e}\left(  \vec{r}_{i}, \vec{r}_{i'} \right) &=& \frac{1}{2} \frac{ \varsigma }{ \left| \vec{r}_{i} - \vec{r}_{i'} \right| }, \nonumber \\
v_{n}\left( \vec{R}_{k},\vec{R}_{k'} \right)   &=& \frac{1}{2} \frac{ Z_{k} Z_{k'} \varsigma }{ \left| \vec{R}_{k} - \vec{R}_{k'} \right| },
\label{eq:HamiltonianEq_3}
\end{eqnarray}
where $\varsigma = e^{2}/ \left( 4 \pi \epsilon_{0} \right)$. The system consists of $N_{e}$ electrons and $N_{n}$ nuclei. The position coordinates of these particles
\begin{equation} 
\vec{r}_{i}, \quad i = 1,\ldots,N_{e}, \quad \vec{R}_{k}, \quad k = 1,\ldots,N_{n},
\label{eq:HamiltonianEq_1_1}
\end{equation}
as they appear in Eqs. \ref{eq:HamiltonianEq_1}-\ref{eq:HamiltonianEq_3}, refer to the laboratory frame. The Hamiltonian $H$ given by Eq. \ref{eq:HamiltonianEq_1} is translationally and rotationally invariant. This has a number of important consequences which will be discussed in the following. The eigenstates in position representation of the Hamiltonian $H$ are the solutions of the stationary Schr\"{o}dinger equation
\begin{equation} 
H \Psi_{s} = E_{s} \Psi_{s},
\label{eq:IntroductionEq_1}
\end{equation}
where $s$ labels the eigenstates. The translational invariance of the Hamiltonian implies that it commutes with the total momentum operator. Hence the eigenstates $\Psi_{s} = \Psi_{s}\left(\vec{r}_{1},\ldots,\vec{r}_{N_{e}},\vec{R}_{1},\ldots,\vec{R}_{N_{n}}\right)$ can be chosen to be simultaneous eigenstates of the Hamiltonian and the total momentum operator:
\begin{equation} 
\Psi_{s} = e^{i \vec{P}_{s} \cdot \vec{R}_{cm} } \Phi_{s}\left(\tilde{\vec{r}}_{2},\ldots,\tilde{\vec{r}}_{N_{e}},\tilde{\vec{R}}_{1},\ldots,\tilde{\vec{R}}_{N_{n}}\right),
\label{eq:IntroductionEq_2}
\end{equation}
where $\vec{R}_{cm}$ is the center-of-mass position vector of the complete system, and $\vec{P}_{s}$ is the total momentum of state $\Psi_{s}$. The coordinates $\tilde{\vec{r}}_{i}$ and $\tilde{\vec{R}}_{k}$ are given by:
\begin{eqnarray} 
\tilde{\vec{r}}_{i} &=& \vec{r}_{i} - \vec{R}_{cm}, \quad i = 2,\ldots,N_{e}, \nonumber \\
\tilde{\vec{R}}_{k} &=& \vec{R}_{k} - \vec{R}_{cm}, \quad k = 1,\ldots,N_{n}.
\label{eq:IntroductionEq_3}
\end{eqnarray}
We now calculate the standard probability density $n_{s}\left(\vec{r}_{1}\right)$ of finding an electron at point $\vec{r}_{1}$ (in the laboratory frame) if the system is in the eigenstate $\Psi_{s}$:
\begin{equation} 
n_{s}\left(\vec{r}_{1}\right) = \int d\vec{r}_{2} \cdots \int d\vec{r}_{N_{e}} \int d\vec{R} \left| \Psi_{s}\left(\vec{r}_{1},\ldots,\vec{r}_{N_{e}},\vec{R}\right) \right|^{2},
\label{eq:IntroductionEq_4}
\end{equation}
where we denote all the nuclear variables by $\vec{R}$. Plugging Eq. \ref{eq:IntroductionEq_2} into Eq. \ref{eq:IntroductionEq_4} and substituting the integration variables $\vec{r}_{2},\ldots, \vec{r}_{N_{e}}, \vec{R}_{1},\ldots,\vec{R}_{N_{n}}$ by the variables given by Eq. \ref{eq:IntroductionEq_3}, we immediately realize that the density $n_{s}\left(\vec{r}_{1}\right)$ does not depend on $\vec{r}_{1}$. This is true for any atom, molecule or solid and is perfectly reasonable: The Hamiltonian $H$ does not localize the system anywhere in space and hence finding an electron is equally likely everywhere. A similar feature appears for the electronic one-body Green's function when defined with respect to the ground state $\Psi_{0}$ of the Hamiltonian $H$:
\begin{equation} 
G_{gs}\left(\vec{r}t,\vec{r}'t'\right) = -\frac{i}{\hbar} \braket{\Psi_{0}|\mathcal{T}\left\{ \hat{\psi}\left(\vec{r}t\right) \hat{\psi}^{\dagger}\left(\vec{r}'t'\right) \right\}|\Psi_{0}}.
\label{eq:IntroductionEq_5}
\end{equation}
Here $\mathcal{T}$ is the usual time-ordering operator and $\hat{\psi}^{\dagger}\left(\vec{r}'t'\right), \hat{\psi}\left(\vec{r}t\right)$ are electron creation and annihilation operators in the Heisenberg picture, the latter referring to the full Hamiltonian given by Eq. \ref{eq:HamiltonianEq_1}. It is easy to verify that the so-defined Green's function $G_{gs}\left(\vec{r}t,\vec{r}'t'\right)$ only depends on $\vec{r}-\vec{r}'$. Again this feature holds true for all atoms, molecules and solids. This property of the Green's function, and likewise the constancy of the probability density, are consequences of the translational invariance of the underlying Hamiltonian $H$ and are, as such, not at all surprising. If one wants to develop a density functional theory or a Green's-function-based many-body theory for the complete system of electrons and nuclei then, clearly, the probability density of Eq. \ref{eq:IntroductionEq_4} and the Green's function of Eq. \ref{eq:IntroductionEq_5} are not useful quantities because they do not reveal any features characteristic of the system, for example the molecule, to be described. We need to find other densities or Green's functions on which the theory can be built, densities or Green's functions that reflect the internal features of the molecule or crystal.

Eigenfunctions do not necessarily have the same symmetry as the Hamiltonian they come from, they may have lower symmetry. In our case, having the eigenstates $\Psi_{s}$ to be simultaneous eigenstates of the total momentum was a choice we made, and it is exactly this choice that produced the constant density. However, it is important to realize that we can always separate off the center-of-mass motion, and hence the eigenstates of the Hamiltonian given by Eq. \ref{eq:HamiltonianEq_1} can always be written as
\begin{equation} 
\Psi_{s} = \Phi_{cm}\left(\vec{R}_{cm}\right) \Phi_{s}\left(\tilde{\vec{r}}_{2},\ldots,\tilde{\vec{r}}_{N_{e}},\tilde{\vec{R}}_{1},\ldots,\tilde{\vec{R}}_{N_{n}}\right),
\label{eq:IntroductionEq_6}
\end{equation}
where $\Phi_{cm}$ satisfies the equation
\begin{equation} 
- \frac{ \hbar^{2} }{2 M } \nabla^{2}_{\vec{R}_{cm}} \Phi_{cm}\left(\vec{R}_{cm}\right) = E_{cm} \Phi_{cm}\left(\vec{R}_{cm}\right),
\label{eq:IntroductionEq_7}
\end{equation}
where $M$ is the total mass of electrons and nuclei. Equation \ref{eq:IntroductionEq_2} was a particular solution of Eq. \ref{eq:IntroductionEq_7} that preserves the translational symmetry of the Hamiltonian. But other solutions can be chosen, for instance angular-momentum eigenstates (i.e. partial waves) of the form
\begin{equation} 
\Phi_{cm}\left(\vec{R}_{cm}\right) = j_{l}\left( R_{cm} \right) Y_{lm}\left(\vartheta,\varphi\right),
\label{eq:IntroductionEq_8}
\end{equation}
as well as linear combinations of these functions with the same absolute total momentum $P_{s}$, as these span the subspace of degenerate eigenfunctions with energy $E_{cm} = P^{2}_{cm} /2 M$. Clearly, an eigenstate of the form \ref{eq:IntroductionEq_6} with $\Phi_{cm}$ being of the form \ref{eq:IntroductionEq_8} will not have a constant probability density $n_{s}\left(\vec{r}_{1}\right)$. There will be a genuine $\vec{r}_{1}$-dependence. However, this $\vec{r}_{1}$-dependence reflects the shape of the angular-momentum eigenstate \ref{eq:IntroductionEq_8} of the center of mass but, again, will not be characteristic of the internal structure of the molecule. So, for building a density functional theory or Green's function theory, such choice will not be helpful. Alternatively, one might consider the possibility of localizing the whole system with an external confining potential. While this will avoid a constant density, the shape of $n_{s}\left(\vec{r}_{1}\right)$ will reflect the shape of the confining potential but will not reveal the internal structure of the molecule. So, once again, this choice is not suitable for constructing a density functional or Green's function approach to the complete system of electrons and nuclei.

There are essentially two distinctly different choices one can make for the density or Green's function that are useful in constructing a density-functional or Green's function approach: One either refers the coordinates on which the density/Green's function depends to a body-fixed coordinate frame, or one works with conditional probabilities. The second option, i.e. the choice of conditional probabilities, is taken in the BO framework: The standard BO-based Green's function for electrons is defined in terms of an electronic many-body wave function $\Phi_{\vec{R}}\left( \vec{r}_{1},\ldots,\vec{r}_{N_{e}} \right)$ which is a conditional probability amplitude for finding the electrons at $\vec{r}_{1},\ldots,\vec{r}_{N_{e}}$, given the nuclei are at $\vec{R}$. Likewise, the standard Hohenberg-Kohn-Sham ground-state density functional theory is based on the conditional probability density $n_{\vec{R}}\left(\vec{r}_{1}\right)$ of finding an electron at $\vec{r}_{1}$, given the nuclei are at $\vec{R}$. Using the framework of the exact factorization \cite{Abedi-ExactFactorization-PhysRevLett.105.123002-2010,Abedi-CorrelatedElectronNuclearDynamicsEF-2012}, a density functional framework beyond the BO limit has been developed \cite{Gidopoulos-Gross-ElectronicNonAdiabaticStates-2014,Requist-ExactFactorBasedDFTofElectronsAndNuclei-PhysRevLett.117.193001-2016,Li-DFTofElectronTransferBeyondTheBOapproximationCaseStudyOfLiF-2018,Requist-ExactFactorizationBasedDFTofElectronPhononSystems-PhysRevB.99.165136-2019} on the basis of the exact (rather than BO) conditional density. A Green's function approach based on the exact factorization has not been attempted so far.

In this article we follow the first option: Our goal is to develop a Green's function-based many-body theory for the complete system of electrons and nuclei where the Green's functions are defined in terms of coordinates that refer to a body-fixed coordinate frame. First steps in this direction were taken with the formulation of a multicomponent density-functional theory \cite{Kreibich-MulticompDFTForElectronsAndNuclei-PhysRevLett.86.2984-2001,Kreibich-MulticompDFTForElectronsAndNuclei-PhysRevA.78.022501-2008,Butriy-MulticomponentDFTforTimeDependentSystems-PhysRevA.76.052514-2007} and with the derivation of a Green's-function framework \cite{vanLeeuwen-FirstPrincElectronPhonon-PhysRevB.69.115110-2004}. In the latter, the electronic degrees of freedom are treated in a fully consistent way. However, no determining equation for the nuclear Green's function was derived. A nuclear density-density correlation function must be imported from outside the theory. Furthermore, the earlier work considered only crystalline solids excluding some important terms when the theory is imposed in the study of molecules, for example. The purpose of the present article is to deduce, from the general Hamiltonian of Eq. \ref{eq:HamiltonianEq_1}, a coupled set of self-consistent equations for the electronic and nuclear Green's functions that does not require any input from outside.

This paper is organized as follows. In Sec. \ref{cha:Hamiltonian} we discuss in detail the coordinate transformation to the body-fixed frame and how the Hamiltonian looks when expressed in terms of these coordinates. The equation of motion (EOM) for the electronic Green's functions are presented in Sec \ref{cha:EquationsOfMotionForElectrons}, and for the nuclear Green's function is Sec. \ref{cha:EquationsOfMotionForNuclei}. Hedin-like equations for the electrons are deduced in Sec. \ref{cha:HedinsEquations}. We discuss how to determine some parameters related to the coordinate transformations in Sec. \ref{cha:ChoiceOfReferencePositions}. The general normal mode frequencies are derived in Sec. \ref{cha:NucleiVibrations} and an expression for the nuclear self-energy (SE) due to the Coulombic interactions to the lowest order is given in Sec. \ref{cha:NucleiSelfEnergy}. The special case of crystalline solids is considered in Sec. \ref{cha:CrystallineSolids}. Phonons and their interactions are discussed in Sec. \ref{cha:PhononsAndTheirInteractions}.

\section{Coordinate transformation and the transformed Hamiltonian}
\label{cha:Hamiltonian}

\subsection{Hamiltonian}

As demonstrated above, the density $n_{s}\left(\vec{r}_{1}\right)$ is constant for the wave function given by Eq. \ref{eq:IntroductionEq_2} if $\vec{r}_{1}$ denotes a coordinate vector in the laboratory frame. By contrast, the density $n_{s}\left(\tilde{\vec{r}}_{1}\right)$ with $\tilde{\vec{r}}_{1}$ being the position measured from the center of mass is not a constant and reveals to some extent the internal structure of the system. However, the density $n_{s}\left(\tilde{\vec{r}}_{1}\right)$ is not good enough yet because it is spherical for all atoms, molecules and solids. Clearly, we still need to deal with the rotational degrees of freedom. This is achieved by the following transformation \cite{Sutcliffe-TheDecouplingOfElectronicAndNuclearMotions-2000,Sutcliffe-CoordinateSystemsAndTransformations-2003,Kreibich-MulticompDFTForElectronsAndNuclei-PhysRevLett.86.2984-2001,vanLeeuwen-FirstPrincElectronPhonon-PhysRevB.69.115110-2004,Kreibich-MulticompDFTForElectronsAndNuclei-PhysRevA.78.022501-2008}:
\begin{eqnarray} 
\vec{r}'_{i} &=& \mathcal{R}\left(\boldsymbol{\theta}\right) \left( \vec{r}_{i} - \vec{R}_{cmn} \right), \quad i = 1,\ldots,N_{e}, \nonumber \\
\vec{R}'_{k} &=& \vec{R}_{k} - \vec{R}_{cmn}, \quad k = 1,\ldots,N_{n}-1, \nonumber \\
\vec{R}''_{N_{n}} &=& \vec{R}_{cm},
\label{eq:HamiltonianEq_4}
\end{eqnarray}
where $\mathcal{R} = \mathcal{R}\left(\boldsymbol{\theta}\right)$ is a rotation matrix and $\boldsymbol{\theta} = \left(\theta_{1},\theta_{2},\theta_{3}\right)$ is the associated vector of Euler angles which, for now, are assumed to be functions of all the $N_{n}-1$ nuclear variables denoted by $\vec{R}'$. We used $\vec{R}''_{N_{n}}$ for one of the transformed coordinates in order to keep our notation consistent in writing the transformed Coulomb potential terms. Furthermore, $\vec{R}_{cmn}$ is the nuclear center of mass given by
\begin{equation} 
\vec{R}_{cmn} = \frac{1}{ M_{nuc} } \sum^{ N_{n} }_{ k = 1 } M_{k} \vec{R}_{k}, \quad M_{nuc} \equiv \sum^{ N_{n} }_{ k = 1 } M_{k}.
\label{eq:HamiltonianEq_5}
\end{equation}
We can think of the coordinate transformation given by Eq. \ref{eq:HamiltonianEq_4} as two sequential steps. First, the coordinates are written relative to the nuclear center of mass and the center of mass of the total system is separated. Secondly, the rotation of the electronic coordinates shown in Eq. \ref{eq:HamiltonianEq_4} is established to fix an orientation of the electronic subsystem relative to the nuclear center-of-mass frame. Such body-fixed-frame transformations are also used in the BO approximation when it is formulated for molecules \cite{Born-Oppenheimer-Adiabatic-Approx.1927,Scivetti-GeneralLocalAndRectilVibrCoordConsistWithEckartsCond-PhysRevA.79.032516-2009}. In terms of the new coordinates given by Eq. \ref{eq:HamiltonianEq_4}, the Hamiltonian reads
\begin{eqnarray}
H &=& T_{cm} + T'_{e} + T'_{n} + T'_{mpe}  + T'_{mpn}  + T_{cvr} + T'_{cvr} \nonumber \\
&&+ V'_{ee} + V'_{en} + V'_{nn} + V_{ext}.
\label{eq:HamiltonianEq_6_3}
\end{eqnarray}
Here $V_{ext}$ is an external potential added to the Hamiltonian (not originating from the coordinate transformation) which is introduced in order to use the functional derivative techniques \cite{Schwinger-OnTheGreensFunctionsOfQuantizedFieldsI-1951} when we derive the EOM. We will specify its explicit form later. After the EOM are obtained we put $V_{ext} \equiv 0$. The transformed kinetic energies in Eq. \ref{eq:HamiltonianEq_6_3} are
\begin{eqnarray}
T_{cm} &=& - \frac{ \hbar^{2}  }{ 2 M }   \nabla^{2}_{\vec{R}_{cm}}, \nonumber \\
T'_{e} &=& - \frac{ \hbar^{2} }{ 2 m_{e} } \sum^{N_{e}}_{i = 1}  \nabla^{2}_{\vec{r}'_{i}}, \quad \nonumber \\
T'_{n} &=& - \sum^{N_{n}-1}_{k = 1}  \frac{ \hbar^{2} }{ 2 M_{k} } \nabla^{2}_{\vec{R}'_{k}}, \nonumber \\
T'_{mpe} &=& - \frac{ \hbar^{2} }{ 2 M_{nuc} } \sum^{N_{e}}_{i,j = 1} \nabla_{\vec{r}'_{i}} \cdot \nabla_{\vec{r}'_{j}}, \nonumber \\
T'_{mpn} &=& \frac{ \hbar^{2} }{ 2 M_{nuc} } \sum^{N_{n}-1}_{k,k' = 1} \nabla_{\vec{R}'_{k}} \cdot \nabla_{\vec{R}'_{k'}},
\label{eq:HamiltonianEq_7}
\end{eqnarray}
where the total mass is $M = M_{nuc} + m_{e} N_{e}$ and the transformed Coulomb potential terms are
\begin{eqnarray} 
V'_{ee} &=& \sum^{ N_{e} }_{ i,j= 1 }{}^{'} v_{e}\left(\vec{r}'_{i}, \vec{r}'_{j} \right), \nonumber \\
V'_{nn} &=& \sum^{N_{n}}_{ k,k'= 1 }{}^{'} v_{n}\left(\vec{R}'_{k}, \vec{R}'_{k'} \right), \nonumber \\
V'_{en} &=& \sum^{ N_{e} }_{ i = 1 } \sum^{N_{n}}_{ k = 1 } v_{en}\left(\vec{r}'_{i}, \mathcal{R} \vec{R}'_{k} \right), \nonumber \\
\vec{R}'_{N_{n}} &\equiv& -\frac{1}{ M_{N_{n}} } \sum^{N_{n}-1}_{ k= 1 } M_{k} \vec{R}'_{k}.
\label{eq:HamiltonianEq_8}
\end{eqnarray}
In Eq. \ref{eq:HamiltonianEq_6_3}, $T_{cvr}$ and $T'_{cvr}$ include the Coriolis and rotational-vibrational coupling terms. The explicit form of these quantities is given in Appendix \ref{CoriolisAndVibrationalRotationalCouplingTerms}. The center-of-mass kinetic energy $T_{cm}$ commutes with all the other terms in the Hamiltonian and does not enter to the EOM and, hence, can be disregarded without loss of generality. There are only $N_{n}-1$ primed nuclear coordinates appearing in the Hamiltonian. However, the number of degrees of freedom is still the same as before since one of the coordinates is the total center-of-mass coordinate of the system. The potential terms $V'_{nn}$ and $V'_{en}$ involving $\vec{R}'_{N_{n}}$ are no longer translation invariant in the new coordinates. The mass-polarization terms $T'_{mpe}$ and $T'_{mpn}$ are expected to be rather small in the case of crystals and larger molecules since they are proportional to the inverse of the total nuclear mass.

Next we perform yet another transformation of the nuclear coordinates and write the position coordinates as a sum of reference positions $\vec{x}_{k}$ (which act as parameters) and displacements $\vec{u}_{k}$ (quantum variables), namely
\begin{equation} 
\vec{R}'_{k} = \vec{x}_{k} + \vec{u}_{k}, \quad k = 1,\ldots, N_{n}-1.
\label{eq:HamiltonianEq_9}
\end{equation}
We will consider the determination of the reference positions $\vec{x}_{k}$ in detail in Sec. \ref{cha:ChoiceOfReferencePositions}. For now these are taken as arbitrary real parameters. Depending on the situation, we use the following notations interchangeably $\vec{u}_{k} = \vec{u}\left(k\right)$ for the displacements and, analogously, for $\vec{R}'_{k}$ and $\vec{x}_{k}$. This allows us to denote the $\alpha$th Cartesian components of $\vec{x}_{k}$ and $\vec{u}_{k}$ by $x_{\alpha}\left(k\right)$ and $u_{\alpha}\left(k\right)$. Further, we sometimes denote all the $N_{n}-1$ nuclear reference positions and displacements by $\vec{x}$ and $\vec{u}$, respectively. We emphasize that the theory is still exact after the transformation given by Eq. \ref{eq:HamiltonianEq_9}. The position variables in the Hamiltonian $H$ transform as Eq. \ref{eq:HamiltonianEq_9} shows and the derivatives in the kinetic energy terms (Eq. \ref{eq:HamiltonianEq_7}) transform as $\nabla_{\vec{R}'_{k}} \rightarrow \nabla_{\vec{u}_{k}}$. We now have all the necessary prerequisites in place to write our Hamiltonian. We choose to write the electronic parts of the Hamiltonian in second quantization. This can be done in a similar way as in the laboratory frame formulation since the transformed Hamiltonian has the same permutation symmetry with respect to the electronic variables as the original one. With respect to the nuclear coordinates $\vec{R}'$, the transformed Hamiltonian does not necessarily have the same permutation symmetry as the original one (for identical nuclei). Therefore, the nuclear part is kept in first quantization. Note that the transformed Hamiltonian only contains $N_{n} - 1$ independent nuclear coordinate and the total (electron-nuclear) center of mass. After using Eq. \ref{eq:HamiltonianEq_9} and ignoring the center-of-mass kinetic energy, we obtain the final form of the Hamiltonian
\begin{equation} 
\hat{H} = \hat{T}_{tot} + \hat{V}'_{ee} + \hat{V}'_{en} + \hat{V}'_{nn} + \hat{V}_{ext},
\label{eq:HamiltonianEq_10}
\end{equation}
where
\begin{equation} 
\hat{T}_{tot} = \hat{T}'_{e} + \hat{T}'_{n}  + \hat{T}'_{mpe} + \hat{T}'_{mpn} + \hat{T}_{cvr} + \hat{T}'_{cvr}.
\label{eq:HamiltonianEq_11}
\end{equation}
Here $\hat{T}_{cvr}$ and $\hat{T}'_{cvr}$ are given in Appendix \ref{CoriolisAndVibrationalRotationalCouplingTerms} and the other kinetic energy terms are
\begin{eqnarray}
\hat{T}'_{n} &=& \sum^{N_{n}-1}_{k = 1}  \frac{ \hat{\vec{p}}_{k} \cdot \hat{\vec{p}}_{k} }{ 2 M_{k} }, \nonumber \\ 
\hat{T}'_{mpn} &=& - \sum^{N_{n}-1}_{k,k' = 1} \frac{ \hat{\vec{p}}_{k} \cdot \hat{\vec{p}}_{k'} }{ 2 M_{nuc} }, \nonumber \\ 
\hat{T}'_{e}  &=& - \frac{ \hbar^{2} }{ 2 m_{e} } \int d\vec{r} \hat{\psi}^{\dagger}\left(\vec{r}\right)  \nabla^{2}_{\vec{r}} \hat{\psi}\left(\vec{r}\right), \nonumber \\
\hat{T}'_{mpe} &=& - \frac{ \hbar^{2} }{ 2 M_{nuc} } \int d\vec{r} \int d\vec{r}' \hat{\psi}^{\dagger}\left(\vec{r}\right) \hat{\psi}^{\dagger}\left(\vec{r}'\right)  \nabla_{\vec{r}} \cdot \nabla_{\vec{r}'} \nonumber \\ 
&&\times \hat{\psi}\left(\vec{r}'\right) \hat{\psi}\left(\vec{r}\right).
\label{eq:HamiltonianEq_13}
\end{eqnarray}
The energy terms associated with the particle-particle interactions are
\begin{eqnarray} 
\hat{V}'_{ee} &=& \frac{1}{2} \int d\vec{r} \int d\vec{r}' v\left(\vec{r},\vec{r}'\right) \hat{\psi}^{\dagger}\left(\vec{r}\right) \hat{\psi}^{\dagger}\left(\vec{r}'\right) \hat{\psi}\left(\vec{r}'\right) \hat{\psi}\left(\vec{r}\right), \nonumber \\
\hat{V}'_{en} &=& \int d\vec{r} \int d\vec{r}'  v\left(\vec{r}, \vec{r}'\right) \hat{n}_{n}\left(\vec{r}'\right) \hat{n}_{e}\left(\vec{r}\right) \nonumber \\
&=& \int d\vec{r} \hat{V}_{en}\left(\vec{r}\right) \hat{n}_{e}\left(\vec{r}\right), \nonumber \\
\hat{V}'_{nn} &=& \sum^{N_{n}}_{ k,k'= 1 }{}^{'} v_{n}\left(\vec{x}_{k} + \hat{\vec{u}}_{k}, \vec{x}_{k'} + \hat{\vec{u}}_{k'} \right),
\label{eq:HamiltonianEq_14}
\end{eqnarray}
and the external potential part is
\begin{eqnarray} 
\hat{V}_{ext} &=& \int d\vec{r} U\left(\vec{r}t\right) \left[ \hat{n}_{e}\left(\vec{r}\right) + \hat{n}_{n}\left(\vec{r}\right) \right] \nonumber \\
&&+ \int d\vec{r} U'\left(\vec{r}t\right) \hat{n}_{n}\left(\vec{r}\right) + \sum^{N_{n}-1}_{k = 1} \vec{J}\left(kt\right) \cdot \hat{\vec{u}}_{k} \nonumber \\
&&+ \int d\vec{r} \varphi\left(\vec{r}t\right) \hat{n}_{e}\left(\vec{r}\right),
\label{eq:HamiltonianEq_15}
\end{eqnarray}
where we use $\hat{n}_{e}\left(\vec{r}\right) = \hat{\psi}^{\dagger}\left(\vec{r}\right) \hat{\psi}\left(\vec{r}\right)$. Furthermore, the quantities $U\left(\vec{r}t\right)$, $U'\left(\vec{r}t\right)$, $\vec{J}\left(kt\right)$ and $\varphi\left(\vec{r}t\right)$ are time-dependent external potentials. In these equations $v\left(  \vec{r}, \vec{r}' \right) =\varsigma / \left| \vec{r} - \vec{r}' \right|$ and
\begin{eqnarray} 
\hat{n}_{n}\left(\vec{r}\right) &\equiv & - \sum^{N_{n}}_{ k = 1 } Z_{k} \delta\left( \vec{r} - \mathcal{R}\left[\boldsymbol{\theta}\left(\hat{\vec{R}}'\right)\right] \hat{\vec{R}}'_{k} \right), \nonumber \\
\hat{V}_{en}\left(\vec{r}\right) &=& \sum^{N_{n}}_{ k= 1 } \frac{ - Z_{k} \varsigma }{ \left| \vec{r} - \mathcal{R}\left[\boldsymbol{\theta}\left(\hat{\vec{R}}'\right)\right] \hat{\vec{R}}'_{k} \right| }.
\label{eq:HamiltonianEq_16}
\end{eqnarray}
The operators $\hat{u}_{\alpha}\left(k\right)$, $\hat{p}_{\alpha'}\left(k'\right)$, $\hat{\psi}^{\dagger}\left(\vec{r}\right)$ and $\hat{\psi}\left(\vec{r}\right)$ satisfy the following commutation and anti-commutation relations
\begin{eqnarray}
\left[\hat{u}_{\alpha}\left(k\right),\hat{p}_{\alpha'}\left(k'\right)\right]_{-} &=& i \hbar \delta_{\alpha\alpha'}\delta_{kk'}, \nonumber \\
\left[\hat{\psi}\left(\vec{r}\right),\hat{\psi}^{\dagger}\left(\vec{r}'\right)\right]_{+} &=& \delta\left(\vec{r} - \vec{r}'\right).
\label{eq:HamiltonianEq_17}
\end{eqnarray}

So far the Euler angles in Eq. \ref{eq:HamiltonianEq_4} were assumed to be generic functions of the nuclear coordinates $\vec{R}'$. We have not introduced any defining relations for them. There are many ways to choose these angles. Without loss of generality we assume that the Euler angles are defined through an implicit equation of the form
\begin{equation}
\vec{F}\left(\boldsymbol{\theta},\vec{R}'\right) = \vec{0}.
\label{eq:HamiltonianEq_8_1}
\end{equation}
For example, one possible form of Eq. \ref{eq:HamiltonianEq_8_1} is the Eckart condition which can be written as \cite{Eckart-SomeStudiesConcerningRotatingAxesAndPolyatomicMolecules-PhysRev.47.552-1935,Wilson-MolecularVibrationsTheTheoryOfInfraredAndRamanVibrationalSpectra-1955,Littlejohn-GaugeFieldsInTheSeparOfRotatAndIntMotionsIntheNbodyProb-RevModPhys.69.213-1997,Sutcliffe-TheDecouplingOfElectronicAndNuclearMotions-2000,vanLeeuwen-FirstPrincElectronPhonon-PhysRevB.69.115110-2004}
\begin{equation}
\sum^{ N_{n} - 1 }_{k = 1} M_{k} \vec{x}_{k} \times \mathcal{R} \vec{R}'_{k} = \vec{0}.
\label{eq:HamiltonianEq_8_2}
\end{equation}
The relations given by Eqs. \ref{eq:HamiltonianEq_8_1} and \ref{eq:HamiltonianEq_8_2} define the Euler angles and thus the rotation matrix as a function of the $N_{n} - 1$ nuclear variables, $\mathcal{R} = \mathcal{R}\left[\boldsymbol{\theta}\left(\vec{R}'\right)\right]$. Some implicit conditions, like the Eckart condition, can change the permutation symmetry present in the original Hamiltonian \cite{Sutcliffe-TheDecouplingOfElectronicAndNuclearMotions-2000,vanLeeuwen-FirstPrincElectronPhonon-PhysRevB.69.115110-2004} with respect to the old $\vec{R}$ and new nuclear variables $\vec{R}'$ leading to possible complications when (anti)symmetrizing the nuclear subsystem(s) properly. However, the implicit condition does not change the canonical commutation relations (Eq. \ref{eq:HamiltonianEq_17}) between the nuclear position and momentum operators. This means that the EOM for the electron and nuclear Green's functions remain the same whether or not the implicit condition changes the permutation symmetry of the Hamiltonian with respect to the nuclear variables. In the case in which the implicit condition changes the aforementioned permutation symmetries and it turns out to be difficult to apply the theory because of the complicated symmetry, we may assume, as in the previous works \cite{Baym-field-1961,Giustino-ElectronPhononInteractFromFirstPrinc-RevModPhys.89.015003-2017}, that the nuclei are distinguishable. By doing so we avoid the problems potentially caused by a more complicated permutation symmetry. There are few physical effects in which the permutation symmetry of the nuclei is crucial. Some implicit conditions are not suitable for describing arbitrary systems, for example, in the case of linear molecules, the Eckart conditions are not well defined \cite{Eckart-SomeStudiesConcerningRotatingAxesAndPolyatomicMolecules-PhysRev.47.552-1935,Sayvetz-TheKineticEnergyOfPolyatomicMolecules-1939}. Thus, while our theory is general, the specific choice of $\vec{F}\left(\boldsymbol{\theta},\vec{R}'\right)$ in Eq. \ref{eq:HamiltonianEq_8_1} may restrict the range of applicability of the theory.

Up to this point, we have not introduced any approximations. We have set up the Hamiltonian to work with and derive the EOM for the electronic and nuclear Green's functions by using it in Secs. \ref{cha:EquationsOfMotionForElectrons} and \ref{cha:EquationsOfMotionForNuclei}. Before that, we discuss in more detail how to actually obtain the Euler angles appearing in the Hamiltonian.

\subsection{Euler angles}
\label{EulerAngles}

The condition given by Eq. \ref{eq:HamiltonianEq_8_1} defines the Euler angles by an implicit equation of the form $\vec{F}\left(\boldsymbol{\theta},\vec{x}+ \vec{u}\right) = \vec{0}$. Explicit solutions of such equations do not always exist \cite{Guzman-DerivativesAndIntegralsOfMultivariableFunctions-2003,Krantz-TheImplicitFunctionTheoremHistoryTheoryAndApplications-2013,Krasnoshchekov-DetermOfTheEckartMolecFixedFrameByUseOfTheApparatusOfQuaternionAlgebra-2014}, that is, it may be impossible to write $\boldsymbol{\theta}$ as an explicit function of the nuclear variables $\vec{x},\vec{u}$. However, for given $\vec{x},\vec{u}$ and given nuclear masses, a numerical solution can always be obtained \cite{Krasnoshchekov-DetermOfTheEckartMolecFixedFrameByUseOfTheApparatusOfQuaternionAlgebra-2014}. We are particularly interested in how to obtain the rotation matrix $\mathcal{R}\left[\boldsymbol{\theta}\left(\vec{x}\right)\right]$ and its derivatives. These quantities appear in the Hamiltonian when we do the Taylor expansions of some parts of the Hamiltonian (see Appendix \ref{TotalEnergy}) and thus they will ultimately appear in the EOM.

The rotation matrix $\mathcal{R}\left[\boldsymbol{\theta}\left(\vec{x} + \vec{u}\right)\right]$ can be obtained from a generic condition given by Eq. \ref{eq:HamiltonianEq_8_1}. In turn, $\mathcal{R}\left[\boldsymbol{\theta}\left(\vec{x}\right)\right]$ can be obtained by solving Eq. \ref{eq:HamiltonianEq_8_1} with $\vec{u}_{k} = \vec{0}$ for all $k$. Thus, $\mathcal{R}\left[\boldsymbol{\theta}\left(\vec{x}\right)\right]$ is obtained from the implicit equation $\vec{F}\left(\boldsymbol{\theta},\vec{x}\right) = \vec{0}$. For instance, in the case of Eckart conditions given by Eq. \ref{eq:HamiltonianEq_8_2}
\begin{equation}
\sum^{ N_{n} - 1 }_{k = 1} M_{k} \vec{x}_{k} \times \mathcal{R}\left(\boldsymbol{\theta}\right) \vec{x}_{k}  = \vec{0},
\label{eq:EulerAnglesEq_2}
\end{equation}
with the Euler angles $\boldsymbol{\theta}\left(\vec{x}\right)$, such that $0 < \theta_{\beta}\left(\vec{x}\right) < 2 \pi$ for $\beta = 1,2,3$. What we need next are the derivatives of $\mathcal{R}\left[\boldsymbol{\theta}\left(\vec{x}\right)\right]$ appearing in the Hamiltonian (see Eqs. \ref{eq:UsefulRelationsEq_9} and \ref{eq:UsefulRelationsEq_10} of Appendix \ref{TotalEnergy}). By the chain rule, we write
\begin{eqnarray}
\frac{\partial{ \mathcal{R} }}{\partial{x_{\alpha}\left(k\right)}} &=& \sum^{3}_{\beta = 1} \frac{\partial{ \mathcal{R} }}{\partial{ \theta_{\beta} }} \frac{\partial{ \theta_{\beta} }}{\partial{x_{\alpha}\left(k\right)}}, \nonumber \\
\frac{\partial^{2}{ \mathcal{R} }}{\partial{x_{\beta}\left(k'\right)} \partial{x_{\alpha}\left(k\right)}} &=& \sum^{3}_{\beta',\beta'' = 1} \frac{\partial^{2}{ \mathcal{R} }}{\partial{ \theta_{\beta'} } \partial{ \theta_{\beta''} } } \frac{\partial{ \theta_{\beta''} }}{\partial{x_{\beta}\left(k'\right)}} \frac{\partial{ \theta_{\beta'} }}{\partial{x_{\alpha}\left(k\right)}} \nonumber \\
&&+ \sum^{3}_{\beta' = 1} \frac{\partial{ \mathcal{R} }}{\partial{ \theta_{\beta'} }} \frac{\partial^{2}{ \theta_{\beta'} }}{\partial{x_{\alpha}\left(k\right)} \partial{x_{\beta}\left(k'\right)} }.
\label{eq:EulerAnglesEq_4}
\end{eqnarray}
Given the explicit form for the rotation matrix as function of the Euler angles we can calculate the derivatives like $\partial{ \mathcal{R} }/\partial{ \theta_{\beta'} }$. Next we calculate the derivatives $\partial{ \theta_{\beta'} }/\partial{x_{\alpha}\left(k\right)}$ and $\partial^{2}{ \theta_{\beta'} }/\partial{x_{\alpha}\left(k\right)} \partial{x_{\beta}\left(k'\right)}$, etc. After taking the total derivative of $F_{\sigma}\left(\boldsymbol{\theta},\vec{x},\vec{u}\right)$ with respect to $x_{\alpha}\left(k\right)$ we write
\begin{equation}
\frac{dF_{\sigma}}{dx_{\alpha}\left(k\right)} = \sum^{3}_{\beta' = 1} A_{\sigma \beta'} \frac{\partial{ \theta_{\beta'} }}{\partial{x_{\alpha}\left(k\right)}} + \frac{\partial{ F_{\sigma} }}{\partial{x_{\alpha}\left(k\right)}}  = 0,
\label{eq:EulerAnglesEq_6}
\end{equation}
where $A_{\sigma \beta'} \equiv \partial{ F_{\sigma} } / \partial{ \theta_{\beta'} }$. Suppose that the matrix $A$ is invertible such that
\begin{equation}
\sum^{3}_{\sigma = 1} A^{-1}_{\gamma \sigma } A_{\sigma \beta'} = \delta_{\gamma \beta'}.
\label{eq:EulerAnglesEq_8}
\end{equation}
By multiplying Eq. \ref{eq:EulerAnglesEq_6} with $A^{-1}_{\gamma \sigma }$, taking a sum over $\sigma$ and re-arranging, we obtain
\begin{equation}
\frac{\partial{ \theta_{\gamma} }}{\partial{x_{\alpha}\left(k\right)}} = - \sum^{3}_{\sigma = 1} A^{-1}_{\gamma \sigma } \frac{\partial{ F_{\sigma} }}{\partial{x_{\alpha}\left(k\right)}}.
\label{eq:EulerAnglesEq_9}
\end{equation}
The result given by Eq. \ref{eq:EulerAnglesEq_9} is essentially included in the implicit function theorem \cite{deOliveira-TheImplicitAndInverseFunctionTheoremsEasyProofs-2013}. For the second order derivative, we take the total derivative of Eq. \ref{eq:EulerAnglesEq_9} with respect to $x_{\beta}\left(k'\right)$ and after some algebra find that
\begin{eqnarray}
&&\frac{\partial^{2}{ \theta_{\gamma} }}{\partial{x_{\alpha}\left(k\right)} \partial{x_{\beta}\left(k'\right)} } \nonumber \\
&&= - \sum^{3}_{\sigma = 1} A^{-1}_{\gamma \sigma} \frac{\partial^{2}{ F_{\sigma} }}{\partial{x_{\alpha}\left(k\right)} \partial{x_{\beta}\left(k'\right)} } - \sum^{3}_{\beta',\sigma = 1} A^{-1}_{\gamma \sigma} \frac{\partial{ \theta_{\beta'} }}{\partial{x_{\alpha}\left(k\right)}} \nonumber \\
&&\ \ \ \times \left[ \sum^{3}_{\beta'' = 1} \frac{\partial^{2}{ F_{\sigma} }}{ \partial{ \theta_{\beta'} } \partial{ \theta_{\beta''} } } \frac{\partial{ \theta_{\beta''} }}{\partial{x_{\beta}\left(k'\right)}} + \frac{\partial^{2}{ F_{\sigma} }}{ \partial{ \theta_{\beta'} } \partial{x_{\beta}\left(k'\right)} } \right]  \nonumber \\
&& \ \ \  - \sum^{3}_{\beta'',\sigma = 1} A^{-1}_{\gamma \sigma} \frac{\partial^{2}{ F_{\sigma} }}{ \partial{x_{\alpha}\left(k\right)} \partial{ \theta_{\beta''} } } \frac{\partial{ \theta_{\beta''} }}{\partial{x_{\beta}\left(k'\right)}}.
\label{eq:EulerAnglesEq_10}
\end{eqnarray}
All the quantities in Eqs. \ref{eq:EulerAnglesEq_9} and \ref{eq:EulerAnglesEq_10} are to be evaluated at $\boldsymbol{\theta} = \boldsymbol{\theta}\left(\vec{x}\right), \vec{u}_{k} = \vec{0}$ for all $k$ when applied in solving the relations of Eq. \ref{eq:EulerAnglesEq_4} aiming to evaluate $\mathcal{R}\left[\boldsymbol{\theta}\left(\vec{x}\right)\right]$. For instance, when the condition $F_{\sigma}\left(\boldsymbol{\theta},\vec{x},\vec{u}\right)$ is the Eckart condition given by Eq. \ref{eq:HamiltonianEq_8_2}, we have
\begin{eqnarray}
\frac{\partial{ F_{\sigma} }}{\partial{ \theta_{\beta'} }} &=& \sum^{ N_{n} - 1 }_{k = 1} M_{k} \sum^{3}_{\eta,\nu = 1 } \epsilon_{\sigma\eta\nu} x_{\eta}\left(k\right) \sum^{3}_{\beta = 1 } \frac{\partial{ \mathcal{R}_{\nu \beta} }}{\partial{ \theta_{\beta'} }}  R'_{\beta}\left(k\right), \nonumber \\
\frac{\partial{ F_{\sigma} }}{ \partial{ x_{\alpha}\left(k\right) } } &=&  M_{k} \sum^{3}_{\nu = 1 } \epsilon_{\sigma\alpha\nu} \sum^{3}_{\beta = 1 }  \mathcal{R}_{\nu \beta}  R'_{\beta}\left(k\right) \nonumber \\
&&+ M_{k} \sum^{3}_{\eta,\nu = 1 } \epsilon_{\sigma\eta\nu} x_{\eta}\left(k\right) \mathcal{R}_{\nu \alpha},
\label{eq:EulerAnglesEq_11}
\end{eqnarray}
and analogously for the derivatives w.r.t. the other components. Here, $\epsilon_{\sigma\eta\nu} $ is the Levi-Civita symbol.

\section{Equation of motion for the electronic Green's Function}
\label{cha:EquationsOfMotionForElectrons}

Here we derive the exact EOM for the electronic Green's function. We denote an ensemble average for an operator $\hat{o}\left(t\right)$ (in the Heisenberg picture) as
\begin{equation} 
\left\langle \hat{o}\left(t\right) \right\rangle = \sum_{n} \braket{\phi_{n}|\hat{\rho} \hat{o}\left(t\right) |\phi_{n}} = \text{Tr}\left[\hat{\rho} \hat{o}\left(t\right) \right],
\label{eq:EquationsOfMotionForElectronsEq_1_1}
\end{equation}
where $\left\{\ket{\phi_{n}}\right\}$ is an orthonormal basis of the electron-nuclear Hilbert space. The density operator is the standard grand canonical statistical operator
\begin{equation} 
\hat{\rho} =  \frac{e^{-\beta \hat{H}^{M} }}{Z}, \quad Z= \text{Tr}\left[e^{-\beta \hat{H}^{M} }\right],
\label{eq:EquationsOfMotionForElectronsEq_1_2}
\end{equation}
where
\begin{equation} 
\hat{H}^{M} = \hat{H} - \mu_{e} \hat{N}_{e}, \quad \hat{N}_{e} = \int d\vec{r} \hat{n}_{e}\left(\vec{r}\right),
\label{eq:EquationsOfMotionForElectronsEq_1_3}
\end{equation}
and $\mu_{e}$ is the chemical potential of the electrons. One has to emphasize that the grand canonical Hamiltonian (Eq. \ref{eq:EquationsOfMotionForElectronsEq_1_3}) refers to a fixed number of nuclei, while for the electrons the chemical potential is fixed (rather than the particle number) \cite{Baym-field-1961}. Here the time variable $t$ could be taken as a variable on a general time-contour such as the real time axis or on a more general time contour such as the Keldysh contour \cite{Keldysh-Contour-1965,Stefanucci-Leeuwen-many-body-book-2013} allowing for a non-equilibrium formulation. This is justifiable since in both cases the EOM are the same \cite{Stefanucci-Leeuwen-many-body-book-2013}. In the following, we assume that the time variables are on the Keldysh contour $\gamma$ and denote the time variables on the contour with $z$.

We start by writing the EOM for the field operator $\hat{\psi}\left(\vec{r}z\right)$ (in the Heisenberg picture, that refers to the full electron-nuclear Hamiltonian), namely
\begin{eqnarray} 
i \hbar \frac{\partial{ \hat{\psi}\left(\vec{r}z\right)} }{\partial{z}} &=& \hat{\mathcal{D}}\left(\vec{r}z\right) \hat{\psi}\left(\vec{r}z\right) \nonumber \\
&&+ \int d\vec{r}'  v\left(\vec{r},\vec{r}'\right) \hat{\psi}^{\dagger}\left(\vec{r}'z\right) \hat{\psi}\left(\vec{r}'z\right) \hat{\psi}\left(\vec{r}z\right) \nonumber \\
&&- \frac{ \hbar^{2} }{ M_{nuc} }  \int d\vec{r}'  \hat{\psi}^{\dagger}\left(\vec{r}'z\right) \nabla_{\vec{r}} \cdot \nabla_{\vec{r}'} \hat{\psi}\left(\vec{r}'z\right) \hat{\psi}\left(\vec{r}z\right) \nonumber \\
&&+ 2 \sum_{ \beta,\sigma,\gamma,\gamma' } \hat{L}^{\left(3\right)}_{\gamma\sigma \gamma' \beta} \int d\vec{r}'   \nonumber \\
&&\times \hat{\psi}^{\dagger}\left(\vec{r}'z\right) r_{\gamma'} r'_{\gamma} \frac{\partial^{2}{ }}{\partial{r'_{\sigma} } \partial{r_{\beta} }} \hat{\psi}\left(\vec{r}'z\right) \hat{\psi}\left(\vec{r}z\right), 
\label{eq:EquationsOfMotionForElectronsEq_1_3_2}
\end{eqnarray}
where
\begin{eqnarray} 
\hat{\mathcal{D}}\left(\vec{r}z\right) &=& -\frac{ \hbar^{2} }{ 2 m } \nabla^{2}_{\vec{r}} +  U\left(\vec{r}z\right) + \varphi\left(\vec{r}z\right)  \nonumber \\
&&+ \int d\vec{r}'  v\left(\vec{r}, \vec{r}'\right) \hat{n}_{n}\left(\vec{r}'z\right) + \sum_{\beta,\gamma } \hat{L}^{\left(1\right)}_{\beta\gamma} r_{\gamma} \frac{\partial{ }}{\partial{r_{\beta} }}  \nonumber \\
&&+ \sum^{N_{n}-1}_{k = 1} \sum_{ \alpha, \beta ,\gamma  } \hat{L}^{\left(2\right)}_{\alpha\beta\gamma}\left(k\right) \hat{p}_{\alpha}\left(kz\right) r_{\gamma} \frac{\partial{ }}{ \partial{r_{\beta} } },
\label{eq:EquationsOfMotionForElectronsEq_1_3_4}
\end{eqnarray}
and
\begin{eqnarray} 
\hat{L}^{\left(1\right)}_{\beta\gamma} &\equiv& \hat{T}^{\left(1\right)}_{\beta\gamma} + \hat{M}^{\left(1\right)}_{\beta\gamma},  \nonumber \\
\hat{L}^{\left(2\right)}_{\alpha\beta\gamma}\left(k\right) &\equiv& \hat{T}^{\left(2\right)}_{\alpha\beta\gamma}\left(k\right) + \hat{M}^{\left(2\right)}_{\alpha\beta\gamma}\left(k\right), \nonumber \\
\hat{L}^{\left(3\right)}_{\gamma\sigma \gamma' \beta} &\equiv& \hat{M}^{\left(3\right)}_{\gamma\sigma \gamma' \beta} + \hat{T}^{\left(3\right)}_{\gamma\sigma \gamma' \beta}.
\label{eq:EquationsOfMotionForElectronsEq_1_3_5}
\end{eqnarray}
The quantities in Eq. \ref{eq:EquationsOfMotionForElectronsEq_1_3_5} originate from the Coriolis and vibrational-rotational coupling terms, see Appendix \ref{CoriolisAndVibrationalRotationalCouplingTerms}. Next we define the electron Green's function as \cite{Martin-SchwingerPhysRev.115.1342-Theory-Many-Part.Syst.-1959,Hedin-NewMethForCalcTheOneParticleGreensFunctWithApplToTheElectronGasProb-PhysRev.139.A796-1965,Stefanucci-Leeuwen-many-body-book-2013,Fetter-Walencka-q-theory-of-many-particle-1971,Mahan-many-particle-1990}
\begin{equation} 
G\left(1,2\right) \equiv - \frac{i}{\hbar} \frac{\text{Tr}\left[ \mathcal{T}_{\gamma} \left\{ e^{-\frac{i}{\hbar} \int_{\gamma} dz' \hat{H}\left(z'\right) } \hat{\psi}\left(1\right) \hat{\psi}^{\dagger}\left(2\right) \right\} \right]}{\text{Tr}\left[ \mathcal{T}_{\gamma} \left\{ e^{-\frac{i}{\hbar} \int_{\gamma} dz' \hat{H}\left(z'\right) } \right\} \right]},
\label{eq:EquationsOfMotionForElectronsEq_1}
\end{equation} 
where $\mathcal{T}_{\gamma}$ is the contour time ordering operator such that
\begin{eqnarray} 
\left\langle  \mathcal{T}_{\gamma} \left\{ \hat{\psi}\left(1\right) \hat{\psi}^{\dagger}\left(2\right) \right\} \right\rangle &=& \theta\left(1-2\right) \left\langle  \hat{\psi}\left(1\right) \hat{\psi}^{\dagger}\left(2\right) \right\rangle \nonumber \\
&&- \theta\left(2-1\right) \left\langle \hat{\psi}^{\dagger}\left(2\right) \hat{\psi}\left(1\right) \right\rangle, \nonumber \\
\label{eq:EquationsOfMotionForElectronsEq_2_1}
\end{eqnarray}
and the Green's function $G\left(1,2\right)$ satisfies the Kubo-Martin-Schwinger boundary conditions \cite{Martin-SchwingerPhysRev.115.1342-Theory-Many-Part.Syst.-1959,Stefanucci-Leeuwen-many-body-book-2013}. Here and from now on, whenever convenient, we use the following shorthand notations
\begin{eqnarray} 
i &\equiv& \vec{r}_{i} z_{i}, \nonumber \\
\theta\left(i-j\right) &\equiv& \theta\left(z_{i}-z_{j}\right), \nonumber \\
\delta\left(i-j\right) &\equiv& \delta\left(z_{i}-z_{j}\right)\delta\left(\vec{r}_{i} - \vec{r}_{j}\right), \nonumber \\
v\left(1,2\right) &\equiv& \delta\left(z_{1} - z_{2}\right) v\left(\vec{r}_{1},\vec{r}_{2}\right), \nonumber \\
\int di &\equiv& \int d\vec{r}_{i}  \int_{\gamma} dz_{i}.
\label{eq:NotationEq_6_1}
\end{eqnarray}
The EOM for the Green's function reads
\begin{equation} 
i \hbar\frac{\partial}{\partial{z_{1}}} G\left(1,2\right) = \left\langle  \mathcal{T}_{\gamma} \left\{ \left[\frac{\partial}{\partial{z_{1}}} \hat{\psi}\left(1\right) \right] \hat{\psi}^{\dagger}\left(2\right) \right\} \right\rangle +  \delta\left(1 - 2\right).
\label{eq:EquationsOfMotionForElectronsEq_2_2}
\end{equation}
After using Eq. \ref{eq:EquationsOfMotionForElectronsEq_1_3_2} in Eq. \ref{eq:EquationsOfMotionForElectronsEq_2_2} we find that
\begin{eqnarray} 
\delta\left(1 - 2\right) &=& \left[ i \hbar\frac{\partial}{\partial{z_{1}}} - Y\left(1\right)\right] G\left(1,2\right) - S\left(1,2\right)  \nonumber \\
&&-  S'\left(1,2\right), 
\label{eq:EquationsOfMotionForElectronsEq_2_3}
\end{eqnarray}
where
\begin{eqnarray} 
S'\left(1,2\right) &\equiv& \frac{1}{i \hbar} \int d3 v\left(1,3\right) \left\langle  \mathcal{T}_{\gamma} \left\{ \hat{n}\left(3\right) \hat{\psi}\left(1\right) \hat{\psi}^{\dagger}\left(2\right) \right\} \right\rangle, \nonumber \\
Y\left(1\right) &\equiv& - \frac{\hbar^{2} }{2 m_{e} } \nabla^{2}_{\vec{r}_{1}}  + U\left(1\right) + \varphi\left(1\right),
\label{eq:EquationsOfMotionForElectronsEq_2_4}
\end{eqnarray}
and the total density is defined by $\hat{n}\left(1\right) \equiv \hat{n}_{e}\left(1\right) + \hat{n}_{n}\left(1\right)$. All the terms originating from the transformed kinetic energies $\hat{T}'_{mpe}, \hat{T}_{cvr}$ and $\hat{T}'_{cvr}$ are included in the quantity $S\left(1,2\right)$ given by
\begin{equation} 
S\left(1,2\right) = \sum^{3}_{ c = 1} S_{c}\left(1,2\right),
\label{eq:EquationsOfMotionForElectronsEq_2_6}
\end{equation}
where $S_{c}\left(1,2\right)$ for each $c$ is given in Sec. \ref{cha:HedinsEquations}. By subtracting the quantity
\begin{equation} 
\int d3  v\left(1,3\right) \left\langle  \hat{n}\left(3\right) \right\rangle G\left(1,2\right),
\label{eq:EquationsOfMotionForElectronsEq_2_7}
\end{equation}
from both sides of Eq. \ref{eq:EquationsOfMotionForElectronsEq_2_3} we find
\begin{eqnarray} 
&&\left[ i \hbar\frac{\partial}{\partial{z_{1}}} + \frac{\hbar^{2} }{2 m_{e}} \nabla^{2}_{\vec{r}_{1}} - V_{tot}\left(1\right) \right] G\left(1,2\right) \nonumber \\
&&= \delta\left(1 - 2\right) + S\left(1,2\right) + \tilde{S}\left(1,2\right), 
\label{eq:EquationsOfMotionForElectronsEq_2_3_2}
\end{eqnarray}
where
\begin{eqnarray} 
\tilde{S}\left(1,2\right) &\equiv& S'\left(1,2\right) - \int d3  v\left(1,3\right) \left\langle  \hat{n}\left(3\right) \right\rangle G\left(1,2\right), \nonumber \\
V_{tot}\left(1\right) &\equiv& \varphi\left(1\right) + U\left(1\right)  + \int d3  v\left(1,3\right) \left\langle  \hat{n}\left(3\right) \right\rangle.
\label{eq:EquationsOfMotionForElectronsEq_2_10}
\end{eqnarray}
Next we write the quantities $S\left(1,2\right)$ and $\tilde{S}\left(1,2\right)$ in terms of the corresponding SE's, define (here $c = 1,2,3$)
\begin{eqnarray} 
\Sigma\left(1,3\right) &\equiv& \int d2 \tilde{S}\left(1,2\right) G^{-1}\left(2,3\right), \nonumber \\
\Sigma_{c}\left(1,3\right) &\equiv& \int d2 S_{c}\left(1,2\right) G^{-1}\left(2,3\right).
\label{eq:EquationsOfMotionForElectronsEq_2_21}
\end{eqnarray}
By inverting \eqref{eq:EquationsOfMotionForElectronsEq_2_21}, we find that
\begin{eqnarray} 
\tilde{S}\left(1,2\right) &=& \int d3 \Sigma\left(1,3\right) G\left(3,1\right), \nonumber \\
S_{c}\left(1,2\right) &=& \int d3 \Sigma_{c}\left(1,3\right) G\left(3,1\right).
\label{eq:EquationsOfMotionForElectronsEq_2_23}
\end{eqnarray}
By using Eq. \ref{eq:EquationsOfMotionForElectronsEq_2_23} in Eq. \ref{eq:EquationsOfMotionForElectronsEq_2_3_2}
\begin{eqnarray} 
\delta\left(1 - 2\right) &=& \left[i \hbar\frac{\partial}{\partial{z_{1}}} + \frac{\hbar^{2} }{2 m_{e}} \nabla^{2}_{\vec{r}_{1}} - V_{tot}\left(1\right) \right] G\left(1,2\right) \nonumber \\
&&- \int d3  \Sigma_{t}\left(1,3\right) G\left(3,2\right),
\label{eq:EquationsOfMotionForElectronsEq_2}
\end{eqnarray}
where
\begin{equation} 
\Sigma_{t}\left(1,2\right) \equiv \Sigma\left(1,2\right) + \sum^{3}_{c = 1} \Sigma_{c}\left(1,2\right). 
\label{eq:EquationsOfMotionForElectronsEq_2_25}
\end{equation}
We can also write Eq. \ref{eq:EquationsOfMotionForElectronsEq_2_3} in the form of a Dyson equation
\begin{equation} 
G\left(1,2\right) = G_{0}\left(1,2\right) + \int d3 \int d4  G_{0}\left(1,3\right) \Sigma_{t}\left(3,4\right) G\left(4,2\right),
\label{eq:EquationsOfMotionForElectronsEq_3}
\end{equation}
the function $G_{0}\left(1,2\right)$ being a solution of
\begin{equation} 
\left[i \hbar\frac{\partial}{\partial{z_{1}}} + \frac{\hbar^{2} }{2 m_{e}} \nabla^{2}_{\vec{r}_{1}} - V_{tot}\left(1\right) \right] G_{0}\left(1,2\right) = \delta\left(1 - 2\right).
\label{eq:EquationsOfMotionForElectronsEq_4}
\end{equation}
Most of the previous literature on the electron-nuclear many-body problem employed the laboratory frame formulation \cite{Baym-field-1961,Giustino-ElectronPhononInteractFromFirstPrinc-RevModPhys.89.015003-2017}. By contrast the present article works with a body-fixed frame. This allows for an explicit inclusion of rotational and vibrational degrees of freedom and their coupling, in a consistent way. A first important step towards a consistent body fixed-frame formulation was taken in Ref. \cite{vanLeeuwen-FirstPrincElectronPhonon-PhysRevB.69.115110-2004}. In the present formulation, the self-energy $\Sigma_{t}$ also contains the kinetic energy contributions $T'_{mpe}, T_{cvr}$ and $T'_{cvr}$. Neglecting these contributions leads to Eqs. \ref{eq:EquationsOfMotionForElectronsEq_2} and \ref{eq:EquationsOfMotionForElectronsEq_3} with $\Sigma_{c}\left(1,3\right) \equiv 0$ for $c = 1,2,3$ and thus $\Sigma_{t}\left(1,3\right) = \Sigma\left(1,3\right)$. The resulting equations are analogous to those of Ref. \cite{vanLeeuwen-FirstPrincElectronPhonon-PhysRevB.69.115110-2004}. All the rather complicated kinetic energy terms, originating from the body-fixed frame transformation, are hidden in the SE's $\Sigma_{c}\left(1,3\right)$. We will give an explicit, approximate form of each $\Sigma_{c}\left(1,3\right)$ in Eqs. \ref{eq:CoriolisAndVibrationalRotCouplingTermsEq_16} and \ref{eq:CoriolisAndVibrationalRotCouplingTermsEq_17} of Sec. \ref{cha:HedinsEquations}. In the present chapter we have obtained the necessary EOM for the electrons. We will deduce the Hedin-like equations for the electrons in Sec. \ref{cha:HedinsEquations}.

\section{Equation of motion for the nuclear Green's function}
\label{cha:EquationsOfMotionForNuclei}

Next we set out to derive the EOM for the nuclear Green's function. The connection of the nuclear and electronic equations will be addressed in more detail in Sec. \ref{cha:HedinsEquations}. We start by writing the Heisenberg EOM for the displacements, namely
\begin{equation}
i \hbar \frac{\partial{}}{\partial{z}} \hat{u}_{\alpha}\left(kz\right) = \left[\hat{u}_{\alpha}\left(kz\right), \hat{H}\left(z\right)\right]_{-}.
\label{eq:EquationsOfMotionForDisplacementsEq_1}
\end{equation}
After computing the commutators by using Eq. \ref{eq:HamiltonianEq_10} [see also Eq. \ref{eq:EquationsOfMotionForElectronsEq_1_3_5}], we obtain
\begin{eqnarray}
M_{k} \frac{\partial{}}{\partial{z}} \hat{u}_{\alpha}\left(kz\right) &=& \hat{p}_{\alpha}\left(kz\right) - \frac{ M_{k} }{  M_{nuc} } \sum^{N_{n}-1}_{k'' = 1} \hat{p}_{\alpha}\left(k''z\right) \nonumber \\
&&+ M_{k} \sum_{ \beta,\gamma } \int d\vec{r} r_{\gamma} \frac{\partial{ }}{ \partial{r_{\beta} }} \hat{L}^{\left(2\right)}_{\alpha\beta\gamma}\left(k\right) \hat{n}_{e}\left(\vec{r}z\right). \nonumber \\
\label{eq:EquationsOfMotionForDisplacementsEq_2}
\end{eqnarray}
Differentiating Eq. \ref{eq:EquationsOfMotionForDisplacementsEq_2} with respect to time and taking an ensemble average leads to
\begin{equation}
M_{k} \frac{\partial^{2}{}}{\partial{z^{2}}} \left\langle  \hat{u}_{\alpha}\left(kz\right) \right\rangle = K_{\alpha}\left(kz\right) + \sum^{4}_{ c = 1 } K^{\left(c\right)}_{\alpha}\left(kz\right) - J_{ \alpha }\left(kz\right),
\label{eq:EquationsOfMotionForDisplacementsEq_3}
\end{equation}
where (we use $\hat{T}''_{cvr} \equiv \hat{T}_{cvr} + \hat{T}'_{cvr}$)
\begin{eqnarray}
K_{\alpha}\left(kz\right) &\equiv& \frac{ 1 }{i\hbar} \left\langle  \left[ \hat{p}_{\alpha}\left(kz\right) , \hat{V}'_{en} +  \hat{V}'_{nn} \right]_{-} \right\rangle + \cdots, \nonumber \\
K^{\left(1\right)}_{\alpha}\left(kz\right) &\equiv& \frac{ i M_{k} }{ \hbar M_{nuc} } \sum^{N_{n}-1}_{k' = 1} \left\langle  \left[ \hat{p}_{\alpha}\left(k'z\right) , \hat{V}'_{en} + \hat{V}'_{nn} \right]_{-} \right\rangle \nonumber \\
&&+ \cdots, \nonumber \\
K^{\left(2\right)}_{\alpha}\left(kz\right) &\equiv& \frac{ 1 }{i\hbar} \left\langle  \left[ \hat{p}_{\alpha}\left(kz\right) , \hat{T}''_{cvr} \right]_{-} \right\rangle, \nonumber \\
K^{\left(3\right)}_{\alpha}\left(kz\right) &\equiv& \frac{ i M_{k} }{ \hbar M_{nuc} } \sum^{N_{n}-1}_{k' = 1} \left\langle  \left[ \hat{p}_{\alpha}\left(k'z\right) , \hat{T}''_{cvr} \right]_{-} \right\rangle, \nonumber \\
K^{\left(4\right)}_{\alpha}\left(kz\right) &\equiv& \frac{ 1 }{i\hbar} \sum_{ \beta,\gamma } \int d\vec{r} r_{\gamma} \frac{\partial{ }}{ \partial{r_{\beta} }} \nonumber \\
&&\times \left\langle  \left[ \hat{L}^{\left(2\right)}_{\alpha\beta\gamma}\left(k\right) \hat{n}_{e}\left(\vec{r}z\right) , \hat{H}\left(z\right) \right]_{-} \right\rangle.
\label{eq:EquationsOfMotionForDisplacementsEq_3_5}
\end{eqnarray}
In the first two quantities of Eq. \ref{eq:EquationsOfMotionForDisplacementsEq_3_5}, all the other contributions except those originating from the external potential terms are explicitly shown.

In order to obtain the EOM for the nuclear Green's function $D_{\alpha\beta}\left(kz,k'z'\right)$, we take a functional derivative of Eq. \ref{eq:EquationsOfMotionForDisplacementsEq_2} with respect to $J_{\beta}\left(k'z'\right)$ and write
\begin{eqnarray}
&&M_{k} \frac{\partial^{2}{}}{\partial{z^{2}}} D_{\alpha\beta}\left(kz,k'z'\right) \nonumber \\
&&= K_{\alpha\beta}\left(kz,k'z'\right) + \sum^{4}_{ c = 1 } K^{\left(c\right)}_{\alpha\beta}\left(kz,k'z'\right) \nonumber \\
&&\ \ \ - \delta_{\alpha\beta} \delta_{kk'} \delta\left(z-z'\right),
\label{eq:EquationsOfMotionForDisplacementsEq_4}
\end{eqnarray}
where
\begin{eqnarray} 
\frac{\delta{\left\langle \hat{u}_{\alpha}\left(kz\right) \right\rangle}}{\delta{J_{\beta}\left(k' z'\right)}} &=& \frac{1}{i \hbar} \left\langle \mathcal{T}_{\gamma} \left\{ \hat{u}_{\alpha}\left(k z\right) \hat{u}_{\beta}\left(k' z'\right) \right\} \right\rangle \nonumber \\
&&- \frac{1}{i \hbar} \left\langle \hat{u}_{\alpha}\left(k z\right) \right\rangle \left\langle \hat{u}_{\beta}\left(k' z'\right) \right\rangle \nonumber \\
&\equiv& D_{\alpha \beta}\left(kz,k'z'\right), \nonumber \\
K_{\alpha\beta}\left(kz,k'z'\right) &\equiv& \frac{\delta{ K_{\alpha}\left(kz\right) }}{\delta{J_{\beta}\left(k' z'\right)}}, \nonumber \\
K^{\left(c\right)}_{\alpha\beta}\left(kz,k'z'\right) &\equiv& \frac{\delta{ K^{\left(c\right)}_{\alpha}\left(kz\right) }}{\delta{J_{\beta}\left(k' z'\right)}}.
\label{eq:EquationsOfMotionForDisplacementsEq_5}
\end{eqnarray}
We write Eq. \ref{eq:EquationsOfMotionForDisplacementsEq_4} in yet another form by using the inverse of $D_{\alpha\beta}\left(k z,k'z'\right)$ that respects the Kubo-Martin-Schwinger boundary conditions, namely, Eq. \ref{eq:EquationsOfMotionForDisplacementsEq_4} can be written in terms of the nuclear SE's
\begin{eqnarray} 
\Pi_{\alpha \alpha'}\left(k z,\bar{k}\bar{z}\right) &\equiv& -\sum_{k',\beta} \int_{\gamma} dz' K_{\alpha \beta}\left(k z,k'z'\right) \nonumber \\
&&\times D^{-1}_{\beta \alpha'}\left(k'z',\bar{k}\bar{z}\right), \nonumber \\
\Pi^{\left(c\right)}_{\alpha \alpha'}\left(k z,\bar{k}\bar{z}\right) &\equiv& -\sum_{k',\beta} \int_{\gamma} dz' K^{\left(c\right)}_{\alpha \beta}\left(k z,k'z'\right) \nonumber \\
&&\times D^{-1}_{\beta \alpha'}\left(k'z',\bar{k}\bar{z}\right), 
\label{eq:EquationsOfMotionForDisplacementsEq_7_2}
\end{eqnarray}
as
\begin{eqnarray}
&&M_{k} \frac{\partial^{2}{}}{\partial{z^{2}}} D_{\alpha\beta}\left(kz,k'z'\right) \nonumber \\
&&= - \sum_{k'',\alpha'} \int_{\gamma} dz'' \Pi^{t}_{\alpha \alpha'}\left(k z,k''z''\right) D_{\alpha' \beta }\left(k''z'',k'z'\right) \nonumber \\
&&\ \ \ - \delta_{\alpha\beta} \delta_{kk'} \delta\left(z-z'\right),
\label{eq:EquationsOfMotionForDisplacementsEq_7}
\end{eqnarray}
where
\begin{equation} 
\Pi^{t}_{\alpha \alpha'}\left(k z,k'z'\right) \equiv \Pi_{\alpha \alpha'}\left(kz,k'z'\right) + \sum^{4}_{c = 1} \Pi^{\left(c\right)}_{\alpha \alpha'}\left(kz,k'z'\right).
\label{eq:EquationsOfMotionForDisplacementsEq_8}
\end{equation}
By neglecting the kinetic energy terms $\hat{T}'_{mpn}, \hat{T}_{cvr}$ and $\hat{T}'_{cvr}$ in the Hamiltonian given by Eq. \ref{eq:HamiltonianEq_10} leads to Eq. \ref{eq:EquationsOfMotionForDisplacementsEq_7} with $\Pi^{\left(c\right)}_{\alpha \alpha'}\left(k z,k'z'\right) = 0$ for all $c$ and thus $\Pi^{t}_{\alpha \alpha'}\left(k z,k'z'\right) = \Pi_{\alpha \alpha'}\left(k z,k'z'\right)$. In other words, all the effects of the mass-polarization, Coriolis and vibrational-rotational coupling terms on the EOM of the function $D_{\alpha\beta}\left(kz,k'z'\right)$ are hidden in the SE's $\Pi^{\left(c\right)}_{\alpha \alpha'}\left(k z,k'z'\right)$.

In the remaining part of this section, we deduce the EOM for the momentum-displacement and momentum-momentum Green's functions since these functions are required in the calculation of the total energy, see Appendix \ref{TotalEnergy}. In order to use the functional derivative technique we add the following potential to the external potential part of the Hamiltonian
\begin{equation} 
\sum^{N_{n}-1}_{ k = 1 } \sum^{3}_{ \beta = 1 } P_{\beta}\left(kz\right) \hat{p}_{\beta}\left(kz\right).
\label{eq:EquationsOfMotionForDisplacementsEq_8_2}
\end{equation}
We note that this term would appear in the EOM of $\left\langle  \hat{u}_{\alpha}\left(kz\right) \right\rangle$, but would not appear in the corresponding equations of $D_{\alpha\beta}\left(kz,k'z'\right)$. We start by writing the EOM for momentum ensemble average, namely
\begin{equation}
\frac{\partial{}}{\partial{z}} \left\langle  \hat{p}_{\alpha}\left(kz\right) \right\rangle = K_{\alpha}\left(kz\right) +  K^{\left(2\right)}_{\alpha}\left(kz\right) - J_{ \alpha }\left(kz\right).
\label{eq:EquationsOfMotionForDisplacementsEq_8_3}
\end{equation}
By taking a functional derivative of Eq. \ref{eq:EquationsOfMotionForDisplacementsEq_8_3} with respect to $P_{\beta}\left(k'z'\right)$ we find that
\begin{equation}
\frac{\partial{}}{\partial{z}} D^{pp}_{\alpha \beta}\left(kz,k'z'\right) = K'_{\alpha\beta}\left(kz,k'z'\right) + K''_{\alpha\beta}\left(kz,k'z'\right),
\label{eq:EquationsOfMotionForDisplacementsEq_8_4}
\end{equation}
where
\begin{eqnarray} 
D^{pp}_{\alpha \beta}\left(kz,k'z'\right) &\equiv& \frac{\delta{\left\langle \hat{p}_{\alpha}\left(k z\right) \right\rangle}}{\delta{P_{\beta}\left(k' z'\right)}}, \nonumber \\
K'_{\alpha\beta}\left(kz,k'z'\right) &\equiv& \frac{\delta{ K_{\alpha}\left(kz\right) }}{\delta{P_{\beta}\left(k' z'\right)}}, \nonumber \\
K''_{\alpha\beta}\left(kz,k'z'\right) &\equiv& \frac{\delta{ K^{\left(2\right)}_{\alpha}\left(kz\right) }}{\delta{P_{\beta}\left(k' z'\right)}}.
\label{eq:EquationsOfMotionForDisplacementsEq_8_5}
\end{eqnarray}
Next we take a functional derivative of Eq. \ref{eq:EquationsOfMotionForDisplacementsEq_8_3} with respect to $J_{ \beta }\left(k'z'\right)$ and write
\begin{eqnarray}
\frac{\partial{}}{\partial{z}} D^{pu}_{\alpha \beta}\left(kz,k'z'\right) &=& - \delta_{\alpha\beta} \delta_{kk'} \delta\left(z-z'\right) + K_{\alpha\beta}\left(kz,k'z'\right) \nonumber \\
&&+ K^{\left(2\right)}_{\alpha\beta}\left(kz,k'z'\right),
\label{eq:EquationsOfMotionForDisplacementsEq_8_6}
\end{eqnarray}
where Eq. \ref{eq:EquationsOfMotionForDisplacementsEq_5} was used and $D^{pu}_{\alpha \beta}\left(kz,k'z'\right) \equiv \delta{ \left\langle \hat{p}_{\alpha}\left(kz\right) \right\rangle }/\delta{J_{\beta}\left(k'z'\right)}$. We can use Eq. \ref{eq:EquationsOfMotionForDisplacementsEq_7_2} and write Eq. \ref{eq:EquationsOfMotionForDisplacementsEq_8_6} as
\begin{eqnarray}
&&\frac{\partial{}}{\partial{z}} D^{pu}_{\alpha \beta}\left(kz,k'z'\right) \nonumber \\
&&= - \sum_{k'',\alpha'} \int_{\gamma} dz'' \tilde{\Pi}^{t}_{\alpha \alpha'}\left(kz,k''z''\right) D_{\alpha' \beta }\left(k''z'',k'z'\right) \nonumber \\
&&\ \ \ - \delta_{\alpha\beta} \delta_{kk'} \delta\left(z-z'\right),
\label{eq:EquationsOfMotionForDisplacementsEq_8_8}
\end{eqnarray}
where
\begin{equation} 
\tilde{\Pi}^{t}_{\alpha \alpha'}\left(kz,k'z'\right) \equiv \Pi_{\alpha \alpha'}\left(kz,k'z'\right) + \Pi^{\left(2\right)}_{\alpha \alpha'}\left(kz,k'z'\right).
\label{eq:EquationsOfMotionForDisplacementsEq_8_9}
\end{equation}
Furthermore, we have
\begin{eqnarray}
D^{pu}_{\alpha \beta}\left(kz,k'z'\right) &=& \frac{\delta{ \left\langle  \hat{p}_{\alpha}\left(kz\right) \right\rangle }}{\delta{J_{\beta}\left(k' z'\right)}} = \frac{\delta{ \left\langle  \hat{u}_{\beta}\left(k'z'\right) \right\rangle }}{\delta{P_{\alpha}\left(k z\right)}} \nonumber \\
&\equiv& D^{up}_{\beta \alpha}\left(k'z',kz\right).
\label{eq:EquationsOfMotionForDisplacementsEq_8_10}
\end{eqnarray}
From Eq. \ref{eq:EquationsOfMotionForDisplacementsEq_8_8} we see that provided we have the necessary SE's and the solution for $D_{\alpha \beta }\left(kz,k'z'\right)$, then we can obtain $D^{pu}_{\alpha \beta}\left(kz,k'z'\right)$ without solving the EOM for it.

So far our derivation is exact. We have not made any approximations or simplifying assumptions. We still use the full Hamiltonian without restricting the expansion in displacements $\hat{\vec{u}}$ to some particular order, which is the usual procedure in the existing formulations \cite{Baym-field-1961,Maksimov-SelfConsistentElectronPhonon-1975,Giustino-ElectronPhononInteractFromFirstPrinc-RevModPhys.89.015003-2017}. We can actually write the exact total energy, defined as an ensemble average of the Hamiltonian, in terms of the quantities appearing in the EOM for the electrons and nuclei, see Appendix \ref{TotalEnergy}. Hence we have deduced a formally exact Green's function theory for arbitrary systems of electrons and nuclei given the Hamiltonian of kinetic energies and Coulombic interactions. Clearly the full solution will be hard to obtain for real-world systems and approximations are needed in practice. We discuss a special case of crystalline solids in Sec. \ref{cha:CrystallineSolids} and give an explicit approximate expression for the Fourier transformed SE $\Pi_{\alpha \alpha'}\left(k,k',\omega\right)$ in Sec. \ref{cha:NucleiSelfEnergy}.

\section{Hedin's equations}
\label{cha:HedinsEquations}

Hedin-like equations for the complete system of electrons and nuclei can be derived in a similar way as has been done earlier \cite{vanLeeuwen-FirstPrincElectronPhonon-PhysRevB.69.115110-2004}. Namely, the equations can be written as
\begin{eqnarray}
G\left(1,2\right) &=& G_{0}\left(1,2\right) \nonumber  \\
                            &&+ \int d3 \int d4  G_{0}\left(1,3\right) \Sigma_{t}\left(3,4\right) G\left(4,2\right), \nonumber \\
\Gamma\left(1,2,3\right) &=& \delta\left(1 - 2\right) \delta\left(1 - 3\right) + \int d4\int d5 \int d6 \int d7 \nonumber  \\
                            &&\times \frac{\delta{\Sigma_{t}\left(1,2\right)}}{\delta{G\left(4,5\right)}} G\left(4,6\right) G\left(7,5\right) \Gamma\left(6,7,3\right), \nonumber \\
W\left(1,2\right) &=& W_{e}\left(1,2\right) + W_{ph}\left(1,2\right),	\nonumber \\	
P_{e}\left(1,2\right) &=& -i \hbar \int d3 \int d4  G\left(1,3\right) G\left(4,1^{+}\right) \Gamma\left(3,4,2\right),	\nonumber \\
\Sigma\left(1,4\right) &=& \Sigma_{e}\left(1,4\right) + \Sigma_{ph}\left(1,4\right).									
\label{eq:EquationsOfMotionForElectronsEq_7}
\end{eqnarray}
Here, the vertex function $\Gamma\left(1,2,3\right)$ is almost the same as in Ref. \cite{vanLeeuwen-FirstPrincElectronPhonon-PhysRevB.69.115110-2004}, but contains contributions from the Coriolis, vibrational-rotational coupling and mass-polarization terms. The electronic polarization $P_{e}\left(1,2\right)$ is formally of the same form as in the earlier works, but also includes the aforementioned additional terms through the vertex function. The electronic part, $\Sigma_{e}\left(1,2\right)$, and the nuclear conribution, $\Sigma_{ph}\left(1,2\right)$ to the SE's are
\begin{eqnarray}
\Sigma_{e}\left(1,4\right) &=& i \hbar \int d3 \int d5 W_{e}\left(1,5\right)  G\left(1,3\right) \Gamma\left(3,4,5\right),	\nonumber \\
\Sigma_{ph}\left(1,4\right) &=& i \hbar \int d3 \int d5 W_{ph}\left(1,5\right)  G\left(1,3\right) \Gamma\left(3,4,5\right).	\nonumber \\							
\label{eq:EquationsOfMotionForElectronsEq_6_16_2}
\end{eqnarray}
The contributions to the screened Coulomb interaction can be written as
\begin{equation} 
W_{ph}\left(1,2\right) = \int d3 \int d4 \tilde{W}_{e}\left(1,3\right) \rho_{n}\left(3,4\right) W_{e}\left(4,2\right), 
\label{eq:EquationsOfMotionForElectronsEq_6_16}
\end{equation}
and
\begin{eqnarray} 
W_{e}\left(1,2\right) &=& \int d3 \epsilon^{-1}_{e}\left(2,3\right) v\left(3,1\right), \nonumber \\
\tilde{W}_{e}\left(1,2\right) &=& \int d3 v\left(1,3\right) \tilde{\epsilon}^{-1}_{e}\left(3,2\right),
\label{eq:EquationsOfMotionForElectronsEq_6_16_1}
\end{eqnarray}
where
\begin{eqnarray} 
\epsilon_{e}\left(1,2\right) &=& \delta\left(1-2\right) - \int d3 v\left(1,3\right) P_{e}\left(3,2\right), \nonumber \\
\tilde{\epsilon}_{e}\left(1,2\right) &=& \delta\left(1-2\right) - \int d3 P_{e}\left(1,3\right)   v\left(3,2\right).
\label{eq:EquationsOfMotionForElectronsEq_6_16_11}
\end{eqnarray}
Furthermore, the nuclear density-density correlation function appearing in Eq. \eqref{eq:EquationsOfMotionForElectronsEq_6_16} is
\begin{equation} 
\rho_{n}\left(1,2\right) = -\frac{i}{ \hbar} \left\langle \mathcal{T}_{\gamma}\left\{ \Delta \hat{n}_{n}\left(1\right) \Delta \hat{n}_{n}\left(2\right)  \right\} \right\rangle,
\label{eq:EquationsOfMotionForElectronsEq_6_9}
\end{equation}
where $\Delta \hat{n}_{n}\left(1\right) = \hat{n}_{n}\left(1\right) - \left\langle \hat{n}_{n}\left(1\right) \right\rangle$.

Formally, the present theory is similar to the conventional ones \cite{Hedin-NewMethForCalcTheOneParticleGreensFunctWithApplToTheElectronGasProb-PhysRev.139.A796-1965,Giustino-ElectronPhononInteractFromFirstPrinc-RevModPhys.89.015003-2017} and thus many of the approximations, like the GW-approximation \cite{Hedin-NewMethForCalcTheOneParticleGreensFunctWithApplToTheElectronGasProb-PhysRev.139.A796-1965,Aryasetiawan-TheGWMethod-1998,Stan-LevelsOfSelfConsistencyInTheGWApproximation-2009,Rostgaard-FullySelfConsistentGWCalculationsForMolecules-PhysRevB.81.085103-2010,Koval-FullySelfConsistentGWandQuasiparticleSelfConsistentGWforMolecules-PhysRevB.89.155417-2014} or other suitable approximations \cite{Lani-ApproximForManyBodyGreensFunctInsightsFromTheFundamentalEquations-2012}, can be established in a similar way as in the existing Green's function theory for electrons within the BO approximation. Namely, in the GW-approximation $\Gamma\left(1,2,3\right) \approx \delta\left(1 - 2\right) \delta\left(1 - 3\right)$ and the electronic polarization and SE become $P_{e}\left(1,2\right) \approx -i \hbar G\left(1,2\right) G\left(2,1^{+}\right)$ and $\Sigma\left(1,2\right) \approx i \hbar W\left(1,2\right)  G\left(1,2\right)$. The nuclear problem enters to the electronic equations, for instance, through the SE's and the nuclear density-density correlation function $\rho_{n}\left(1,2\right)$. We can write $\hat{n}_{n}\left(1\right)$ and $\rho_{n}\left(1,2\right)$ by expanding them in a Taylor series with respect to the displacements and taking into account consistently all terms up to within a given order. It thus follows that in order to evaluate the expanded $\left\langle  \hat{n}_{n}\left(1\right) \right\rangle$ and $\rho_{n}\left(1,2\right)$, we need to evaluate ensemble averages like $\left\langle \hat{u}_{\alpha}\left(kz\right) \right\rangle$, $\left\langle \hat{u}_{\alpha}\left(kz\right) \hat{u}_{\beta}\left(k'z\right) \right\rangle$ and so on. Similar terms appear if we expand the nuclei terms included in the SE's $\Sigma_{c}\left(1,2\right)$. In order to determine these quantities, we can use the EOM for the nuclear Green's functions derived in Sec. \ref{cha:EquationsOfMotionForNuclei}.

Next we give exact and approximate expressions for the SE's arising from the Coriolis, vibrational-rotational coupling and mass-polarization terms. The quantities $S_{c}\left(\vec{r}z,\vec{r}'z'\right)$ appearing in Eq. \ref{eq:EquationsOfMotionForElectronsEq_2_6} are defined as
\begin{eqnarray} 
S_{1}\left(\vec{r}z,\vec{r}'z'\right) &\equiv& - \frac{1}{i \hbar}\frac{ \hbar^{2} }{ M_{nuc} }  \int d\vec{r}''  \nabla_{\vec{r}} \cdot \nabla_{\vec{r}''} \nonumber \\
&&\times \left\langle  \mathcal{T}_{\gamma} \left\{ \hat{n}_{e}\left(\vec{r}''z\right) \hat{\psi}\left(\vec{r}z\right) \hat{\psi}^{\dagger}\left(\vec{r}'z'\right) \right\} \right\rangle, \nonumber \\
S_{2}\left(\vec{r}z,\vec{r}'z'\right) &\equiv& \frac{1}{i \hbar} \sum_{\beta,\gamma } r_{\gamma} \frac{\partial{ }}{\partial{r_{\beta} }} \nonumber \\
&&\times \left\langle  \mathcal{T}_{\gamma} \left\{ \hat{\mathcal{D}}'_{\beta\gamma}\left(z\right) \hat{\psi}\left(\vec{r}z\right) \hat{\psi}^{\dagger}\left(\vec{r}'z'\right) \right\} \right\rangle, \nonumber \\
S_{3}\left(\vec{r}z,\vec{r}'z'\right) &\equiv& \frac{2}{i \hbar} \sum_{ \beta,\sigma,\gamma,\gamma' }  \int d\vec{r}''  r_{\gamma'} r''_{\gamma} \frac{\partial^{2}{ }}{\partial{r''_{\sigma} } \partial{r_{\beta} }}  \nonumber \\
&&\times \left\langle  \mathcal{T}_{\gamma} \left\{ \hat{L}^{\left(3\right)}_{\gamma\sigma \gamma' \beta} \hat{n}_{e}\left(\vec{r}''z\right) \hat{\psi}\left(\vec{r}z\right) \hat{\psi}^{\dagger}\left(\vec{r}'z'\right) \right\} \right\rangle, \nonumber \\
\label{eq:CoriolisAndVibrationalRotCouplingTermsEq_11}
\end{eqnarray}
where
\begin{equation} 
\hat{\mathcal{D}}'_{\beta\gamma}\left(z\right) \equiv \hat{L}^{\left(1\right)}_{\beta\gamma} + \sum^{N_{n}-1}_{k = 1} \sum_{ \alpha } \hat{L}^{\left(2\right)}_{\alpha\beta\gamma}\left(k\right) \hat{p}_{\alpha}\left(kz\right).
\label{eq:CoriolisAndVibrationalRotCouplingTermsEq_12}
\end{equation}
The perhaps simplest approximation to the SE's $\Sigma_{c}\left(\vec{r}z,\vec{r}'z'\right)$ is obtained by introducing a mean-field-like factorization of the nuclear contributions:  
\begin{eqnarray} 
S_{1}\left(\vec{r}z,\vec{r}'z'\right) &=& -i \frac{ \hbar^{3} }{ M_{nuc} }  \int d\vec{r}''  \nabla_{\vec{r}} \cdot \nabla_{\vec{r}''} \nonumber \\
&&\times  G_{2}\left(\vec{r}''z,\vec{r}z;\vec{r}'z',\vec{r}''z\right), \nonumber \\
S_{2}\left(\vec{r}z,\vec{r}'z'\right) &\approx& \sum_{\beta,\gamma } r_{\gamma} \frac{\partial{ }}{\partial{r_{\beta} }} \left\langle  \hat{\mathcal{D}}'_{\beta\gamma}\left(z\right) \right\rangle G\left(\vec{r}z,\vec{r}'z'\right), \nonumber \\
S_{3}\left(\vec{r}z,\vec{r}'z'\right) &\approx& - 2\frac{ \hbar }{i } \sum_{ \beta,\sigma,\gamma,\gamma' }  \int d\vec{r}''  r_{\gamma'} r''_{\gamma} \frac{\partial^{2}{ }}{\partial{r''_{\sigma} } \partial{r_{\beta} }}  \nonumber \\
&&\times \left\langle \hat{L}^{\left(3\right)}_{\gamma\sigma \gamma' \beta} \right\rangle  G_{2}\left(\vec{r}''z,\vec{r}z;\vec{r}'z',\vec{r}''z\right),
\label{eq:CoriolisAndVibrationalRotCouplingTermsEq_13}
\end{eqnarray}
where
\begin{equation} 
G_{2}\left(1,2;1',2'\right) \equiv -\frac{1}{\hbar^{2} } \left\langle  \mathcal{T}_{\gamma} \left\{  \hat{\psi}\left(1\right) \hat{\psi}\left(2\right) \hat{\psi}^{\dagger}\left(2'\right) \hat{\psi}^{\dagger}\left(1'\right) \right\} \right\rangle.
\label{eq:CoriolisAndVibrationalRotCouplingTermsEq_14}
\end{equation}
A similar approximation is introduced in Sec. \ref{cha:NucleiSelfEnergy} when we approximate $\left\langle  \hat{u}_{\alpha_{1}}\left(k_{1}z\right) \hat{n}_{e}\left(\vec{r}z\right) \right\rangle \approx \left\langle  \hat{u}_{\alpha_{1}}\left(k_{1}z\right) \right\rangle \left\langle  \hat{n}_{e}\left(\vec{r}z\right) \right\rangle$. If we further make the Hartree-Fock approximation \cite{Stefanucci-Leeuwen-many-body-book-2013}
\begin{equation} 
G_{2}\left(1,2;1',2'\right) \approx G\left(1,1'\right) G\left(2,2'\right) - G\left(1,2'\right) G\left(2,1'\right),
\label{eq:CoriolisAndVibrationalRotCouplingTermsEq_15}
\end{equation}
we find by using Eq. \ref{eq:EquationsOfMotionForElectronsEq_2_21}
\begin{eqnarray} 
\Sigma_{1}\left(\vec{r}z,\vec{r}''z''\right) &=& \frac{ \hbar^{3} }{ i M_{nuc} }  \int d\bar{\vec{r}}  \nabla_{\vec{r}} \cdot \nabla_{\bar{\vec{r}}} \left[ \delta\left(\bar{\vec{r}} - \vec{r}''\right) G\left(\vec{r}z,\bar{\vec{r}}z\right) \right. \nonumber \\
&&- \left. G\left(\bar{\vec{r}}z,\bar{\vec{r}}z\right) \delta\left(\vec{r} - \vec{r}''\right) \right] \delta\left(z - z''\right), \nonumber \\
\Sigma_{2}\left(\vec{r}z,\vec{r}''z''\right) &=& \sum_{\beta,\gamma } \left\langle  \hat{\mathcal{D}}'_{\beta\gamma}\left(z\right) \right\rangle r_{\gamma} \frac{\partial{ }}{\partial{r_{\beta} }} \delta\left(\vec{r} - \vec{r}''\right) \delta\left(z - z''\right), \nonumber \\
\label{eq:CoriolisAndVibrationalRotCouplingTermsEq_16}
\end{eqnarray}
and
\begin{eqnarray} 
&&\Sigma_{3}\left(\vec{r}z,\vec{r}''z''\right) \nonumber \\
&&= 2 i \hbar \sum_{ \beta,\sigma,\gamma,\gamma' } \left\langle \hat{L}^{\left(3\right)}_{\gamma\sigma \gamma' \beta} \right\rangle  \int d\bar{\vec{r}}  r_{\gamma'} \bar{r}_{\gamma} \frac{\partial^{2}{ }}{\partial{\bar{r}_{\sigma} } \partial{r_{\beta} }}  \nonumber \\
&&\times  \left[ G\left(\vec{r}z,\bar{\vec{r}}z\right) \delta\left(\bar{\vec{r}} - \vec{r}''\right) - G\left(\bar{\vec{r}}z,\bar{\vec{r}}z\right) \delta\left(\vec{r} - \vec{r}''\right) \right] \delta\left(z - z''\right).  \nonumber \\
\label{eq:CoriolisAndVibrationalRotCouplingTermsEq_17}
\end{eqnarray}
After Taylor expanding $\hat{\mathcal{D}}'_{\beta\gamma}\left(t\right)$ and $\hat{L}^{\left(3\right)}_{\gamma\sigma \gamma' \beta}$ in displacements $\hat{\vec{u}}$, $\Sigma_{2}\left(1,2\right)$ and $\Sigma_{3}\left(1,2\right)$ become functions of $\left\langle \hat{u}_{\alpha}\left(k\right) \right\rangle$, $D_{\alpha\beta}\left(kz,k'z'\right)$ and so on. One has to emphasize that, although derived in a BO-like manner, the terms in Eqs. \eqref{eq:CoriolisAndVibrationalRotCouplingTermsEq_16} and \eqref{eq:CoriolisAndVibrationalRotCouplingTermsEq_17} represent physical effects beyond the BO approximation.

\section{Choice of reference positions}
\label{cha:ChoiceOfReferencePositions}

We have not yet discussed in detail how to choose the reference positions $\vec{x}$ and so far these quantities have been considered as arbitrary parameters. Consider, for instance, the total energy of the system, $E_{tot}$, defined as
\begin{equation} 
E_{tot} \equiv \left\langle \hat{H} \right\rangle.
\label{eq:DeterminationOfRestPositionsEq_1}
\end{equation}
The value of $E_{tot}$ must be independent of $\vec{x}$ provided our expansion (in displacements $\hat{\vec{u}}$) of the SE's, or the Hamiltonian ensemble average itself, converges for a given $\vec{x}$. However, in practice we are not able to make such an expansion up to arbitrary order in $\hat{\vec{u}}$ and solve the exact equations. Consequently, the value of $E_{tot}$ will become dependent on $\vec{x}$ due to the approximations made. For instance, if the positions $\vec{x}$ are chosen far away from the positions around which the nuclei would vibrate, then in order to calculate $E_{tot}$ accurately, the displacements $\vec{u}$ would be rather large. This implies that we would need a rather high order nuclear Green's functions in order to describe the system properly and this is expected to be extremely difficult. If we choose the positions $\vec{x}$ rather poorly, i.e. far away from the equilibrium positions, and at the same time take only the lowest order quantities in $\hat{\vec{u}}$ into account, we expect to find rather unrealistic results. Our theory is formally exact and hence it also describes situations where the nuclei do not vibrate around some particular regions of space (like close to the melting point of a crystal, or beyond it in the liquid phase). However, it is beneficial to have a consistent strategy to obtain the parameters $\vec{x}$ such that those systems in which the nuclei perform such rather small vibrations, can be described with reasonable accuracy by using the approximations of lowest orders in $\hat{\vec{u}}$. We note that in these cases, the (anti)symmetrization with respect to those variables $\hat{\vec{R}}'$ that refer to identical nuclei may not be necessary to obtain accurate results. If the lowest order approximations in $\hat{\vec{u}}$ describe the system in sufficient detail, the nuclei are well-localized near the positions $\vec{x}$, which also means that the effects of the (anti)symmetrization on the observables are expected to be rather small.

For the aforementioned systems, like some stable crystal lattices, we use the following method to determine the positions $\vec{x}$. The initial value of $\vec{x}$ is obtained in the same way as within the BO approximation, that is, either from experimental data of known structures or in the case of hypothetical structures by using the methods of structural chemistry \cite{Muller-InorganicStructuralChemistry-2006,Karttunen-StructuralPrinciplesOfSemiconductingGroup14ClathrateFrameworks-2010}. After the initial guess, which serves as starting point of an iteration, we aim to find the positions $\vec{x}$ such that they are equal to the expectation values of the nuclear positions (in the body-fixed frame), that is, we seek the positions such that
\begin{equation} 
\vec{x}_{k} = \left\langle \hat{\vec{R}}'_{k} \right\rangle = \vec{x}_{k} + \left\langle \hat{\vec{u}}_{k} \right\rangle.
\label{eq:DeterminationOfRestPositionsEq_2}
\end{equation}
We note that $\left\langle \hat{\vec{u}}_{k} \right\rangle$ is a function of $\vec{x}$ since the Hamiltonian is, as consequence of truncating the Taylor expansion in $\hat{\vec{u}}$ at a finite order. In the absence of the external potential terms, $\left\langle \hat{\vec{R}}'_{k} \right\rangle$ is independent of time, and hence the equation determining the reference positions $\vec{x}$ becomes
\begin{equation} 
\left\langle \hat{\vec{u}}_{k} \right\rangle = 0.
\label{eq:DeterminationOfRestPositionsEq_2_2}
\end{equation}
That is, if we are able to choose the positions $\vec{x}$ such that Eq. \ref{eq:DeterminationOfRestPositionsEq_2} holds, then the expectation values of the displacements vanish. Next we seek a way how to find the displacements satisfying Eq. \ref{eq:DeterminationOfRestPositionsEq_2_2} and thus the reference positions $\vec{x}$ satisfying Eq. \ref{eq:DeterminationOfRestPositionsEq_2}. The determining equation for $\left\langle \hat{\vec{u}}_{k} \right\rangle$ is given by Eq. \ref{eq:EquationsOfMotionForDisplacementsEq_3}. If we put the external potential terms to zero, truncate the Taylor expansion (w.r.t. $\hat{\vec{u}}$) of the Hamiltonian at a finite order and choose as initial condition the equilibrium ensemble associated with this truncated Hamiltonian, then $\left\langle  \hat{u}_{\alpha}\left(kz\right) \right\rangle = \left\langle  \hat{u}_{\alpha}\left(k\right) \right\rangle$ and thus 
\begin{equation}
M_{k} \frac{\partial^{2}{}}{\partial{z^{2}}} \left\langle  \hat{u}_{\alpha}\left(kz\right) \right\rangle = 0,
\label{eq:DeterminationOfRestPositionsEq_4}
\end{equation}
meaning that the expectation value of the total force on each nucleus $k$ vanishes. Thus, if the external potential terms vanish, Eq. \ref{eq:EquationsOfMotionForDisplacementsEq_3} becomes
\begin{equation}
0 = K_{\alpha}\left(kz\right) + \sum^{4}_{ c = 1 } K^{\left(c\right)}_{\alpha}\left(kz\right).
\label{eq:DeterminationOfRestPositionsEq_5}
\end{equation}
The quantities $K_{\alpha}\left(kz\right)$ and $K^{\left(c\right)}_{\alpha}\left(kz\right)$ are given by Eq. \ref{eq:EquationsOfMotionForDisplacementsEq_3_5} (here without the external potential terms). We define the total force
\begin{equation}
F_{\alpha}\left(kz\right) \equiv K_{\alpha}\left(kz\right) + \sum^{4}_{ c = 1 } K^{\left(c\right)}_{\alpha}\left(kz\right),
\label{eq:DeterminationOfRestPositionsEq_7}
\end{equation}
and decompose it further as $F_{\alpha}\left(kz\right) = F'_{\alpha}\left(kz\right) + F''_{\alpha}\left(kz\right)$, where $F'_{\alpha}\left(kz\right)$ is the part of $F_{\alpha}\left(kz\right)$ which is a function of $\left\langle  \hat{u}_{\alpha}\left(k\right) \right\rangle$ and $F''_{\alpha}\left(kz\right)$ is the remaining part (not a function of $\left\langle  \hat{u}_{\alpha}\left(k\right) \right\rangle$). We can therefore write Eq. \ref{eq:DeterminationOfRestPositionsEq_5} as
\begin{equation}
0 = F'_{\alpha}\left(kz\right) + F''_{\alpha}\left(kz\right).
\label{eq:DeterminationOfRestPositionsEq_8}
\end{equation}
This is the determining equation for $\left\langle  \hat{u}_{\alpha}\left(k\right) \right\rangle$. If we now want to find $\vec{x}$ such that $\left\langle  \hat{u}_{\alpha}\left(k\right) \right\rangle = 0$, then Eq. \eqref{eq:DeterminationOfRestPositionsEq_8} becomes
\begin{equation}
0 = F''_{\alpha}\left(kz\right).
\label{eq:DeterminationOfRestPositionsEq_9}
\end{equation}

We now give an example where an explicit and approximate form of Eq. \ref{eq:DeterminationOfRestPositionsEq_9} is given. Suppose that the only non-zero term in Eq. \ref{eq:DeterminationOfRestPositionsEq_5} is $K_{\alpha}\left(kz\right)$ meaning that we take into account terms which originate from the Coulomb interactions and neglect all the other terms. This is expected to be a relatively good approximation for most crystalline solids provided we define the Euler angles such that the Coriolios and vibrational-rotational terms in the Hamiltonian are small, see Sec. \ref{cha:CrystallineSolids}. We approximate $K_{\alpha}\left(kz\right)$ by expanding $\hat{V}'_{en}$ and $\hat{V}'_{nn}$ in displacements (see Appendix \ref{TotalEnergy}) and retain only the lowest orders. After calculating the commutators, we obtain to the lowest orders in displacements
\begin{eqnarray}
K_{\alpha}\left(kz\right) &\approx& -\frac{ \partial{ V_{nn}\left(\vec{x}\right) }}{\partial{x_{\alpha}\left(k\right)} } - \int d\vec{r} \frac{ \partial{ V_{en}\left(\vec{r},\vec{x}\right) }}{\partial{x_{\alpha}\left(k\right)} } \left\langle \hat{n}_{e}\left(\vec{r}z\right) \right\rangle \nonumber \\
&&- \sum_{k_{2},\alpha_{2}} \int d\vec{r} \frac{ \partial^{2}{ V_{en}\left(\vec{r},\vec{x}\right) }}{\partial{x_{\alpha}\left(k\right)} \partial{x_{\alpha_{2}}\left(k_{2}\right)} } \nonumber \\
&&\times \left\langle  \hat{u}_{\alpha_{2}}\left(k_{2}z\right) \hat{n}_{e}\left(\vec{r}z\right) \right\rangle   \nonumber \\
&&- \sum_{k_{2},\alpha_{2}} \frac{ \partial^{2}{ V_{nn}\left(\vec{x}\right) }}{\partial{x_{\alpha}\left(k\right)} \partial{x_{\alpha_{2}}\left(k_{2}\right)} } \left\langle  \hat{u}_{\alpha_{2}}\left(k_{2}z\right) \right\rangle.
\label{eq:DeterminationOfRestPositionsEq_10}
\end{eqnarray}
Next we use the following result
\begin{eqnarray} 
\frac{\delta{ \left\langle  \hat{n}_{e}\left(\vec{r}z\right) \right\rangle }}{\delta{J_{\beta}\left(k' z'\right)}} &=&  \int d\vec{r}_{2} \int_{\gamma} dz_{2} \int d\vec{r}_{3} \int_{\gamma} dz_{3} P_{e}\left(\vec{r}z,\vec{r}_{3}z_{3}\right) \nonumber \\
&&\times \tilde{W}_{e}\left(\vec{r}_{3}z_{3},\vec{r}_{2}z_{2}\right) \frac{\delta{\left\langle  \hat{n}_{n}\left(\vec{r}_{2}z_{2}\right) \right\rangle}}{\delta{J_{\beta}\left(k'z'\right)}},
\label{eq:DeterminationOfRestPositionsEq_10_2}
\end{eqnarray}
where $\tilde{W}_{e}\left(\vec{r}_{3}z_{3},\vec{r}_{2}z_{2}\right)$ and $\tilde{\epsilon}_{e}\left(\vec{r}_{1}z_{1},\vec{r}_{2}z_{2}\right)$ are given by Eqs. \ref{eq:EquationsOfMotionForElectronsEq_6_16_1} and \ref{eq:EquationsOfMotionForElectronsEq_6_16_11}, respectively. These quantities are obtained from the solution of Hedin's equations. By using Eq. \ref{eq:DeterminationOfRestPositionsEq_10_2} we find that
\begin{eqnarray} 
&&\left\langle \mathcal{T}_{\gamma} \left\{  \hat{u}_{\beta}\left(k'z'\right) \hat{n}_{e}\left(\vec{r}z\right) \right\} \right\rangle \nonumber \\
&&\approx \left\langle \hat{n}_{e}\left(\vec{r}z\right) \right\rangle \left\langle \hat{u}_{\beta}\left(k'z'\right) \right\rangle + i\hbar \int d\vec{r}_{2} \int_{\gamma} dz_{2} \nonumber \\
&&\times \int d\vec{r}_{3} \int_{\gamma} dz_{3} P_{e}\left(\vec{r}z,\vec{r}_{3}z_{3}\right) \tilde{W}_{e}\left(\vec{r}_{3}z_{3},\vec{r}_{2}z_{2}\right)  \nonumber \\
&&\times \sum_{k_{1},\alpha_{1}} \frac{ \partial{ n^{\left(0\right)}_{n}\left(\vec{x},\vec{r}_{2}\right) }}{\partial{x_{\alpha_{1}}\left(k_{1}\right)} } D_{\alpha_{1}\beta}\left(k_{1}z_{2},k'z'\right).
\label{eq:DeterminationOfRestPositionsEq_11}
\end{eqnarray}
Here, the quantity $n^{\left(0\right)}_{n}\left(\vec{x},\vec{r}\right)$ is given by Eq. \ref{eq:UsefulRelationsEq_10} and related to this, we consider how to obtain the rotation matrix $\mathcal{R}\left[\boldsymbol{\theta}\left(\vec{x}\right)\right]$ and its derivatives in Sec. \ref{EulerAngles}. When we use Eq. \ref{eq:DeterminationOfRestPositionsEq_11} in Eq. \ref{eq:DeterminationOfRestPositionsEq_10} to approximate $\left\langle  \hat{u}_{\alpha}\left(kz\right) \hat{n}_{e}\left(\vec{r}z\right) \right\rangle$, it can be seen that if we seek the solution such that $\left\langle \hat{\vec{u}}_{k} \right\rangle = \vec{0}$, meaning that $F'_{\alpha}\left(kz\right)$ vanishes, Eq. \ref{eq:DeterminationOfRestPositionsEq_9} is of the following form 
\begin{eqnarray}
0 &\approx& \frac{ \partial{ V_{nn}\left(\vec{x}\right) }}{\partial{x_{\alpha}\left(k\right)} } + \int d\vec{r} \frac{ \partial{ V_{en}\left(\vec{r},\vec{x}\right) }}{\partial{x_{\alpha}\left(k\right)} } \left\langle \hat{n}_{e}\left(\vec{r}z\right) \right\rangle \nonumber \\
&&+ i\hbar \sum_{k_{2},\alpha_{2}} \int d\vec{r} \frac{ \partial^{2}{ V_{en}\left(\vec{r},\vec{x}\right) }}{\partial{x_{\alpha}\left(k\right)} \partial{x_{\alpha_{2}}\left(k_{2}\right)} }  \int d\vec{r}_{2} \int_{\gamma} dz_{2} \nonumber \\
&&\times \int d\vec{r}_{3} \int_{\gamma} dz_{3} P_{e}\left(\vec{r}z,\vec{r}_{3}z_{3}\right) \tilde{W}_{e}\left(\vec{r}_{3}z_{3},\vec{r}_{2}z_{2}\right)  \nonumber \\
&&\times \sum_{k_{1},\alpha_{1}} \frac{ \partial{ n^{\left(0\right)}_{n}\left(\vec{x},\vec{r}_{2}\right) }}{\partial{x_{\alpha_{1}}\left(k_{1}\right)} } D^{\mathcal{T}}_{\alpha_{1}\beta}\left(k_{1}z_{2},k_{2} z\right),
\label{eq:DeterminationOfRestPositionsEq_12}
\end{eqnarray}
where (Eq. \ref{eq:EquationsOfMotionForDisplacementsEq_5})
\begin{eqnarray} 
D^{\mathcal{T}}_{\alpha \beta}\left(kz,k'z'\right) \equiv \frac{1}{i \hbar} \left\langle \mathcal{T}_{\gamma} \left\{ \hat{u}_{\alpha}\left(k z\right) \hat{u}_{\beta}\left(k' z'\right) \right\} \right\rangle.
\label{eq:DeterminationOfRestPositionsEq_12_2}
\end{eqnarray}
If we further approximate and neglect the last term of Eq. \ref{eq:DeterminationOfRestPositionsEq_12} we obtain
\begin{equation} 
0 = \frac{ \partial{ V_{nn}\left(\vec{x}\right) }}{\partial{x_{\alpha}\left(k\right)} } + \int d\vec{r} \frac{ \partial{ V_{en}\left(\vec{r},\vec{x}\right) }}{\partial{x_{\alpha}\left(k\right)} } \left\langle \hat{n}_{e}\left(\vec{r}\right) \right\rangle.
\label{eq:DeterminationOfRestPositionsEq_13}
\end{equation}
\begin{figure}
\includegraphics[width=0.47\textwidth]{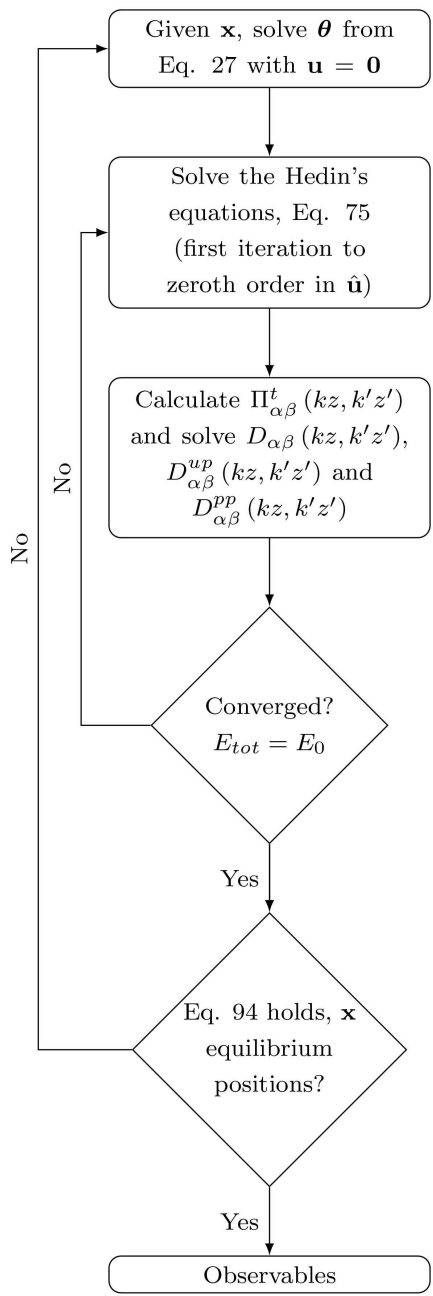}
\caption{Example flowchart for solving the coupled equations for the electronic and nuclear Green's functions. Here the ground state total energy is denoted by $E_{0}$.}
\label{fig:FlowChart}
\end{figure}
Equations \ref{eq:DeterminationOfRestPositionsEq_12} and \ref{eq:DeterminationOfRestPositionsEq_13} are examples of approximate conditions for choosing the parameters $\vec{x}$ such that Eq. \ref{eq:DeterminationOfRestPositionsEq_2} holds. We call the positions $\vec{x}$ the nuclear equilibrium positions. The approximate condition given by Eq. \ref{eq:DeterminationOfRestPositionsEq_13} certainly makes a lot of sense: The equilibrium position of the $k$-th nucleus is the location where the total electrostatic force on this nucleus (coming from the other nuclei and the electron cloud) vanishes. This is formally the same condition as in the BO approximation \cite{Baroni-PhononsAndRelatedCrystalPropFromDFTPT-RevModPhys.73.515-2001}, when the Hellmann-Feynman theorem \cite{Hellmann-EinfuhrungInDieQuantenchemie-1937,Feynman-ForcesInMolecules-PhysRev.56.340-1939} is used to calculate the total force. The difference is that our parameters $\vec{x}$ refer to the body-fixed frame. Further, the electron density $\left\langle \hat{n}_{e}\left(\vec{r}\right) \right\rangle$ is an ensemble average taken with respect to the full electron-nuclear Hamiltonian rather than an expectation value with respect to an eigenstate of the electronic BO Hamiltonian. The electron density can be obtained from the electron Green's function as $\left\langle  \hat{n}_{e}\left(\vec{r}z\right) \right\rangle = -i \hbar G\left(\vec{r}z,\vec{r}z^{+}\right)$. Thus, for a given $\vec{x}$ one solves the coupled set of equations for the electrons and nuclei and from the solutions of these equations we obtain the quantities like $G\left(\vec{r}z,\vec{r}'z'\right)$, $P_{e}\left(\vec{r}z,\vec{r}'z'\right)$, $\tilde{W}_{e}\left(\vec{r}z,\vec{r}'z'\right)$ and $D^{\mathcal{T}}_{\alpha \beta}\left(kz,k'z'\right)$ and we can check whether or not Eq. \ref{eq:DeterminationOfRestPositionsEq_12}, or the corresponding general expression given by Eq. \ref{eq:DeterminationOfRestPositionsEq_9}, holds and indicates that we have found the positions such that Eq. \ref{eq:DeterminationOfRestPositionsEq_2} is satisfied.

Next we give an overview of the work flow how the coupled set of equations for the electrons and nuclei could be solved. The procedure is summarized in Fig. \ref{fig:FlowChart}. First we give an initial guess for the reference positions $\vec{x}$ and solve $\boldsymbol{\theta}$ from Eq. \ref{eq:HamiltonianEq_8_1} with $\vec{u}_{k} = \vec{0}$, see Sec. \ref{EulerAngles}. Given the reference positions we expand all the quantities in Hedin's equations, which are functions of the nuclear variables, into a Taylor series in $\hat{\vec{u}}$. In the first iteration, we retain only the zeroth order terms meaning that $W_{ph}\left(1,2\right) = 0$ from which it follows that $W\left(1,2\right) = W_{e}\left(1,2\right)$ and $\Sigma\left(1,2\right) = \Sigma_{e}\left(1,2\right)$ (see Eq. \ref{eq:EquationsOfMotionForElectronsEq_7}). With the preceding choices, Hedin's equations are actually similar to those in the BO approximation apart from the SE's $\Sigma_{c}\left(1,2\right)$, but here all quantities refer to the body-fixed frame. The zeroth order approximations for the SE's $\Sigma_{c}\left(1,2\right)$ can be obtained by using the relations given by Eqs. \ref{eq:EquationsOfMotionForElectronsEq_2_21} and \ref{eq:CoriolisAndVibrationalRotCouplingTermsEq_13}. After the Hedin equations are solved, we calculate the nuclear SE, see for example  Sec. \ref{cha:NucleiSelfEnergy}. Given the nuclear SE, the EOM for the nuclear Green's function can be solved. Once we have obtained the nuclear Green's function, we can write Hedin's equations with non-zero $W_{ph}\left(1,2\right)$, solve these equations with the SE $\Sigma\left(1,2\right) = \Sigma_{e}\left(1,2\right) + \Sigma_{ph}\left(1,2\right)$ and by including the nuclear contributions beyond the zeroth order in $\hat{\vec{u}}$ to the SE's $\Sigma_{c}\left(1,2\right)$. We iterate the electronic and nuclear equations in order to minimize the grand potential, or in the case of zero temperature formalism, the total energy $E_{tot}$ (see Appendix \ref{TotalEnergy}). After the convergence, we test whether or not the reference positions $\vec{x}$ are the equilibrium positions by checking if Eq. \ref{eq:DeterminationOfRestPositionsEq_9} is satisfied. We iterate this whole process until we have found the equilibrium positions.

\section{Nuclear vibrations and self-energy}
\label{cha:NucleiVibrationsAndSelfEnergy}

\subsection{Nuclear vibrations}
\label{cha:NucleiVibrations}

We start by writing the nuclear EOM in frequency space, but before Fourier transforming, all the external potential terms are set equal to zero. It follows that the nuclear Green's function and the nuclear SE's are then functions of time differences only and we can write Eq. \ref{eq:EquationsOfMotionForDisplacementsEq_7} in terms of the Fourier transformed quantities as
\begin{eqnarray}
\delta_{\alpha\beta} \delta_{kk'} &=& \sum_{k'',\alpha'} \left[ M_{k} \omega^{2}  \delta_{\alpha\alpha'} \delta_{kk''} - \Pi^{t}_{\alpha \alpha'}\left(k,k'',\omega\right) \right] \nonumber \\
&&\times D_{\alpha'\beta}\left(k'',k',\omega\right).
\label{eq:NucleiVibrations_1}
\end{eqnarray}
Here the reader is free to choose, for example, the time-ordered or retarded components of these functions as the Eq. \ref{eq:EquationsOfMotionForDisplacementsEq_7} is defined on a general time contour. We write Eq. \ref{eq:NucleiVibrations_1} in matrix form and re-arrange such that
\begin{equation}
\tilde{\vec{D}}\left(\omega\right) = \left[ \omega^{2} \vec{I} - \vec{C}^{t}\left(\omega\right) \right]^{-1},
\label{eq:NucleiVibrations_2}
\end{equation}
where $\tilde{\vec{D}}\left(\omega\right) \equiv \vec{M}^{1/2} \vec{D}\left(\omega\right) \vec{M}^{1/2}$ and $\vec{C}^{t}\left(\omega\right) \equiv \vec{M}^{-1/2} \vec{\Pi}^{t}\left(\omega\right) \vec{M}^{-1/2}$. The components of these matrices are denoted by $\alpha k$ and so on. Next we write the SE as $\vec{C}^{t}\left(\omega\right) = \vec{C}^{A} + \vec{C}^{NA}\left(\omega\right)$ where we choose the "adiabatic part" $\vec{C}^{A}$ as the zero-frequency-limit $\vec{C}^{A} = \vec{C}^{t}\left(0\right)$. The "non-adiabatic" part $\vec{C}^{NA}$ is the remainder, i.e. $\vec{C}^{NA}\left(\omega\right) = \vec{C}^{t}\left(\omega\right) - \vec{C}^{t}\left(0\right)$. It is easy to see that $\vec{C}^{A}$ is a Hermitian matrix. Hence we can find a complete and orthonormal set of eigenvectors satisfying
\begin{eqnarray} 
\tilde{\omega}^{2}_{j} v_{\alpha}\left(k|j\right) &=& \sum_{k',\beta} C^{A}_{\alpha\beta}\left(kk'\right) v_{\beta}\left(k'|j\right), \nonumber \\
\delta_{j j'} &=& \sum_{k,\alpha} v_{\alpha}\left(k|j'\right) v^{\ast}_{\alpha}\left(k|j\right), \nonumber \\
\delta_{\alpha\beta}\delta_{kk'} &=& \sum_{j} v_{\alpha}\left(k|j\right) v^{\ast}_{\beta}\left(k'|j\right).
\label{eq:NucleiVibrations_4}
\end{eqnarray}
We call the quantities $\tilde{\omega}_{j}$ the adiabatic normal mode frequencies. The eigenvalues $\tilde{\omega}^{2}_{j}$ are real since $\vec{C}^{A}$ is Hermitian and $\tilde{\omega}_{j}$ are real if $\vec{C}^{A}$ is positive definite. We transform Eq. \ref{eq:NucleiVibrations_2} by using the eigenvectors of $\vec{C}^{A}$ such that
\begin{equation}
\boldsymbol{\mathcal{D}}\left(\omega\right) = \left[ \omega^{2} \vec{I} - \tilde{\boldsymbol{\omega}}^{2}  - \boldsymbol{\mathcal{C}}^{NA}\left(\omega\right) \right]^{-1},
\label{eq:NucleiVibrations_5}
\end{equation}
where $\boldsymbol{\mathcal{D}}\left(\omega\right) \equiv \vec{v}^{\dagger} \tilde{\vec{D}}\left(\omega\right) \vec{v}$ and $\boldsymbol{\mathcal{C}}^{NA}\left(\omega\right) \equiv \vec{v}^{\dagger} \vec{C}^{NA}\left(\omega\right) \vec{v}$. If $\boldsymbol{\mathcal{C}}^{NA}\left(\omega\right)$ is sufficiently small, the quasi-particle picture holds and $\boldsymbol{\mathcal{C}}^{NA}\left(\omega\right)$ can be pictured in generating interactions between adiabatic normal modes by shifts of their values and finite lifetimes [imaginary part of $\boldsymbol{\mathcal{C}}^{NA}\left(\omega\right)$]. We can obtain the Green's function of complex frequency by just replacing the real frequency $\omega$ by the complex one \cite{Farid-GroundAndLowLyingExcitedStatesOfInteractElectronSystemsASurveyAndSomeCriticalAnalyses-1999} in Eq. \ref{eq:NucleiVibrations_5}. The usual procedure is to consider the Matsubara Green's functions continued to the complex frequency plane \cite{Baym-DeterminationOfThermodynGreensFunctions-1961,Baym-field-1961,Maradudin-Fein-PhysRev.128.2589-Scat-Neutr.1962,Rickayzen-GreensFunctions-1980,Karlsson-PartialSelfConsistAndAnalytInManyBodyPertTheorParticNumberConservAndAGenerSumRule-PhysRevB.94.125124-2016}. We point out that the present discussion is still completely general. In Sec. \ref{cha:PhononsAndTheirInteractions} we will deduce a representation of $\vec{D}\left(\omega\right)$ in terms of the phonon basis of crystalline solids.

\subsection{Nuclear self-energy}
\label{cha:NucleiSelfEnergy}

In order to derive an approximate form for the SE $\Pi_{\alpha \alpha'}\left(k,k',\omega\right)$, we start from the explicit form of the terms included in $K_{\alpha\beta'}\left(k z,k'z'\right)$ (see Eq. \ref{eq:EquationsOfMotionForDisplacementsEq_5}). We have already calculated the commutators included to the lowest orders in $\hat{u}_{\alpha}\left(kz\right)$ and we can obtain $K_{\alpha\beta'}\left(k z,k'z'\right)$ by taking a functional derivative of $K_{\alpha}\left(k z\right)$ given by Eq. \ref{eq:DeterminationOfRestPositionsEq_10} with respect to $J_{\beta}\left(k'z'\right)$, namely
\begin{eqnarray} 
&&K_{\alpha\beta}\left(k z,k'z'\right) \nonumber \\
&&= - \sum_{k'',\alpha''} \frac{ \partial^{2}{ V_{nn}\left(\vec{x}\right) }}{ \partial{x_{\alpha}\left(k\right)} \partial{x_{\alpha''}\left(k''\right)} } D_{\alpha'' \beta}\left(k''z,k'z'\right)  \nonumber \\
&&\ \ \ - \sum_{k_{1},\alpha_{1}} \int d\vec{r} \frac{ \partial^{2}{ V_{en}\left(\vec{r},\vec{x}\right) }}{\partial{x_{\alpha}\left(k\right)} \partial{x_{\alpha_{1}}\left(k_{1}\right)} } \frac{\delta{ \left\langle  \hat{u}_{\alpha_{1}}\left(k_{1}z\right) \hat{n}_{e}\left(\vec{r}z\right) \right\rangle }}{\delta{J_{\beta}\left(k'z'\right)}} \nonumber \\
&&\ \ \ - \int d\vec{r} \frac{ \partial{ V_{en}\left(\vec{r},\vec{x}\right) }}{\partial{x_{\alpha}\left(k\right)} } \frac{\delta{ \left\langle \hat{n}_{e}\left(\vec{r}z\right) \right\rangle }}{\delta{J_{\beta}\left(k'z'\right)}} + \mathcal{O}\left(\hat{\vec{u}}^{2}\right).
\label{eq:PhononSelfEnergyEq_1}
\end{eqnarray}
The terms visible in Eq. \ref{eq:PhononSelfEnergyEq_1} are all the terms included in $K_{\alpha\beta'}\left(k z,k'z'\right)$ if the Hamiltonian is expanded in displacements up to second order before writing the EOM. This is the usual procedure in the earlier laboratory frame formulations \cite{Baym-field-1961,Giustino-ElectronPhononInteractFromFirstPrinc-RevModPhys.89.015003-2017}. We further evaluate the quantity in the second term on the right hand side of Eq. \ref{eq:PhononSelfEnergyEq_1}
\begin{eqnarray} 
\frac{\delta{ \left\langle  \hat{u}_{\alpha_{1}}\left(k_{1}z\right) \hat{n}_{e}\left(\vec{r}z\right) \right\rangle }}{\delta{J_{\beta}\left(k'z'\right)}} &=& i\hbar \frac{\delta^{2}{\left\langle \hat{n}_{e}\left(\vec{r}z\right) \right\rangle}}{ \delta{J_{\beta}\left(k'z'\right)} \delta{J_{\alpha_{1}}\left(k_{1}z\right)}} \nonumber \\
&&+ \left\langle \hat{u}_{\alpha_{1}}\left(k_{1}z\right) \right\rangle \frac{\delta{ \left\langle \hat{n}_{e}\left(\vec{r}z\right) \right\rangle }}{\delta{J_{\beta}\left(k'z'\right)}} \nonumber \\
&&+ \left\langle \hat{n}_{e}\left(\vec{r}z\right) \right\rangle D_{\alpha_{1}\beta}\left(k_{1}z,k'z'\right). \nonumber \\
\label{eq:PhononSelfEnergyEq_5}
\end{eqnarray} 
Finally, by using Eqs. \ref{eq:DeterminationOfRestPositionsEq_10_2}, \ref{eq:PhononSelfEnergyEq_1} and \ref{eq:PhononSelfEnergyEq_5} in Eq. \ref{eq:EquationsOfMotionForDisplacementsEq_7_2}, we can write the nuclear SE in frequency space as:
\begin{eqnarray} 
&&\Pi_{\alpha \beta}\left(k ,k',\omega\right) \nonumber \\
&&= \frac{ \partial^{2}{ V_{nn}\left(\vec{x}\right) }}{ \partial{x_{\alpha}\left(k\right)} \partial{x_{\beta}\left(k'\right)} } + \int d\vec{r} \frac{ \partial^{2}{ V_{en}\left(\vec{r},\vec{x}\right) }}{\partial{x_{\alpha}\left(k\right)} \partial{x_{\beta}\left(k'\right)} } \left\langle \hat{n}_{e}\left(\vec{r}\right) \right\rangle \nonumber \\
&&+ \int d\vec{r} \int d\vec{r}' \frac{ \partial{ n^{\left(0\right)}_{n}\left(\vec{x}, \vec{r}\right) }}{\partial{x_{\alpha}\left(k\right)} } \left[ \tilde{W}_{e}\left(\vec{r},\vec{r}', \omega\right) - v\left(\vec{r},\vec{r}'\right) \right] \nonumber \\
&&\times  \left[ \frac{\partial{ n^{\left(0\right)}_{n}\left(\vec{x}, \vec{r}'\right) }}{ \partial{x_{\beta}\left(k'\right)} } + \sum_{k_{1}, \alpha_{1}}  \frac{\partial^{2}{ n^{\left(0\right)}_{n}\left(\vec{x}, \vec{r}'\right) }}{\partial{x_{\alpha_{1}}\left(k_{1}\right)}  \partial{x_{\beta}\left(k'\right)} }  \left\langle u_{\alpha_{1}}\left(k_{1}\right) \right\rangle \right] \nonumber \\
&&+ \sum_{k_{1}, \alpha_{1}}  \int d\vec{r} \int d\vec{r}'  \frac{ \partial^{2}{ n^{\left(0\right)}_{n}\left(\vec{x}, \vec{r}\right) }}{\partial{x_{\alpha}\left(k\right)} \partial{x_{\alpha_{1}}\left(k_{1}\right)} } \nonumber \\
&&\times \left[ \tilde{W}_{e}\left(\vec{r},\vec{r}', \omega\right) - v\left(\vec{r},\vec{r}'\right)  \right] \left\langle \hat{u}_{\alpha_{1}}\left(k_{1}\right) \right\rangle  \nonumber \\
&&\times \left[ \frac{\partial{ n^{\left(0\right)}_{n}\left(\vec{x}, \vec{r}'\right) }}{ \partial{x_{\beta}\left(k'\right)} } + \sum_{k_{2}, \alpha_{2}} \frac{\partial^{2}{ n^{\left(0\right)}_{n}\left(\vec{x}, \vec{r}'\right) }}{\partial{x_{\alpha_{2}}\left(k_{2}\right)}  \partial{x_{\beta}\left(k'\right)} }  \left\langle u_{\alpha_{2}}\left(k_{2}\right) \right\rangle \right] \nonumber \\
&&+ \cdots.
\label{eq:PhononSelfEnergyEq_6}
\end{eqnarray}
In Eq. \ref{eq:PhononSelfEnergyEq_6}, $\tilde{W}_{e}\left(\vec{r},\vec{r}',\omega\right)$ is obtained by Fourier transforming $\tilde{W}_{e}\left(\vec{r}z,\vec{r}'z'\right)$ (Eq. \ref{eq:EquationsOfMotionForElectronsEq_6_16_1}) in the relative time variable $z-z'$. Terms explicitly shown in Eq. \ref{eq:PhononSelfEnergyEq_6} are already included in the harmonic approximation, which means that the Hamiltonian is expanded up to second order in displacements (before writing the EOM). Actually not all the terms found in the harmonic approximation appear in Eq. \ref{eq:PhononSelfEnergyEq_6}. Namely, the first term on the right hand side of Eq. \ref{eq:PhononSelfEnergyEq_5} does not appear because we have made the approximation $\left\langle  \hat{u}_{\alpha}\left(kz\right) \hat{n}_{e}\left(\vec{r}z\right) \right\rangle \approx \left\langle  \hat{u}_{\alpha}\left(kz\right) \right\rangle \left\langle  \hat{n}_{e}\left(\vec{r}z\right) \right\rangle$. Further, only some of the second order terms of the right hand side of Eq. \ref{eq:DeterminationOfRestPositionsEq_10_2} are included. The validity of this approximation, i.e. of making $\hat{u}_{\alpha}\left(kz\right)$ and $\hat{n}_{e}\left(\vec{r}z\right)$ uncorrelated in the sense that the expectation value of their product is the product of their expectation values has to be assessed carefully when the states with respect to which we define the Green's functions are not the BO eigenstates.

Sometimes it is convenient to write \cite{Giustino-ElectronPhononInteractFromFirstPrinc-RevModPhys.89.015003-2017} $\Pi_{\alpha \beta}\left(k,k',\omega\right) = \Pi^{A}_{\alpha \beta}\left(k,k'\right) + \Pi^{NA}_{\alpha \beta}\left(k,k',\omega\right)$ where the adiabatic SE $\Pi^{A}_{\alpha \beta}\left(k,k'\right) = \Pi_{\alpha \beta}\left(k,k',0\right)$ is Hermitian while the non-adiabatic part $\Pi^{NA}_{\alpha \beta}\left(k,k',\omega\right) = \Pi_{\alpha \beta}\left(k,k',\omega\right) - \Pi_{\alpha \beta}\left(k,k',0\right)$ is non-Hermitian, in general. In the present case (Eq. \ref{eq:PhononSelfEnergyEq_6}), the contribution from the Coulombic nuclei-nuclei interaction to the SE is $\partial^{2}{ V_{nn}\left(\vec{x}\right) }/ \partial{x_{\alpha}\left(k\right)} \partial{x_{\beta}\left(k'\right)}$, while the rest of the terms visible in Eq. \ref{eq:PhononSelfEnergyEq_6} originates from the Coulomb electron-nuclei interaction. For a better comparison with existing theories, we use Eq. \ref{eq:PhononSelfEnergyEq_6} to write the non-adiabatic contribution as
\begin{eqnarray} 
\Pi^{NA}_{\alpha \beta}\left(k,k',\omega\right) &=& \int d\vec{r} \int d\vec{r}' \frac{ \partial{ n^{\left(0\right)}_{n}\left(\vec{x}, \vec{r}\right) }}{\partial{x_{\alpha}\left(k\right)} } \nonumber \\
&&\times \left[ \tilde{W}_{e}\left(\vec{r},\vec{r}',\omega\right) - \tilde{W}_{e}\left(\vec{r},\vec{r}',0\right) \right] \nonumber \\
&&\times \frac{\partial{ n^{\left(0\right)}_{n}\left(\vec{x}, \vec{r}'\right) }}{ \partial{x_{\beta}\left(k'\right)} } + \cdots,
\label{eq:PhononSelfEnergyEq_8}
\end{eqnarray}
where terms including the quantity $\left\langle \hat{u}_{\alpha}\left(k\right) \right\rangle$ are not explicitly shown. Provided we can choose the parameters $\vec{x}$ such that Eq. \ref{eq:DeterminationOfRestPositionsEq_2} holds, then terms involving the quantities $\left\langle \hat{u}_{\alpha}\left(k\right) \right\rangle$ vanish, see Sec. \ref{cha:ChoiceOfReferencePositions}. The relation given by Eq. \ref{eq:PhononSelfEnergyEq_8} is formally analogous with the result obtained earlier \cite{Giustino-ElectronPhononInteractFromFirstPrinc-RevModPhys.89.015003-2017}. However, the difference is in the densities $n^{\left(0\right)}_{n}\left(\vec{x}, \vec{r}'\right)$ (see Eq. \ref{eq:UsefulRelationsEq_10} of Appendix \ref{TotalEnergy}) and here all the variables refer to the body-fixed frame. We note that if necessary, the higher order expressions for the nuclear and phonon SE's can be obtained by systematically including higher order terms in the expansion of $K_{\alpha\beta}\left(k z,k'z'\right)$. Related to this, some results of a recent study \cite{Marini-FunctApprToTheElectrAndBosonicDynamOfManyBodySystPerturbWithAnArbitStrongElectronBosonInteract-PhysRevB.98.075105-2018} on generic electron-boson Hamiltonians may be useful within the present theory as well.

While we do not give explicit expressions for the SE's $\Pi^{\left(c\right)}_{\alpha \alpha'}\left(k z,k'z'\right)$, it is obvious that these quantities can be obtained with the same procedure as established here for $\Pi_{\alpha \alpha'}\left(kz,k'z'\right)$. Namely, calculate the commutators contained in $K^{\left(c\right)}_{\alpha}\left(kz\right)$ (Eq. \ref{eq:EquationsOfMotionForDisplacementsEq_3_5}) by expanding the quantities involved in $\hat{\vec{u}}$, take the functional derivative with respect to $J_{\beta}\left(k'z'\right)$ in order to obtain $K^{\left(c\right)}_{\alpha\beta}\left(kz,k'z'\right)$ and then use Eq. \ref{eq:EquationsOfMotionForDisplacementsEq_7_2} to calculate $\Pi^{\left(c\right)}_{\alpha \alpha'}\left(k z,k'z'\right)$. In obtaining these expressions, similar approximations can be made as we did in the case of $\Pi_{\alpha \alpha'}\left(kz,k'z'\right)$. We leave a more detailed discussions of these terms for future work.

\section{Crystalline solids}
\label{cha:CrystallineSolids}

In this section, we apply the general theory to crystalline solids. First of all, the electronic and nuclear mass polarization terms $\hat{T}'_{mpe}$, $\hat{T}'_{mpn}$ and $\hat{T}'_{cvr}$ appearing in the transformed kinetic energy are proportional to the inverse of the total nuclear mass and are thus small for crystals. Further, we assume that one can find an implicit condition of the form (Eq. \ref{eq:HamiltonianEq_8_1}) such that the contributions from the kinetic energy $\hat{T}_{cvr}$ become small. Some justifications for their neglect have been given \cite{vanLeeuwen-FirstPrincElectronPhonon-PhysRevB.69.115110-2004}, when the implicit condition is chosen to be the Eckart condition given by Eq. \ref{eq:HamiltonianEq_8_2}. By neglecting the aforementioned terms the EOM for electrons remain otherwise the same, except that we have $\Sigma_{t}\left(1,3\right) = \Sigma\left(1,3\right)$ in Eqs. \ref{eq:EquationsOfMotionForElectronsEq_2} and \ref{eq:EquationsOfMotionForElectronsEq_3} and in the Hedin's equations given by Eq. \ref{eq:EquationsOfMotionForElectronsEq_7}. In turn, the nuclear EOM remains otherwise the same, but in Eq. \ref{eq:EquationsOfMotionForDisplacementsEq_7} the SE is $\Pi^{t}_{\alpha \alpha'}\left(kz,k'z'\right) = \Pi_{\alpha \alpha'}\left(kz,k'z'\right)$.

\subsection{Phonons and their interactions}
\label{cha:PhononsAndTheirInteractions}

Here we use a notation suitable for the description of crystalline solids and impose periodic boundary conditions \cite{born-huang-dynamical-1954,maradudin-harm-appr-1971,maradudin-dyn-prop-solids-1974}. We label the nucleus by $k = l\kappa$, and write $\vec{x}_{k} = \vec{x}_{l\kappa} = \vec{x}_{l} + \vec{x}_{\kappa}$, where $\vec{x}_{\kappa}$ is the position vector of the nuclei $\kappa$ within the unit cell and
\begin{equation} 
\vec{x}_{l} \equiv \vec{x}_{l_{1}l_{2}l_{3}} = l_{1} \vec{a}_{1} + l_{2} \vec{a}_{2} + l_{3} \vec{a}_{3},
\label{eq:NotationEq_3}
\end{equation}
is the lattice translational vector of the $l$th unit cell with integers $l_{1}$, $l_{2}$, $l_{3}$, and the vectors $\vec{a}_{j}$ being the primitive translational vectors of the lattice. With this notation, we write $\vec{R}'_{k} = \vec{R}'_{l\kappa} = \vec{x}_{l\kappa} + \vec{u}_{l\kappa}$, where $l\kappa$ goes over $N_{n}-1$ values in total.

Our aim is to define phonon frequencies beyond the BO approximation as has been established earlier by using the Green's function approach \cite{Baym-field-1961,Giustino-ElectronPhononInteractFromFirstPrinc-RevModPhys.89.015003-2017}. However, we do not assume the harmonic approximation. In the present case, Eq. \ref{eq:NucleiVibrations_1} can be written otherwise the same, but with $\Pi^{t}_{\alpha \alpha'}\left(l\kappa, l''\kappa'',\omega\right) = \Pi_{\alpha \alpha'}\left(l\kappa, l''\kappa'',\omega\right)$. We also note that the quantities like $D_{\alpha\beta}\left(l\kappa,l'\kappa',\omega\right)$ and $\Pi_{\alpha\beta}\left(l\kappa,l'\kappa',\omega\right)$ are dependent only on the difference of the cell indices $l-l'$. Therefore, we write $D_{\alpha\beta}\left(l\kappa,l'\kappa',\omega\right)$ as a discrete Fourier transform in the relative coordinate allowed by the periodic boundary conditions \cite{born-huang-dynamical-1954}, namely
\begin{equation} 
D_{\alpha \alpha'}\left(\kappa\kappa',l,\omega\right) = \frac{1}{N} \sum^{N}_{\vec{q}} D_{\alpha \alpha'}\left(\kappa\kappa',\vec{q},\omega\right) e^{i \vec{q} \cdot \vec{x}_{l} },
\label{eq:PhononsAndTheirInteractionsEq_2}
\end{equation}
and in a similar way for $\Pi_{\alpha \alpha'}\left(\kappa\kappa',l,\omega\right)$. In Eq. \ref{eq:PhononsAndTheirInteractionsEq_2}, $N$ is the number of $\vec{q}$ points and thus the number of unit cells in the Born-von Karman cell \cite{born-huang-dynamical-1954,maradudin-harm-appr-1971}. By using Eq. \ref{eq:PhononsAndTheirInteractionsEq_2} and the analogous expression for $\Pi_{\alpha \alpha'}\left(\kappa\kappa',l'',\omega\right)$ in Eq. \ref{eq:NucleiVibrations_1}, we obtain
\begin{eqnarray} 
&&\sum_{\kappa'',\alpha'} \left[ M_{\kappa} \omega^{2} \delta_{\alpha \alpha'} \delta_{\kappa\kappa''} - \Pi_{\alpha \alpha'}\left(\kappa\kappa'',\vec{q},\omega\right) \right] \nonumber \\
&&\times  D_{\alpha' \beta}\left(\kappa''\kappa',\vec{q},\omega\right) = \delta_{\alpha \beta} \delta_{\kappa\kappa'},
\label{eq:PhononsAndTheirInteractionsEq_3}
\end{eqnarray}
and after some re-arranging and by using matrix notation
\begin{equation} 
\tilde{\vec{D}}\left(\vec{q},\omega\right) = \left[ \omega^{2} \vec{I} - \vec{C}\left(\vec{q},\omega\right) \right]^{-1},
\label{eq:PhononsAndTheirInteractionsEq_4}
\end{equation}
where $\vec{C}\left(\vec{q},\omega\right) \equiv \vec{M}^{-1/2} \vec{\Pi}\left(\vec{q},\omega\right) \vec{M}^{-1/2}$ and $\tilde{\vec{D}}\left(\vec{q},\omega\right) \equiv \vec{M}^{1/2} \vec{D}\left(\vec{q},\omega\right) \vec{M}^{1/2}$. The components of the matrix like $\vec{C}\left(\vec{q},\omega\right)$ are labeled by $\alpha \kappa$ and $\beta \kappa'$, etc. Following the procedure of Sec. \ref{cha:NucleiVibrations}, namely, we write the SE as $\vec{C}\left(\vec{q},\omega\right) = \vec{C}^{A}\left(\vec{q}\right) + \vec{C}^{NA}\left(\vec{q},\omega\right)$, where $\vec{C}^{A}\left(\vec{q}\right) = \vec{C}\left(\vec{q},0\right)$ and $\vec{C}^{NA}\left(\vec{q},\omega\right) = \vec{C}\left(\vec{q},\omega\right) - \vec{C}\left(\vec{q},0\right)$. The eigenvalue equation for $\vec{C}^{A}\left(\vec{q}\right)$ can be written as 
\begin{equation} 
\omega^{2}_{\vec{q}j} e_{\alpha}\left(\kappa|\vec{q}j\right) = \sum_{\kappa',\beta} C^{A}_{\alpha\beta}\left(\kappa\kappa'|\vec{q}\right) e_{\beta}\left(\kappa'|\vec{q}j\right).
\label{eq:PhononsAndTheirInteractionsEq_10}
\end{equation}
We call the quantities $\omega_{\vec{q}j}$ the adiabatic phonon frequencies. The eigenvalue equation given by Eq. \ref{eq:PhononsAndTheirInteractionsEq_10} is analogous to the eigenvalue equation written for the dynamical matrix in the conventional theory of lattice dynamics \cite{born-huang-dynamical-1954,maradudin-harm-appr-1971}. As the adiabatic phonon frequencies $\omega_{\vec{q}j}$ are defined by Eq. \ref{eq:PhononsAndTheirInteractionsEq_10}, in contrast to the conventional theory and existing theories beyond the BO approximation \cite{Baym-field-1961,Giustino-ElectronPhononInteractFromFirstPrinc-RevModPhys.89.015003-2017}, already the non-interacting phonons potentially contain terms up to arbitrary order in $\hat{\vec{u}}$. In order for $\omega^{2}_{\vec{q}j}$ to be positive and thus $\omega_{\vec{q}j}$ to be real, the matrix $\vec{C}^{A}\left(\vec{q}\right)$ has to be positive definite, which is not in general the case for a given $\vec{x}$. This is also the case in the BO theory of lattice dynamics \cite{born-huang-dynamical-1954}, where the positive definiteness of the dynamical matrix implies the minimum of the BO energy surface and the stability of the crystal lattice \cite{Born-OnTheStabilityOfCrystalLatticesI-1940}. Provided the matrix $\vec{C}^{A}\left(\vec{q}\right)$ is Hermitian (as it is, for example, if Eq. \ref{eq:PhononSelfEnergyEq_6} is used), the components of the eigenvector $e_{\alpha}\left(\kappa|\vec{q}j\right)$ can be chosen to satisfy the orthonormality and completeness conditions
\begin{eqnarray}
\sum_{\kappa,\alpha} e_{\alpha}\left(\kappa|\vec{q}j'\right) e^{\ast}_{\alpha}\left(\kappa|\vec{q}j\right) &=& \delta_{j j'}, \nonumber \\
\sum_{j} e_{\alpha}\left(\kappa|\vec{q}j\right) e^{\ast}_{\beta}\left(\kappa'|\vec{q}j\right) &=& \delta_{\alpha\beta}\delta_{\kappa\kappa'}.
\label{eq:PhononsAndTheirInteractionsEq_12}
\end{eqnarray}
We use the eigenvectors of the adiabatic SE to rewrite Eq. \ref{eq:PhononsAndTheirInteractionsEq_4} in terms of the adiabatic phonon basis. After the transformation we obtain
\begin{equation} 
\boldsymbol{\mathcal{D}}\left(\vec{q},\xi\right) = \left[ \left( \xi^{2} -  \boldsymbol{\omega}^{2}_{\vec{q}} \right) \vec{I} - \boldsymbol{\mathcal{C}}^{NA}\left(\vec{q},\xi\right) \right]^{-1},
\label{eq:PhononsAndTheirInteractionsEq_15}
\end{equation}
where $\boldsymbol{\mathcal{D}}\left(\vec{q},\xi\right) \equiv \vec{e}^{\dagger}\left(\vec{q}\right) \tilde{\vec{D}}\left(\vec{q},\xi\right) \vec{e}\left(\vec{q}\right)$ and $\boldsymbol{\mathcal{C}}^{NA}\left(\vec{q},\xi\right) \equiv \vec{e}^{\dagger}\left(\vec{q}\right) \vec{C}^{NA}\left(\vec{q},\xi\right) \vec{e}\left(\vec{q}\right)$. We call $\boldsymbol{\mathcal{D}}\left(\vec{q},\xi\right)$ the phonon Green's function and $\boldsymbol{\mathcal{C}}^{NA}\left(\vec{q},\xi\right)$ the non-adiabatic phonon SE. Here, $\xi$ denotes a complex frequency variable \cite{Farid-GroundAndLowLyingExcitedStatesOfInteractElectronSystemsASurveyAndSomeCriticalAnalyses-1999}. If the non-adiabatic SE vanishes, we recover the adiabatic phonon Green's function $\mathcal{D}^{A}_{jj}\left(\vec{q},\omega\right) = 1/\left( \omega^{2}  - \omega^{2}_{\vec{q}j} \right)$ which is of the usual form and appears also in the BO theory of lattice dynamics \cite{Maradudin-Fein-PhysRev.128.2589-Scat-Neutr.1962}. The adiabatic phonons have infinite lifetimes if the non-adiabatic SE vanishes. The non-adiabatic part can be pictured in generating interactions between the adiabatic phonons which appear as shifts to the adiabatic eigenvalues and finite lifetimes of these quasiparticles. This picture is valid if the non-adiabatic SE is sufficiently small in comparison to the adiabatic part. The shifts to the adiabatic phonon frequencies and finite lifetimes of adiabatic phonons can be obtained as in the laboratory frame formulation \cite{Allen-NeutronSpectroscopyOfSuperconductors-PhysRevB.6.2577-1972,Grimvall-ElectronPhononInteractionInMetals-Book-1981,Giustino-ElectronPhononInteractFromFirstPrinc-RevModPhys.89.015003-2017}.

By using Eqs. \ref{eq:EquationsOfMotionForElectronsEq_6_16_1} and \ref{eq:EquationsOfMotionForElectronsEq_6_16_11} in Eq. \ref{eq:PhononSelfEnergyEq_8} and then Fourier transforming and changing the representation, we obtain the following approximate form for the non-adiabatic phonon SE
\begin{eqnarray} 
\mathcal{C}^{NA}_{jj'}\left(\vec{q},\xi\right) &=& \frac{1}{N} \int d\vec{r} \int d\vec{r}' \left[ g_{\vec{q}j}\left(\vec{r},\xi\right) P_{e}\left(\vec{r},\vec{r}',\xi\right) \tilde{g}^{\ast}_{\vec{q}j'}\left(\vec{r}'\right) \right.  \nonumber \\  
&&- \left. g_{\vec{q}j}\left(\vec{r},0\right) P_{e}\left(\vec{r},\vec{r}',0\right) \tilde{g}^{\ast}_{\vec{q}j'}\left(\vec{r}'\right) \right].
\label{eq:PhononSelfEnergyEq_9}
\end{eqnarray}
In Eq. \ref{eq:PhononSelfEnergyEq_9}
\begin{eqnarray}
\tilde{g}^{\ast}_{\vec{q}j}\left(\vec{r}\right) &=& \sum_{l,\kappa,\alpha} M^{-1/2}_{\kappa} e_{\alpha}\left(\kappa|\vec{q}j\right) e^{ i \vec{q} \cdot \vec{x}_{l} } \frac{ \partial{ V_{en}\left(\vec{x},\vec{r}\right) }}{\partial{x_{\alpha}\left(\kappa l\right)} }, \nonumber \\
g_{\vec{q}j}\left(\vec{r},\xi\right) &=& \int d\vec{r}' \tilde{g}_{\vec{q}j}\left(\vec{r}'\right) \tilde{\epsilon}^{-1}_{e}\left(\vec{r}',\vec{r},\xi\right), 
\label{eq:PhononSelfEnergyEq_10}
\end{eqnarray}
where $\tilde{g}_{\vec{q}j}\left(\vec{r}\right)$ is the complex conjugate of $\tilde{g}^{\ast}_{\vec{q}j}\left(\vec{r}\right)$. The diagrams corresponding to $\boldsymbol{\mathcal{D}}\left(\vec{q},\xi\right)$, $\boldsymbol{\mathcal{C}}^{NA}\left(\vec{q},\xi\right)$ are of a similar form as in the laboratory frame formulation \cite{Giustino-ElectronPhononInteractFromFirstPrinc-RevModPhys.89.015003-2017,Giustino-ErratumElectronPhononInteractFromFirstPrinciples-RevModPhys.91.019901-2019}.

\subsection{Momentum functions}
\label{cha:MomentumFunctions}

By making the approximations discussed at the beginning of this section and then comparing Eqs. \ref{eq:EquationsOfMotionForDisplacementsEq_7} and \ref{eq:EquationsOfMotionForDisplacementsEq_8_8} we see that
\begin{equation}
\frac{\partial{}}{\partial{z}} D^{pu}_{\alpha \beta}\left(kz,k'z'\right) = M_{k} \frac{\partial^{2}{}}{\partial{z^{2}}} D_{\alpha\beta}\left(kz,k'z'\right),
\label{eq:MomentumFunctionsEq_1}
\end{equation}
and thus for the Fourier transforms
\begin{equation}
D^{pu}_{\alpha \beta}\left(k,k',\omega\right) = - i M_{k} \omega D_{\alpha \beta}\left(k,k',\omega\right).
\label{eq:MomentumFunctionsEq_2}
\end{equation}
Therefore, after we have obtained a solution for $D_{\alpha \beta}\left(l\kappa,l'\kappa',\omega\right)$ we can also obtain $D^{pu}_{\alpha \beta}\left(l\kappa,l'\kappa',\omega\right)$. The last function we need is the momentum-momentum Green's function $D^{pp}_{\alpha \beta}\left(kz,k'z'\right)$ and in the present case Eq. \ref{eq:EquationsOfMotionForDisplacementsEq_8_4} becomes
\begin{equation}
\frac{\partial{}}{\partial{z}} D^{pp}_{\alpha \beta}\left(kz,k'z'\right) = K'_{\alpha\beta}\left(kz,k'z'\right),
\label{eq:MomentumFunctionsEq_3}
\end{equation}
where we approximate
\begin{eqnarray}
K'_{\alpha\beta}\left(kz,k'z'\right) &\approx& - \sum_{k_{2},\alpha_{2}} \frac{ \partial^{2}{ V_{nn}\left(\vec{x}\right) }}{\partial{x_{\alpha}\left(k\right)} \partial{x_{\alpha_{2}}\left(k_{2}\right)} } \frac{\delta{ \left\langle  \hat{u}_{\alpha_{2}}\left(k_{2}z\right) \right\rangle }}{\delta{P_{\beta}\left(k' z'\right)}} \nonumber \\
&&- \int d\vec{r} \frac{ \partial{ V_{en}\left(\vec{r},\vec{x}\right) }}{\partial{x_{\alpha}\left(k\right)} } \frac{\delta{ \left\langle \hat{n}_{e}\left(\vec{r}z\right) \right\rangle }}{\delta{P_{\beta}\left(k'z'\right)}} \nonumber \\
&&- \sum_{k_{2},\alpha_{2}} \int d\vec{r} \frac{ \partial^{2}{ V_{en}\left(\vec{r},\vec{x}\right) }}{\partial{x_{\alpha}\left(k\right)} \partial{x_{\alpha_{2}}\left(k_{2}\right)} } \nonumber \\
&&\times \frac{\delta{ \left\langle  \hat{u}_{\alpha_{2}}\left(k_{2}z\right) \hat{n}_{e}\left(\vec{r}z\right) \right\rangle }}{\delta{P_{\beta}\left(k'z'\right)}}  + \mathcal{O}\left(\hat{\vec{u}}^{2}\right),
\label{eq:MomentumFunctionsEq_4}
\end{eqnarray}
which can be obtained directly from Eq. \ref{eq:DeterminationOfRestPositionsEq_10}. By employing the same approximations for $K'_{\alpha\beta}\left(kz,k'z'\right)$, as used in Sec. \ref{cha:NucleiSelfEnergy} in writing $K_{\alpha\beta}\left(kz,k'z'\right)$, we find that Eq. \ref{eq:MomentumFunctionsEq_3} can be written as
\begin{eqnarray}
\frac{\partial{}}{\partial{z}} D^{pp}_{\alpha \beta}\left(kz,k'z'\right) &=& -\sum_{k'',\alpha'} \int_{\gamma} dz'' \Pi_{\alpha \alpha'}\left(kz,k''z''\right) \nonumber \\
&&\times D^{up}_{\alpha' \beta }\left(k''z'',k'z'\right).
\label{eq:MomentumFunctionsEq_5}
\end{eqnarray}
The Fourier transform of this equation is
\begin{equation}
i \omega D^{pp}_{\alpha \beta}\left(k,k',\omega\right) = \sum_{k'',\alpha'}  \Pi_{\alpha \alpha'}\left(k,k'',\omega\right) D^{up}_{\alpha' \beta}\left(k'',k',\omega\right).
\label{eq:MomentumFunctionsEq_6}
\end{equation}
Since by Eq. \ref{eq:EquationsOfMotionForDisplacementsEq_8_10}, $D^{up}_{\beta \alpha}\left(k'z',kz\right) = D^{pu}_{\alpha \beta}\left(kz,k'z'\right)$, we can write
\begin{eqnarray}
D^{pp}_{\alpha \beta}\left(k,k',\omega\right) &=& - M_{k} \sum_{k'',\alpha'}  \Pi_{\alpha \alpha'}\left(k,k'',\omega\right) \nonumber \\
&&\times D_{\beta \alpha'}\left(k',k'',-\omega\right),
\label{eq:MomentumFunctionsEq_7}
\end{eqnarray}
where Eq. \ref{eq:MomentumFunctionsEq_2} was used. Now we have all the necessary equations in place to calculate the total energy of the system, e.g.


\section{Conclusions}
\label{cha:Conclusions}

In this work, we have derived a coupled and self-consistent set of exact equations for the electronic and nuclear Green's functions following from the Hamiltonian of Coulomb interactions and kinetic energies as a starting point. The present theory, when applied to crystalline solids, resembles in some aspects the previous ones \cite{Baym-field-1961,Maksimov-SelfConsistentElectronPhonon-1975,Giustino-ElectronPhononInteractFromFirstPrinc-RevModPhys.89.015003-2017}. However, we resolve an issue arising from the translational and rotational symmetry of the Hamiltonian. This symmetry prevents the use of the existing many-body Green's function theories beyond the BO approximation in describing systems other than those with constant density eigenstates. The present theory is formally exact and it is not limited to the harmonic approximation. The complexity of our system of EOM is of the same order as the existing ones when applied to crystalline solids.

In addition to the general EOM, we specifically consider the normal modes. For the special case of crystalline solids, phonons and their interactions beyond the BO approximation are rigorously defined and discussed in detail. While it is probably not a realistic goal to obtain the solution of the EOM in general form in arbitrary systems, there is some work left to derive computationally accessible approximations to be used in the actual calculations. For instance, numerically tractable approximations to some parts of the nuclear SE originating from the Coriolis and vibrational-rotational couplings are yet to be derived. Our main emphasis in this work was on crystalline solids and we leave a more detailed treatment of molecules for future work. The practical implementations of the present method are under preparation.

To summarize, the present theory allows one to go beyond the BO approximation in a systematic and formally exact way by using the method of many-body Green's functions. We expect it will become a useful tool in treating beyond-Born-Oppenheimer effects in solids and molecules.

\appendix

\section{Coriolis and vibrational-rotational coupling terms}
\label{CoriolisAndVibrationalRotationalCouplingTerms}

The kinetic energy contributions $\hat{T}_{cvr}$ and $\hat{T}'_{cvr}$ are
\begin{eqnarray} 
\hat{T}_{cvr} &=& \sum_{\beta,\gamma } \hat{T}^{\left(1\right)}_{\beta\gamma} \int d\vec{r} \hat{\psi}^{\dagger}\left(\vec{r}\right)  r_{\gamma} \frac{\partial{ }}{\partial{r_{\beta} }} \hat{\psi}\left(\vec{r}\right)  \nonumber \\
&&+ \sum^{N_{n}-1}_{k = 1} \sum_{ \alpha, \beta ,\gamma  } \hat{T}^{\left(2\right)}_{\alpha\beta\gamma}\left(k\right) \hat{p}_{\alpha}\left(k\right) \nonumber \\
&&\times \int d\vec{r} \hat{\psi}^{\dagger}\left(\vec{r}\right) r_{\gamma} \frac{\partial{ }}{ \partial{r_{\beta} } } \hat{\psi}\left(\vec{r}\right)  \nonumber \\
&&+ \sum_{ \beta,\sigma,\gamma,\gamma' } \hat{T}^{\left(3\right)}_{\gamma\sigma \gamma' \beta} \int d\vec{r} \int d\vec{r}' \nonumber \\
&&\times \hat{\psi}^{\dagger}\left(\vec{r}\right) \hat{\psi}^{\dagger}\left(\vec{r}'\right) r_{\gamma'} r'_{\gamma} \frac{\partial^{2}{ }}{\partial{r'_{\sigma} } \partial{r_{\beta} }}   \hat{\psi}\left(\vec{r}'\right) \hat{\psi}\left(\vec{r}\right),  \nonumber \\
\hat{T}'_{cvr} &=&  \sum_{ \sigma,\gamma  } \hat{M}^{\left(1\right)}_{\sigma\gamma} \int d\vec{r} \hat{\psi}^{\dagger}\left(\vec{r}\right) r_{\gamma} \frac{\partial{ }}{\partial{r_{\sigma} }} \hat{\psi}\left(\vec{r}\right)  \nonumber \\
&&+ \sum^{N_{n}-1}_{k=1} \sum_{ \alpha,\sigma,\gamma } \hat{M}^{\left(2\right)}_{\alpha\sigma\gamma}\left(k\right) \hat{p}_{\alpha}\left(k\right) \nonumber \\
&&\times \int d\vec{r} \hat{\psi}^{\dagger}\left(\vec{r}\right) r_{\gamma} \frac{\partial{ }}{ \partial{r_{\sigma} }} \hat{\psi}\left(\vec{r}\right) \nonumber \\
&&+ \sum_{ \beta, \sigma,\gamma,\gamma'  } \hat{M}^{\left(3\right)}_{\gamma\sigma \gamma' \beta} \int d\vec{r} \int d\vec{r}' \nonumber \\
&&\times \hat{\psi}^{\dagger}\left(\vec{r}\right) \hat{\psi}^{\dagger}\left(\vec{r}'\right) r_{\gamma'} r'_{\gamma} \frac{\partial^{2}{ }}{\partial{r'_{\sigma} } \partial{r_{\beta} }} \hat{\psi}\left(\vec{r}'\right) \hat{\psi}\left(\vec{r}\right).
\label{eq:CoriolisAndVibrationalRotCouplingTermsEq_1}
\end{eqnarray}
In Eq. \ref{eq:CoriolisAndVibrationalRotCouplingTermsEq_1}
\begin{eqnarray} 
\hat{T}^{\left(1\right)}_{\beta\gamma} &\equiv& - \sum^{N_{n}-1}_{k = 1}  \frac{ \hbar^{2} }{ 2 M_{k} } \sum^{3}_{\alpha,\beta' = 1 } \frac{\partial^{2}{ \hat{\mathcal{R}}_{\beta\beta'} }}{ \partial{u^{2}_{\alpha}\left(k\right)} } \hat{\mathcal{R}}^{T}_{\beta'\gamma}, \nonumber \\
\hat{T}^{\left(2\right)}_{\alpha\beta\gamma}\left(k\right) &\equiv& - \frac{ i \hbar }{ M_{k} } \sum^{3}_{ \beta' = 1 } \frac{\partial{ \hat{\mathcal{R}}_{\beta\beta'} }}{\partial{ u_{\alpha}\left(k\right) }} \hat{\mathcal{R}}^{T}_{\beta'\gamma},  \nonumber \\
\hat{T}^{\left(3\right)}_{\gamma\sigma \gamma' \beta} &\equiv& - \sum^{N_{n}-1}_{k = 1}  \frac{ \hbar^{2} }{ 2 M_{k} } \sum^{3}_{ \alpha,\beta',\sigma' = 1 } \frac{\partial{ \hat{\mathcal{R}}_{\sigma\sigma'} }}{\partial{ u_{\alpha}\left(k\right) }}  \hat{\mathcal{R}}^{T}_{\sigma'\gamma}  \nonumber \\
&&\times \frac{\partial{ \hat{\mathcal{R}}_{\beta\beta'} }}{\partial{ u_{\alpha}\left(k\right) }} \hat{\mathcal{R}}^{T}_{\beta'\gamma'}, 
\label{eq:CoriolisAndVibrationalRotCouplingTermsEq_3}
\end{eqnarray}
and
\begin{eqnarray} 
\hat{M}^{\left(1\right)}_{\sigma\gamma} &\equiv&  \frac{ \hbar^{2} }{ 2 M_{nuc} } \sum^{N_{n}-1}_{k,k'=1}  \sum^{3}_{ \alpha,\sigma' = 1 } \frac{\partial^{2}{ \hat{\mathcal{R}}_{\sigma\sigma'} }}{ \partial{u_{\alpha}\left(k\right)} \partial{ u_{\alpha}\left(k'\right) }} \hat{\mathcal{R}}^{T}_{\sigma'\gamma},  \nonumber \\
\hat{M}^{\left(2\right)}_{\alpha\sigma\gamma}\left(k\right) &\equiv& \frac{i \hbar }{  M_{nuc} } \sum^{N_{n}-1}_{k'=1} \sum^{3}_{ \sigma' = 1 } \frac{\partial{ \hat{\mathcal{R}}_{\sigma\sigma'} }}{\partial{ u_{\alpha}\left(k'\right) }} \hat{\mathcal{R}}^{T}_{\sigma'\gamma}, \nonumber \\
\hat{M}^{\left(3\right)}_{\gamma\sigma \gamma' \beta} &\equiv& \frac{ \hbar^{2} }{ 2 M_{nuc} } \sum^{N_{n}-1}_{k,k'=1}  \sum^{3}_{ \alpha,\beta',\sigma' = 1 }  \frac{\partial{ \hat{\mathcal{R}}_{\sigma\sigma'} }}{\partial{ u_{\alpha}\left(k'\right) }}  \hat{\mathcal{R}}^{T}_{\sigma'\gamma} \nonumber \\
&&\times \frac{\partial{ \hat{\mathcal{R}}_{\beta\beta'} }}{\partial{ u_{\alpha}\left(k\right) }} \hat{\mathcal{R}}^{T}_{\beta'\gamma'}. 
\label{eq:CoriolisAndVibrationalRotCouplingTermsEq_4}
\end{eqnarray}
Here we divided the Coriolis and vibrational-rotational coupling terms into two parts: one part, $\hat{T}'_{cvr}$, is proportional to the inverse of the total nuclear mass while the other, $\hat{T}_{cvr}$, is not. All the quantities defined by Eqs. \ref{eq:CoriolisAndVibrationalRotCouplingTermsEq_3} and \ref{eq:CoriolisAndVibrationalRotCouplingTermsEq_4} are functions of $\vec{x}$ and $\hat{\vec{u}}$. Strictly speaking, the derivatives like $\partial{ \hat{\mathcal{R}}_{\sigma\sigma'} }/ \partial{ u_{\alpha}\left(k'\right) }$, are not so well defined objects from a notational point of view. By this notation we mean that the quantities like $\partial{ \mathcal{R}_{\sigma\sigma'} } / \partial{ u_{\alpha}\left(k'\right) }$ and $\partial^{2}{ \mathcal{R}_{\sigma\sigma'} }/ \partial{u_{\alpha}\left(k\right)} \partial{ u_{\alpha}\left(k'\right) }$, are some functions of $\vec{x},\vec{u}$ after differentiation and we can then find the corresponding functions of operators $\hat{\vec{u}}$. For example if $\partial{ \mathcal{R} }/\partial{ u_{\alpha}\left(k\right) } = G_{\alpha}\left(k,\vec{x},\vec{u}\right)$, then in our notation $\partial{ \hat{\mathcal{R}} }/\partial{ u_{\alpha}\left(k\right) } = G_{\alpha}\left(k,\vec{x},\hat{\vec{u}}\right)$.

We also use the following form of the Coriolis and vibrational-rotational coupling terms
\begin{equation}
\hat{T}_{cvr} + \hat{T}'_{cvr} = \sum^{3}_{c=1} \hat{T}^{\left(c\right)}_{cvr},
\label{eq:CoriolisAndVibrationalRotCouplingTermsEq_9}
\end{equation}
where
\begin{eqnarray} 
\hat{T}^{\left(1\right)}_{cvr} &\equiv& \sum_{\beta,\gamma } \int d\vec{r} r_{\gamma} \frac{\partial{ }}{\partial{r_{\beta} }}  \hat{L}^{\left(1\right)}_{\beta\gamma}  \hat{n}_{e}\left(\vec{r}\right), \nonumber \\
\hat{T}^{\left(2\right)}_{cvr} &\equiv& \sum^{N_{n}-1}_{k = 1} \sum_{ \alpha, \beta ,\gamma  } \int d\vec{r} r_{\gamma} \frac{\partial{ }}{ \partial{r_{\beta} } }  \hat{L}^{\left(2\right)}_{\alpha\beta\gamma}\left(k\right) \hat{p}_{\alpha}\left(k\right)    \hat{n}_{e}\left(\vec{r}\right),  \nonumber \\
\hat{T}^{\left(3\right)}_{cvr} &\equiv& \sum_{ \beta,\sigma,\gamma,\gamma' } \int d\vec{r} \int d\vec{r}'  r_{\gamma'} r'_{\gamma} \frac{\partial^{2}{ }}{\partial{r'_{\sigma} } \partial{r_{\beta} }}   \nonumber \\
&&\times \hat{L}^{\left(3\right)}_{\gamma\sigma \gamma' \beta} \hat{\psi}^{\dagger}\left(\vec{r}\right) \hat{\psi}^{\dagger}\left(\vec{r}'\right) \hat{\psi}\left(\vec{r}'\right) \hat{\psi}\left(\vec{r}\right),
\label{eq:CoriolisAndVibrationalRotCouplingTermsEq_10}
\end{eqnarray}
and the quantities $\hat{L}^{\left(1\right)}_{\beta\gamma}$, $\hat{L}^{\left(2\right)}_{\alpha\beta\gamma}\left(k\right)$ and $\hat{L}^{\left(3\right)}_{\gamma\sigma \gamma' \beta}$ are defined by Eq. \ref{eq:EquationsOfMotionForElectronsEq_1_3_5}.

\section{Total Energy}
\label{TotalEnergy}

We expand $\hat{n}_{n}\left(\vec{r}\right)$, $\hat{V}_{en}\left(\vec{r}\right)$ and $\hat{V}'_{nn}$ into Taylor series with respect to the displacements:
\begin{eqnarray} 
\hat{V}'_{nn} &=& \sum^{\infty}_{m = 0} \frac{ 1 }{m!} \sum_{k_{\bar{m}},\alpha_{\bar{m}}} \frac{ \partial^{m}{ V_{nn}\left(\vec{x}\right) }}{ \partial{x_{\alpha_{\bar{m}}}\left(k_{\bar{m}}\right)} } \hat{u}_{\alpha_{\bar{m}}}\left(k_{\bar{m}}\right), \nonumber \\
V_{nn}\left(\vec{x}\right) &=& \sum^{N_{n}}_{ k,k'= 1 }{}^{'} v\left(\vec{x}_{k}, \vec{x}_{k'} \right), \nonumber \\
\vec{x}_{N_{n}} &\equiv& -\frac{1}{ M_{N_{n}} } \sum^{N_{n}-1}_{ k= 1 } M_{k} \vec{x}_{k}, \nonumber \\
\hat{V}_{en}\left(\vec{r}\right) &=& \sum^{\infty}_{m = 0} \frac{ 1 }{m!} \sum_{k_{\bar{m}},\alpha_{\bar{m}}} \frac{ \partial^{m}{ V_{en}\left(\vec{r},\vec{x}\right) }}{ \partial{x_{\alpha_{\bar{m}}}\left(k_{\bar{m}}\right)} } \hat{u}_{\alpha_{\bar{m}}}\left(k_{\bar{m}}\right), \nonumber \\
V_{en}\left(\vec{r},\vec{x}\right) &=& \sum^{N_{n}}_{ k= 1 } \frac{ - Z_{k} \varsigma }{ \left| \vec{r} - \mathcal{R}\left[\boldsymbol{\theta}\left(\vec{x}\right)\right] \vec{x}_{k} \right| },
\label{eq:UsefulRelationsEq_9}
\end{eqnarray}
and
\begin{eqnarray}
\hat{n}_{n}\left(\vec{r}\right) &=& \sum^{\infty}_{m = 0} \frac{1}{m!} \sum_{k_{\bar{m}},\alpha_{\bar{m}}} \frac{ \partial^{m}{ n^{\left(0\right)}_{n}\left(\vec{x},\vec{r}\right) }}{ \partial{x_{\alpha_{\bar{m}}}\left(k_{\bar{m}}\right)} } \hat{u}_{\alpha_{\bar{m}}}\left(k_{\bar{m}}\right), \nonumber \\
n^{\left(0\right)}_{n}\left(\vec{x}, \vec{r}\right) &\equiv& - \sum^{N_{n}}_{ k = 1 } Z_{k} \delta\left( \vec{r} - \mathcal{R}\left[\boldsymbol{\theta}\left(\vec{x}\right)\right] \vec{x}_{k} \right).
\label{eq:UsefulRelationsEq_10}
\end{eqnarray}
Here, the following notations are used
\begin{eqnarray} 
\hat{u}_{\alpha_{\bar{m}}}\left(k_{\bar{m}}\right) &\equiv& \hat{u}_{\alpha_{1}}\left(k_{1}\right) \cdots \hat{u}_{\alpha_{m}}\left(k_{m}\right), \nonumber \\
\sum_{k_{\bar{m}},\alpha_{\bar{m}}} &=& \sum_{k_{1},\alpha_{1}} \cdots \sum_{k_{m},\alpha_{m}},
\label{eq:NotationEq_1}
\end{eqnarray}
and so on. In Sec. \ref{EulerAngles} we discuss how to actually obtain the rotation matrix $\mathcal{R}\left[\boldsymbol{\theta}\left(\vec{x}\right)\right]$ and the necessary derivatives of it.

Next we deduce an exact form of the total energy of the system. We start by writing
\begin{equation}
\hat{V}'_{nn} = \sum^{\infty}_{m = 0} \hat{V}^{\left(m\right)}_{nn}, \quad  \hat{V}_{en}\left(\vec{r}\right) = \sum^{\infty}_{m = 0} \hat{V}^{\left(m\right)}_{en}\left(\vec{r}\right),
\label{eq:UsefulRelationsEq_10_0_1}
\end{equation}
where
\begin{eqnarray} 
\hat{V}^{\left(m\right)}_{nn} &\equiv& \frac{1}{m!} \sum_{k_{\bar{m}},\alpha_{\bar{m}}} \frac{ \partial^{m}{ V_{nn}\left(\vec{x}\right) }}{ \partial{x_{\alpha_{\bar{m}}}\left(k_{\bar{m}}\right)} } \hat{u}_{\alpha_{\bar{m}}}\left(k_{\bar{m}}\right), \nonumber \\
\hat{V}^{\left(m\right)}_{en} &\equiv& \frac{1}{m!} \sum_{k_{\bar{m}},\alpha_{\bar{m}}} \frac{ \partial^{m}{ V_{en}\left(\vec{r},\vec{x}\right) }}{ \partial{x_{\alpha_{\bar{m}}}\left(k_{\bar{m}}\right)} } \hat{u}_{\alpha_{\bar{m}}}\left(k_{\bar{m}}\right),
\label{eq:UsefulRelationsEq_10_0_2}
\end{eqnarray}
such that
\begin{equation}
\hat{V}'_{en} = \int d\vec{r} \sum^{\infty}_{m = 0} \hat{V}^{\left(m\right)}_{en}\left(\vec{r}\right) \hat{n}_{e}\left(\vec{r}\right).
\label{eq:UsefulRelationsEq_10_0_3}
\end{equation}
We note that from these expansions it follows that
\begin{eqnarray} 
\left\langle  \hat{V}'_{en} \right\rangle &=& -i\hbar \int d\vec{r} V_{en}\left(\vec{r},\vec{x}\right) G\left(\vec{r}z,\vec{r}z^{+}\right) \nonumber \\
&&- \frac{ 1 }{i \hbar } \sum^{\infty}_{m = 1} \frac{1 }{m} \sum_{k,\alpha} \int d\vec{r} \nonumber \\
&&\times \left\langle  \left[\hat{p}_{\alpha}\left(k\right), \hat{V}^{\left(m\right)}_{en}\left(\vec{r}\right) \right]_{-} \hat{u}_{\alpha}\left(k\right) \hat{n}_{e}\left(\vec{r}\right) \right\rangle, \nonumber \\
\label{eq:UsefulRelationsEq_10_1}
\end{eqnarray}
and
\begin{eqnarray}
\left\langle  \hat{V}'_{nn} \right\rangle &=& V_{nn}\left(\vec{x}\right) - \frac{ 1 }{i \hbar } \sum^{\infty}_{m = 1} \frac{1 }{m} \sum_{k,\alpha} \nonumber \\
&&\times \left\langle  \left[\hat{p}_{\alpha}\left(k\right), \hat{V}^{\left(m\right)}_{nn} \right]_{-} \hat{u}_{\alpha}\left(k\right) \right\rangle,
\label{eq:UsefulRelationsEq_10_2}
\end{eqnarray}
where we used $\left\langle  \hat{n}_{e}\left(\vec{r}z\right) \right\rangle = -i\hbar G\left(\vec{r}z,\vec{r}z^{+}\right)$. We write the quantity defined by Eq. \ref{eq:EquationsOfMotionForDisplacementsEq_3_5} as $K_{\alpha}\left(kz\right) = K^{en}_{\alpha}\left(kz\right) + K^{nn}_{\alpha}\left(kz\right)$ such that
\begin{equation}
K^{y}_{\alpha}\left(kz\right) \equiv \frac{ 1 }{i\hbar} \left\langle  \left[ \hat{p}_{\alpha}\left(kz\right) , \hat{V}'_{y} \right]_{-} \right\rangle,
\label{eq:UsefulRelationsEq_10_3}
\end{equation}
where $y = en,nn$. In a similar way we write the quantity defined by Eq. \ref{eq:EquationsOfMotionForDisplacementsEq_5} as $K_{\alpha\beta}\left(kz,k'z'\right) = K^{en}_{\alpha\beta}\left(kz,k'z'\right) + K^{nn}_{\alpha\beta}\left(kz,k'z'\right)$, where
\begin{equation}
K^{y}_{\alpha\beta}\left(kz,k'z'\right) \equiv \frac{\delta{ K^{y}_{\alpha}\left(kz\right) }}{\delta{J_{\beta}\left(k'z'\right)}}.
\label{eq:UsefulRelationsEq_10_4}
\end{equation}
We define
\begin{eqnarray} 
\Pi^{y}_{\alpha \alpha'}\left(kz,\bar{k}\bar{z}\right) &\equiv& -\sum_{k',\beta} \int_{\gamma} dz' K^{y}_{\alpha \beta}\left(k z,k'z'\right) \nonumber \\
&&\times D^{-1}_{\beta \alpha'}\left(k'z',\bar{k}\bar{z}\right).
\label{eq:UsefulRelationsEq_10_5}
\end{eqnarray}
Next we use Eq. \ref{eq:UsefulRelationsEq_10_0_1} and write $K^{en}_{\alpha}\left(kz\right)$ and $K^{nn}_{\alpha}\left(kz\right)$ such that
\begin{equation}
K^{y}_{\alpha}\left(kz\right) = \sum^{\infty}_{m = 1} K^{y,\left(m\right)}_{\alpha}\left(kz\right), 
\label{eq:UsefulRelationsEq_10_5_2}
\end{equation}
and in a similar way for all the quantities derived by using $K^{en}_{\alpha}\left(kz\right)$ and $K^{nn}_{\alpha}\left(kz\right)$. By using Eq. \ref{eq:UsefulRelationsEq_10_5_2} in Eqs. \ref{eq:UsefulRelationsEq_10_4} and \ref{eq:UsefulRelationsEq_10_5} and these results with Eqs. \ref{eq:UsefulRelationsEq_10_1} and \ref{eq:UsefulRelationsEq_10_2} we eventually find that
\begin{eqnarray}
\left\langle  \hat{V}'_{nn} \right\rangle &=& V_{nn}\left(\vec{x}\right) - \sum^{\infty}_{m = 1} \frac{1 }{m} \sum_{k,\alpha}  K^{nn,\left(m\right)}_{\alpha}\left(kz\right) \left\langle \hat{u}_{\alpha}\left(kz\right) \right\rangle \nonumber \\
&&+ i \hbar \sum^{\infty}_{m = 1} \frac{1 }{m} \sum_{k,\alpha} \sum_{k'',\alpha'} \int_{\gamma} dz'' \Pi^{nn,\left(m\right)}_{\alpha \alpha'}\left(k z,k''z''\right) \nonumber \\
&&\times D_{\alpha' \alpha }\left(k''z'',kz\right),
\label{eq:UsefulRelationsEq_10_6}
\end{eqnarray}
and
\begin{eqnarray} 
\left\langle  \hat{V}'_{en} \right\rangle &=& -i\hbar \int d\vec{r} V_{en}\left(\vec{r},\vec{x}\right) G\left(\vec{r}z,\vec{r}z^{+}\right) \nonumber \\
&&- \sum^{\infty}_{m = 1} \frac{1 }{m} \sum_{k,\alpha}  K^{en,\left(m\right)}_{\alpha}\left(kz\right) \left\langle \hat{u}_{\alpha}\left(kz\right) \right\rangle  \nonumber \\
&&+ i \hbar \sum^{\infty}_{m = 1} \frac{1 }{m} \sum_{k,\alpha} \sum_{k'',\alpha'} \int_{\gamma} dz'' \Pi^{en,\left(m\right)}_{\alpha \alpha'}\left(k z,k''z''\right) \nonumber \\
&&\times D_{\alpha' \alpha }\left(k''z'',kz\right).
\label{eq:UsefulRelationsEq_10_7}
\end{eqnarray}
We use these results to write the total energy $E_{tot}$ in terms of the Green's functions and related quantities as
\begin{equation} 
E_{tot} = \left\langle  \hat{T}_{tot} \right\rangle + \left\langle  \hat{V}'_{ee} \right\rangle + \left\langle  \hat{V}'_{en} \right\rangle + \left\langle  \hat{V}'_{nn} \right\rangle,
\label{eq:TotalEnergyAndGrandPotentialEq_7}
\end{equation}
where the external potential terms are put to zero and $\hat{T}_{tot}$ is given by Eq. \ref{eq:HamiltonianEq_11}. We have
\begin{eqnarray} 
\left\langle  \hat{T}'_{e} \right\rangle &=& i \frac{ \hbar^{3} }{ 2 m_{e} } \int d\vec{r} \nabla^{2}_{\vec{r}} G\left(\vec{r}z,\vec{r}z^{+}\right), \nonumber \\
\left\langle  \hat{T}'_{n} \right\rangle &=& \sum^{N_{n}-1}_{k = 1} \sum^{3}_{\alpha = 1} \frac{  i \hbar }{ 2 M_{k} } D^{pp,\mathcal{T}}_{\alpha \alpha}\left(kz,kz^{+}\right), \nonumber \\
\left\langle  \hat{T}'_{mpn} \right\rangle &=& - \frac{  i \hbar }{ 2 M_{nuc} } \sum^{N_{n}-1}_{k = 1} \sum^{3}_{\alpha = 1}  D^{pp,\mathcal{T}}_{\alpha \alpha}\left(kz,k'z^{+}\right). \nonumber \\
\label{eq:TotalEnergyAndGrandPotentialEq_8}
\end{eqnarray}
Here
\begin{equation} 
D^{pp,\mathcal{T}}_{\alpha \beta}\left(kz,k'z'\right) \equiv \frac{1}{i \hbar} \left\langle \mathcal{T}_{\gamma} \left\{ \hat{p}_{\alpha}\left(kz\right) \hat{p}_{\beta}\left(k'z'\right) \right\} \right\rangle,
\label{eq:TotalEnergyAndGrandPotentialEq_9}
\end{equation}
and this function can be obtained from the solution of Eq. \ref{eq:EquationsOfMotionForDisplacementsEq_8_4} since with the external potentials put to zero $\left\langle \hat{p}_{\alpha}\left(k z\right) \right\rangle = \left\langle \hat{p}_{\alpha}\left(k\right) \right\rangle$. By using the EOM for the field operator given by Eq. \ref{eq:EquationsOfMotionForElectronsEq_1_3_2} without the external potentials we find that
\begin{eqnarray} 
&& \frac{ i \hbar }{2} \int d\vec{r}  \left[- i \hbar \frac{\partial}{\partial{z}} - \frac{ \hbar^{2} }{ 2 m_{e} } \nabla^{2}_{\vec{r}} \right] G\left(\vec{r}z,\vec{r}z^{+}\right)  \nonumber \\
&&=  \frac{1}{2} \left[ \left\langle  \hat{V}'_{en} \right\rangle + \left\langle  \hat{T}^{\left(1\right)}_{cvr} \right\rangle + \left\langle  \hat{T}^{\left(2\right)}_{cvr} \right\rangle \right] + \left\langle  \hat{T}^{\left(3\right)}_{cvr} \right\rangle \nonumber \\
&&\ \ \ + \left\langle  \hat{V}'_{ee} \right\rangle + \left\langle  \hat{T}'_{mpe} \right\rangle,
\label{eq:TotalEnergyAndGrandPotentialEq_10}
\end{eqnarray} 
where $\hat{T}^{\left(c\right)}_{cvr}$ are given by Eqs. \ref{eq:CoriolisAndVibrationalRotCouplingTermsEq_9} and \ref{eq:CoriolisAndVibrationalRotCouplingTermsEq_10}. This result can be found \cite{Hedin-EffectsOfElectElectrAndElectrPhononInteractOnTheOneElectrStatesOfSolids-1970} by multiplying Eq. \ref{eq:EquationsOfMotionForElectronsEq_1_3_2} from the left with $\hat{\psi}^{\dagger}\left(\vec{r}z\right)/2$, integrating over $\vec{r}$, taking an ensemble average and then establishing some rearranging. From Eqs. \ref{eq:EquationsOfMotionForElectronsEq_2_21}, \ref{eq:CoriolisAndVibrationalRotCouplingTermsEq_11} and \ref{eq:CoriolisAndVibrationalRotCouplingTermsEq_10}, it follows that
\begin{eqnarray} 
\left\langle  \hat{T}^{\left(1\right)}_{cvr} \right\rangle + \left\langle  \hat{T}^{\left(2\right)}_{cvr} \right\rangle &=& -i \hbar \int d\vec{r} \int d\vec{r}' \int_{\gamma} dz'  \Sigma_{2}\left(\vec{r}z,\vec{r}'z'\right) \nonumber \\
&&\times G\left(\vec{r}'z',\vec{r}z\right).
\label{eq:TotalEnergyAndGrandPotentialEq_11}
\end{eqnarray}
Now we have all the necessary ingredients in place to express the exact total energy in terms of the Green's functions and SE's. By using Eqs. \ref{eq:UsefulRelationsEq_10_6}, \ref{eq:UsefulRelationsEq_10_7}, \ref{eq:TotalEnergyAndGrandPotentialEq_10} and \ref{eq:TotalEnergyAndGrandPotentialEq_11}, the total energy can be written as
\begin{eqnarray} 
&&E_{tot} \nonumber \\
&&= \frac{i \hbar}{2}\sum^{N_{n}-1}_{k,k' = 1} \sum^{3}_{\alpha = 1} \left( M^{-1}_{k} \delta_{kk'} - M^{-1}_{nuc} \right) D^{pp,\mathcal{T}}_{\alpha \alpha}\left(kz,k'z^{+}\right) \nonumber \\
&&- \frac{1}{2}\sum^{\infty}_{m = 1} \frac{1 }{m} \sum_{k,\alpha} \left[ K^{en,\left(m\right)}_{\alpha}\left(kz\right) + 2 K^{nn,\left(m\right)}_{\alpha}\left(kz\right) \right] \left\langle \hat{u}_{\alpha}\left(kz\right) \right\rangle \nonumber \\
&&+ V_{nn}\left(\vec{x}\right) + \frac{i \hbar}{2} \sum^{\infty}_{m = 1} \frac{1 }{m} \sum_{k,\alpha} \sum_{k'',\alpha'} \int_{\gamma} dz'' \nonumber \\
&&\times \left[ \Pi^{en,\left(m\right)}_{\alpha \alpha'}\left(kz,k''z''\right) + 2 \Pi^{nn,\left(m\right)}_{\alpha \alpha'}\left(k z,k''z''\right) \right] \nonumber \\
&&\times D_{\alpha' \alpha }\left(k''z'',kz\right) \nonumber \\
&&-\frac{ i \hbar }{2} \int d\vec{r}  \left[ i \hbar \frac{\partial}{\partial{z}} - \frac{ \hbar^{2} }{ 2 m } \nabla^{2}_{\vec{r}} + V_{en}\left(\vec{r},\vec{x}\right) \right] G\left(\vec{r}z,\vec{r}z^{+}\right) \nonumber \\
&&- \frac{i \hbar}{2} \int d\vec{r} \int d\vec{r}' \int_{\gamma} dz'  \Sigma_{2}\left(\vec{r}z,\vec{r}'z'\right) G\left(\vec{r}'z',\vec{r}z\right).
\label{eq:TotalEnergyAndGrandPotentialEq_12}
\end{eqnarray}
This is an exact form of $E_{tot}$. Provided $\vec{x}$ are the equilibrium positions, the second term on the right hand side of Eq. \ref{eq:TotalEnergyAndGrandPotentialEq_12} involving $\left\langle \hat{u}_{\alpha}\left(kz\right) \right\rangle$ vanishes and $D_{\beta \alpha }\left(k'z',kz\right) = D^{\mathcal{T}}_{\beta \alpha }\left(k'z',kz\right)$.



\begin{acknowledgments}
V.J.H. thanks Dr. Ivan Gonoskov for useful discussions on various aspects of the present work. E.K.U.G. acknowledges financial support by the European Research Council Advanced Grant FACT (ERC-2017-AdG-788890).
\end{acknowledgments}
\bibliography{bibfile}

\begin{thebibliography}{90}%
\makeatletter
\providecommand \@ifxundefined [1]{%
 \@ifx{#1\undefined}
}%
\providecommand \@ifnum [1]{%
 \ifnum #1\expandafter \@firstoftwo
 \else \expandafter \@secondoftwo
 \fi
}%
\providecommand \@ifx [1]{%
 \ifx #1\expandafter \@firstoftwo
 \else \expandafter \@secondoftwo
 \fi
}%
\providecommand \natexlab [1]{#1}%
\providecommand \enquote  [1]{``#1''}%
\providecommand \bibnamefont  [1]{#1}%
\providecommand \bibfnamefont [1]{#1}%
\providecommand \citenamefont [1]{#1}%
\providecommand \href@noop [0]{\@secondoftwo}%
\providecommand \href [0]{\begingroup \@sanitize@url \@href}%
\providecommand \@href[1]{\@@startlink{#1}\@@href}%
\providecommand \@@href[1]{\endgroup#1\@@endlink}%
\providecommand \@sanitize@url [0]{\catcode `\\12\catcode `\$12\catcode
  `\&12\catcode `\#12\catcode `\^12\catcode `\_12\catcode `\%12\relax}%
\providecommand \@@startlink[1]{}%
\providecommand \@@endlink[0]{}%
\providecommand \url  [0]{\begingroup\@sanitize@url \@url }%
\providecommand \@url [1]{\endgroup\@href {#1}{\urlprefix }}%
\providecommand \urlprefix  [0]{URL }%
\providecommand \Eprint [0]{\href }%
\providecommand \doibase [0]{http://dx.doi.org/}%
\providecommand \selectlanguage [0]{\@gobble}%
\providecommand \bibinfo  [0]{\@secondoftwo}%
\providecommand \bibfield  [0]{\@secondoftwo}%
\providecommand \translation [1]{[#1]}%
\providecommand \BibitemOpen [0]{}%
\providecommand \bibitemStop [0]{}%
\providecommand \bibitemNoStop [0]{.\EOS\space}%
\providecommand \EOS [0]{\spacefactor3000\relax}%
\providecommand \BibitemShut  [1]{\csname bibitem#1\endcsname}%
\let\auto@bib@innerbib\@empty
\bibitem [{\citenamefont {Born}\ and\ \citenamefont
  {Oppenheimer}(1927)}]{Born-Oppenheimer-Adiabatic-Approx.1927}%
  \BibitemOpen
  \bibfield  {author} {\bibinfo {author} {\bibfnamefont {M.}~\bibnamefont
  {Born}}\ and\ \bibinfo {author} {\bibfnamefont {R.}~\bibnamefont
  {Oppenheimer}},\ }\href {\doibase 10.1002/andp.19273892002} {\bibfield
  {journal} {\bibinfo  {journal} {Ann. Phys. (Leipzig)}\ }\textbf {\bibinfo
  {volume} {389}},\ \bibinfo {pages} {457} (\bibinfo {year}
  {1927})}\BibitemShut {NoStop}%
\bibitem [{\citenamefont {Huang}\ and\ \citenamefont
  {Born}(1954)}]{born-huang-dynamical-1954}%
  \BibitemOpen
  \bibfield  {author} {\bibinfo {author} {\bibfnamefont {K.}~\bibnamefont
  {Huang}}\ and\ \bibinfo {author} {\bibfnamefont {M.}~\bibnamefont {Born}},\
  }\href@noop {} {\emph {\bibinfo {title} {Dynamical Theory of Crystal
  Lattices}}}\ (\bibinfo  {publisher} {Clarendon Press Oxford},\ \bibinfo
  {year} {1954})\BibitemShut {NoStop}%
\bibitem [{\citenamefont {Hohenberg}\ and\ \citenamefont
  {Kohn}(1964)}]{Hohenberg-DFT-PhysRev.136.B864-1964}%
  \BibitemOpen
  \bibfield  {author} {\bibinfo {author} {\bibfnamefont {P.}~\bibnamefont
  {Hohenberg}}\ and\ \bibinfo {author} {\bibfnamefont {W.}~\bibnamefont
  {Kohn}},\ }\href {http://link.aps.org/doi/10.1103/PhysRev.136.B864}
  {\bibfield  {journal} {\bibinfo  {journal} {Phys. Rev.}\ }\textbf {\bibinfo
  {volume} {136}},\ \bibinfo {pages} {B864} (\bibinfo {year}
  {1964})}\BibitemShut {NoStop}%
\bibitem [{\citenamefont {Kohn}\ and\ \citenamefont
  {Sham}(1965)}]{KohnSham-DFT-PhysRev.140.A1133-1965}%
  \BibitemOpen
  \bibfield  {author} {\bibinfo {author} {\bibfnamefont {W.}~\bibnamefont
  {Kohn}}\ and\ \bibinfo {author} {\bibfnamefont {L.~J.}\ \bibnamefont
  {Sham}},\ }\href {http://link.aps.org/doi/10.1103/PhysRev.140.A1133}
  {\bibfield  {journal} {\bibinfo  {journal} {Phys. Rev.}\ }\textbf {\bibinfo
  {volume} {140}},\ \bibinfo {pages} {A1133} (\bibinfo {year}
  {1965})}\BibitemShut {NoStop}%
\bibitem [{\citenamefont {Parr}\ and\ \citenamefont
  {Yang}(1989)}]{Parr-DFTbook-1989}%
  \BibitemOpen
  \bibfield  {author} {\bibinfo {author} {\bibfnamefont {R.~G.}\ \bibnamefont
  {Parr}}\ and\ \bibinfo {author} {\bibfnamefont {W.}~\bibnamefont {Yang}},\
  }\href@noop {} {\emph {\bibinfo {title} {Density Functional Theory of Atoms
  and Molecules}}}\ (\bibinfo  {publisher} {Oxford University Press},\ \bibinfo
  {year} {1989})\BibitemShut {NoStop}%
\bibitem [{\citenamefont {Dreizler}\ and\ \citenamefont
  {Gross}(1990)}]{DreizlerGross-DFTbook-1990}%
  \BibitemOpen
  \bibfield  {author} {\bibinfo {author} {\bibfnamefont {R.~M.}\ \bibnamefont
  {Dreizler}}\ and\ \bibinfo {author} {\bibfnamefont {E.~K.~U.}\ \bibnamefont
  {Gross}},\ }\href@noop {} {\emph {\bibinfo {title} {Density Functional
  Theory: An Approach to the Quantum Many-Body Problem}}}\ (\bibinfo
  {publisher} {Springer-Verlag},\ \bibinfo {year} {1990})\BibitemShut {NoStop}%
\bibitem [{\citenamefont {Martin}\ and\ \citenamefont
  {Schwinger}(1959)}]{Martin-SchwingerPhysRev.115.1342-Theory-Many-Part.Syst.-1959}%
  \BibitemOpen
  \bibfield  {author} {\bibinfo {author} {\bibfnamefont {P.~C.}\ \bibnamefont
  {Martin}}\ and\ \bibinfo {author} {\bibfnamefont {J.}~\bibnamefont
  {Schwinger}},\ }\href {\doibase 10.1103/PhysRev.115.1342} {\bibfield
  {journal} {\bibinfo  {journal} {Phys. Rev.}\ }\textbf {\bibinfo {volume}
  {115}},\ \bibinfo {pages} {1342} (\bibinfo {year} {1959})}\BibitemShut
  {NoStop}%
\bibitem [{\citenamefont {Baym}(1961)}]{Baym-field-1961}%
  \BibitemOpen
  \bibfield  {author} {\bibinfo {author} {\bibfnamefont {G.}~\bibnamefont
  {Baym}},\ }\href@noop {} {\bibfield  {journal} {\bibinfo  {journal} {Ann.
  Phys.}\ }\textbf {\bibinfo {volume} {14}},\ \bibinfo {pages} {1} (\bibinfo
  {year} {1961})}\BibitemShut {NoStop}%
\bibitem [{\citenamefont
  {Maksimov}(1975)}]{Maksimov-SelfConsistentElectronPhonon-1975}%
  \BibitemOpen
  \bibfield  {author} {\bibinfo {author} {\bibfnamefont {E.~G.}\ \bibnamefont
  {Maksimov}},\ }\href@noop {} {\bibfield  {journal} {\bibinfo  {journal} {Zh.
  Eksp. Teor. Fiz}\ }\textbf {\bibinfo {volume} {69}},\ \bibinfo {pages} {2236}
  (\bibinfo {year} {1975})}\BibitemShut {NoStop}%
\bibitem [{\citenamefont
  {Hedin}(1965)}]{Hedin-NewMethForCalcTheOneParticleGreensFunctWithApplToTheElectronGasProb-PhysRev.139.A796-1965}%
  \BibitemOpen
  \bibfield  {author} {\bibinfo {author} {\bibfnamefont {L.}~\bibnamefont
  {Hedin}},\ }\href {\doibase 10.1103/PhysRev.139.A796} {\bibfield  {journal}
  {\bibinfo  {journal} {Phys. Rev.}\ }\textbf {\bibinfo {volume} {139}},\
  \bibinfo {pages} {A796} (\bibinfo {year} {1965})}\BibitemShut {NoStop}%
\bibitem [{\citenamefont {Mahan}(1990)}]{Mahan-many-particle-1990}%
  \BibitemOpen
  \bibfield  {author} {\bibinfo {author} {\bibfnamefont {G.}~\bibnamefont
  {Mahan}},\ }\href@noop {} {\emph {\bibinfo {title} {Many-Particle Physics}}}\
  (\bibinfo  {publisher} {Plenum Press},\ \bibinfo {year} {1990})\BibitemShut
  {NoStop}%
\bibitem [{\citenamefont {Fetter}\ and\ \citenamefont
  {Walecka}(1971)}]{Fetter-Walencka-q-theory-of-many-particle-1971}%
  \BibitemOpen
  \bibfield  {author} {\bibinfo {author} {\bibfnamefont {A.}~\bibnamefont
  {Fetter}}\ and\ \bibinfo {author} {\bibfnamefont {J.}~\bibnamefont
  {Walecka}},\ }\href@noop {} {\emph {\bibinfo {title} {Quantum Theory of
  Many-Particle Systems}}}\ (\bibinfo  {publisher} {McGraw-Hill},\ \bibinfo
  {year} {1971})\BibitemShut {NoStop}%
\bibitem [{\citenamefont {Stefanucci}\ and\ \citenamefont {van
  Leeuwen}(2013)}]{Stefanucci-Leeuwen-many-body-book-2013}%
  \BibitemOpen
  \bibfield  {author} {\bibinfo {author} {\bibfnamefont {G.}~\bibnamefont
  {Stefanucci}}\ and\ \bibinfo {author} {\bibfnamefont {R.}~\bibnamefont {van
  Leeuwen}},\ }\href@noop {} {\emph {\bibinfo {title} {Nonequilibrium Many-Body
  Theory of Quantum Systems}}}\ (\bibinfo  {publisher} {Cambridge University
  Press},\ \bibinfo {year} {2013})\BibitemShut {NoStop}%
\bibitem [{\citenamefont {Golze}\ \emph {et~al.}(2019)\citenamefont {Golze},
  \citenamefont {Dvorak},\ and\ \citenamefont
  {Rinke}}]{Golze-TheGWcompendiumApracticalGuideToTheoreticalPhotoemissionSpectroscopy-2019}%
  \BibitemOpen
  \bibfield  {author} {\bibinfo {author} {\bibfnamefont {D.}~\bibnamefont
  {Golze}}, \bibinfo {author} {\bibfnamefont {M.}~\bibnamefont {Dvorak}}, \
  and\ \bibinfo {author} {\bibfnamefont {P.}~\bibnamefont {Rinke}},\
  }\href@noop {} {\bibfield  {journal} {\bibinfo  {journal} {Frontiers Chem.}\
  }\textbf {\bibinfo {volume} {7}} (\bibinfo {year} {2019})}\BibitemShut
  {NoStop}%
\bibitem [{\citenamefont {Maradudin}\ and\ \citenamefont
  {Fein}(1962)}]{Maradudin-Fein-PhysRev.128.2589-Scat-Neutr.1962}%
  \BibitemOpen
  \bibfield  {author} {\bibinfo {author} {\bibfnamefont {A.~A.}\ \bibnamefont
  {Maradudin}}\ and\ \bibinfo {author} {\bibfnamefont {A.~E.}\ \bibnamefont
  {Fein}},\ }\href {\doibase 10.1103/PhysRev.128.2589} {\bibfield  {journal}
  {\bibinfo  {journal} {Phys. Rev.}\ }\textbf {\bibinfo {volume} {128}},\
  \bibinfo {pages} {2589} (\bibinfo {year} {1962})}\BibitemShut {NoStop}%
\bibitem [{\citenamefont {Cowley}(1963)}]{Cowley-lattice-1963}%
  \BibitemOpen
  \bibfield  {author} {\bibinfo {author} {\bibfnamefont {R.~A.}\ \bibnamefont
  {Cowley}},\ }\href@noop {} {\bibfield  {journal} {\bibinfo  {journal} {Adv.
  Phys.}\ }\textbf {\bibinfo {volume} {12}},\ \bibinfo {pages} {421} (\bibinfo
  {year} {1963})}\BibitemShut {NoStop}%
\bibitem [{\citenamefont {Shukla}\ and\ \citenamefont
  {Cowley}(1971)}]{Shukla-Helmholtz-1971}%
  \BibitemOpen
  \bibfield  {author} {\bibinfo {author} {\bibfnamefont {R.~C.}\ \bibnamefont
  {Shukla}}\ and\ \bibinfo {author} {\bibfnamefont {E.~R.}\ \bibnamefont
  {Cowley}},\ }\href@noop {} {\bibfield  {journal} {\bibinfo  {journal} {Phys.
  Rev. B}\ }\textbf {\bibinfo {volume} {3}},\ \bibinfo {pages} {4055} (\bibinfo
  {year} {1971})}\BibitemShut {NoStop}%
\bibitem [{\citenamefont {Maradudin}(1974)}]{maradudin-dyn-prop-solids-1974}%
  \BibitemOpen
  \bibfield  {author} {\bibinfo {author} {\bibfnamefont {A.~A.}\ \bibnamefont
  {Maradudin}},\ }\href@noop {} {\emph {\bibinfo {title} {Elements of The
  Theory of Lattice Dynamics}}},\ Vol.~\bibinfo {volume} {1}\ (\bibinfo
  {publisher} {North-Holland Publishing Company},\ \bibinfo {year} {1974})\
  pp.\ \bibinfo {pages} {1--82}\BibitemShut {NoStop}%
\bibitem [{\citenamefont {Rickayzen}(1980)}]{Rickayzen-GreensFunctions-1980}%
  \BibitemOpen
  \bibfield  {author} {\bibinfo {author} {\bibfnamefont {G.}~\bibnamefont
  {Rickayzen}},\ }\href@noop {} {\emph {\bibinfo {title} {Green's Functions and
  Condensed Matter, Acad}}}\ (\bibinfo  {publisher} {Academic Press},\ \bibinfo
  {year} {1980})\BibitemShut {NoStop}%
\bibitem [{\citenamefont {Giannozzi}\ \emph {et~al.}(1991)\citenamefont
  {Giannozzi}, \citenamefont {de~Gironcoli}, \citenamefont {Pavone},\ and\
  \citenamefont
  {Baroni}}]{Giannozzi-AbInitioCalcOfPhononDispersInSemicond-PhysRevB.43.7231-1991}%
  \BibitemOpen
  \bibfield  {author} {\bibinfo {author} {\bibfnamefont {P.}~\bibnamefont
  {Giannozzi}}, \bibinfo {author} {\bibfnamefont {S.}~\bibnamefont
  {de~Gironcoli}}, \bibinfo {author} {\bibfnamefont {P.}~\bibnamefont
  {Pavone}}, \ and\ \bibinfo {author} {\bibfnamefont {S.}~\bibnamefont
  {Baroni}},\ }\href {\doibase 10.1103/PhysRevB.43.7231} {\bibfield  {journal}
  {\bibinfo  {journal} {Phys. Rev. B}\ }\textbf {\bibinfo {volume} {43}},\
  \bibinfo {pages} {7231} (\bibinfo {year} {1991})}\BibitemShut {NoStop}%
\bibitem [{\citenamefont {Baroni}\ \emph {et~al.}(2001)\citenamefont {Baroni},
  \citenamefont {de~Gironcoli}, \citenamefont {Dal~Corso},\ and\ \citenamefont
  {Giannozzi}}]{Baroni-PhononsAndRelatedCrystalPropFromDFTPT-RevModPhys.73.515-2001}%
  \BibitemOpen
  \bibfield  {author} {\bibinfo {author} {\bibfnamefont {S.}~\bibnamefont
  {Baroni}}, \bibinfo {author} {\bibfnamefont {S.}~\bibnamefont
  {de~Gironcoli}}, \bibinfo {author} {\bibfnamefont {A.}~\bibnamefont
  {Dal~Corso}}, \ and\ \bibinfo {author} {\bibfnamefont {P.}~\bibnamefont
  {Giannozzi}},\ }\href {\doibase 10.1103/RevModPhys.73.515} {\bibfield
  {journal} {\bibinfo  {journal} {Rev. Mod. Phys.}\ }\textbf {\bibinfo {volume}
  {73}},\ \bibinfo {pages} {515} (\bibinfo {year} {2001})}\BibitemShut
  {NoStop}%
\bibitem [{\citenamefont {Togo}\ and\ \citenamefont
  {Tanaka}(2015)}]{Togo-FirstPrincPhononCalcInMaterScience-2015}%
  \BibitemOpen
  \bibfield  {author} {\bibinfo {author} {\bibfnamefont {A.}~\bibnamefont
  {Togo}}\ and\ \bibinfo {author} {\bibfnamefont {I.}~\bibnamefont {Tanaka}},\
  }\href@noop {} {\bibfield  {journal} {\bibinfo  {journal} {Scr. Mater.}\
  }\textbf {\bibinfo {volume} {108}},\ \bibinfo {pages} {1} (\bibinfo {year}
  {2015})}\BibitemShut {NoStop}%
\bibitem [{\citenamefont {Ribeiro}\ \emph {et~al.}(2018)\citenamefont
  {Ribeiro}, \citenamefont {Paulatto}, \citenamefont {Bianco}, \citenamefont
  {Errea}, \citenamefont {Mauri},\ and\ \citenamefont
  {Calandra}}]{Ribeiro-StrongAnharmonInThePhononSpectraOfPbTeAndSnTeFromFirstPrinc-PhysRevB.97.014306-2018}%
  \BibitemOpen
  \bibfield  {author} {\bibinfo {author} {\bibfnamefont {G.~A.~S.}\
  \bibnamefont {Ribeiro}}, \bibinfo {author} {\bibfnamefont {L.}~\bibnamefont
  {Paulatto}}, \bibinfo {author} {\bibfnamefont {R.}~\bibnamefont {Bianco}},
  \bibinfo {author} {\bibfnamefont {I.}~\bibnamefont {Errea}}, \bibinfo
  {author} {\bibfnamefont {F.}~\bibnamefont {Mauri}}, \ and\ \bibinfo {author}
  {\bibfnamefont {M.}~\bibnamefont {Calandra}},\ }\href {\doibase
  10.1103/PhysRevB.97.014306} {\bibfield  {journal} {\bibinfo  {journal} {Phys.
  Rev. B}\ }\textbf {\bibinfo {volume} {97}},\ \bibinfo {pages} {014306}
  (\bibinfo {year} {2018})}\BibitemShut {NoStop}%
\bibitem [{\citenamefont {Baroni}\ \emph {et~al.}(2010)\citenamefont {Baroni},
  \citenamefont {Giannozzi},\ and\ \citenamefont
  {Isaev}}]{Baroni-DFPTForQuasiHarmonicCalculations-2010}%
  \BibitemOpen
  \bibfield  {author} {\bibinfo {author} {\bibfnamefont {S.}~\bibnamefont
  {Baroni}}, \bibinfo {author} {\bibfnamefont {P.}~\bibnamefont {Giannozzi}}, \
  and\ \bibinfo {author} {\bibfnamefont {E.}~\bibnamefont {Isaev}},\
  }\href@noop {} {\bibfield  {journal} {\bibinfo  {journal} {Rev. Mineral.
  Geochem.}\ }\textbf {\bibinfo {volume} {71}},\ \bibinfo {pages} {39}
  (\bibinfo {year} {2010})}\BibitemShut {NoStop}%
\bibitem [{\citenamefont {H{\"a}rk{\"o}nen}\ and\ \citenamefont
  {Karttunen}(2014)}]{Harkonen-NTE-2014}%
  \BibitemOpen
  \bibfield  {author} {\bibinfo {author} {\bibfnamefont {V.~J.}\ \bibnamefont
  {H{\"a}rk{\"o}nen}}\ and\ \bibinfo {author} {\bibfnamefont {A.~J.}\
  \bibnamefont {Karttunen}},\ }\href@noop {} {\bibfield  {journal} {\bibinfo
  {journal} {Phys. Rev. B}\ }\textbf {\bibinfo {volume} {89}},\ \bibinfo
  {pages} {024305} (\bibinfo {year} {2014})}\BibitemShut {NoStop}%
\bibitem [{\citenamefont {Mittal}\ \emph {et~al.}(2018)\citenamefont {Mittal},
  \citenamefont {Gupta},\ and\ \citenamefont
  {Chaplot}}]{Mittal-PhononsAndAnomalousThermalExpBhaviourInCrystSolids-2018}%
  \BibitemOpen
  \bibfield  {author} {\bibinfo {author} {\bibfnamefont {R.}~\bibnamefont
  {Mittal}}, \bibinfo {author} {\bibfnamefont {M.}~\bibnamefont {Gupta}}, \
  and\ \bibinfo {author} {\bibfnamefont {S.}~\bibnamefont {Chaplot}},\
  }\href@noop {} {\bibfield  {journal} {\bibinfo  {journal} {Prog. Mater.
  Sci.}\ }\textbf {\bibinfo {volume} {92}},\ \bibinfo {pages} {360} (\bibinfo
  {year} {2018})}\BibitemShut {NoStop}%
\bibitem [{\citenamefont {H\"ark\"onen}\ and\ \citenamefont
  {Karttunen}(2016{\natexlab{a}})}]{Harkonen-Tcond-II-VIII-PhysRevB.93.024307-2016}%
  \BibitemOpen
  \bibfield  {author} {\bibinfo {author} {\bibfnamefont {V.~J.}\ \bibnamefont
  {H\"ark\"onen}}\ and\ \bibinfo {author} {\bibfnamefont {A.~J.}\ \bibnamefont
  {Karttunen}},\ }\href {\doibase 10.1103/PhysRevB.93.024307} {\bibfield
  {journal} {\bibinfo  {journal} {Phys. Rev. B}\ }\textbf {\bibinfo {volume}
  {93}},\ \bibinfo {pages} {024307} (\bibinfo {year}
  {2016}{\natexlab{a}})}\BibitemShut {NoStop}%
\bibitem [{\citenamefont {Linnera}\ and\ \citenamefont
  {Karttunen}(2017)}]{Linnera-AbInitStudyOfTheLattTCondOfCu2OUsingTheGGAAndHybridDFMethodsPhysRevB.96.014304-2017}%
  \BibitemOpen
  \bibfield  {author} {\bibinfo {author} {\bibfnamefont {J.}~\bibnamefont
  {Linnera}}\ and\ \bibinfo {author} {\bibfnamefont {A.~J.}\ \bibnamefont
  {Karttunen}},\ }\href {\doibase 10.1103/PhysRevB.96.014304} {\bibfield
  {journal} {\bibinfo  {journal} {Phys. Rev. B}\ }\textbf {\bibinfo {volume}
  {96}},\ \bibinfo {pages} {014304} (\bibinfo {year} {2017})}\BibitemShut
  {NoStop}%
\bibitem [{\citenamefont {H\"ark\"onen}\ and\ \citenamefont
  {Karttunen}(2016{\natexlab{b}})}]{Harkonen-AbInitioComputStudyOnTheLattThermalCondOfZintlClathrates-PhysRevB.94.054310-2016}%
  \BibitemOpen
  \bibfield  {author} {\bibinfo {author} {\bibfnamefont {V.~J.}\ \bibnamefont
  {H\"ark\"onen}}\ and\ \bibinfo {author} {\bibfnamefont {A.~J.}\ \bibnamefont
  {Karttunen}},\ }\href {\doibase 10.1103/PhysRevB.94.054310} {\bibfield
  {journal} {\bibinfo  {journal} {Phys. Rev. B}\ }\textbf {\bibinfo {volume}
  {94}},\ \bibinfo {pages} {054310} (\bibinfo {year}
  {2016}{\natexlab{b}})}\BibitemShut {NoStop}%
\bibitem [{\citenamefont {Norouzzadeh}\ \emph {et~al.}(2017)\citenamefont
  {Norouzzadeh}, \citenamefont {Krasinski},\ and\ \citenamefont
  {Tadano}}]{Norouzzadeh-TCondOfTypeITypeIIAndTypeVIIIPristineSiliconClathAFirstPrincStudy-PhysRevB.96.245201-2017}%
  \BibitemOpen
  \bibfield  {author} {\bibinfo {author} {\bibfnamefont {P.}~\bibnamefont
  {Norouzzadeh}}, \bibinfo {author} {\bibfnamefont {J.~S.}\ \bibnamefont
  {Krasinski}}, \ and\ \bibinfo {author} {\bibfnamefont {T.}~\bibnamefont
  {Tadano}},\ }\href {\doibase 10.1103/PhysRevB.96.245201} {\bibfield
  {journal} {\bibinfo  {journal} {Phys. Rev. B}\ }\textbf {\bibinfo {volume}
  {96}},\ \bibinfo {pages} {245201} (\bibinfo {year} {2017})}\BibitemShut
  {NoStop}%
\bibitem [{\citenamefont {Euchner}\ \emph {et~al.}(2018)\citenamefont
  {Euchner}, \citenamefont {Pailh\`es}, \citenamefont {Giordano},\ and\
  \citenamefont
  {de~Boissieu}}]{Euchner-UnderstLatTCondInThermoelClathrADFTStudyOnBinarySiBasedTypeVClathratesPhysRevB.97.014304-2018}%
  \BibitemOpen
  \bibfield  {author} {\bibinfo {author} {\bibfnamefont {H.}~\bibnamefont
  {Euchner}}, \bibinfo {author} {\bibfnamefont {S.}~\bibnamefont {Pailh\`es}},
  \bibinfo {author} {\bibfnamefont {V.~M.}\ \bibnamefont {Giordano}}, \ and\
  \bibinfo {author} {\bibfnamefont {M.}~\bibnamefont {de~Boissieu}},\ }\href
  {\doibase 10.1103/PhysRevB.97.014304} {\bibfield  {journal} {\bibinfo
  {journal} {Phys. Rev. B}\ }\textbf {\bibinfo {volume} {97}},\ \bibinfo
  {pages} {014304} (\bibinfo {year} {2018})}\BibitemShut {NoStop}%
\bibitem [{\citenamefont {Tadano}\ and\ \citenamefont
  {Tsuneyuki}(2018)}]{Tadano-QuarticAnharmOfRattlersAndItsEffOnLatTCondOfClathrFromFirstPrinciplesPhysRevLett.120.105901-2018}%
  \BibitemOpen
  \bibfield  {author} {\bibinfo {author} {\bibfnamefont {T.}~\bibnamefont
  {Tadano}}\ and\ \bibinfo {author} {\bibfnamefont {S.}~\bibnamefont
  {Tsuneyuki}},\ }\href {\doibase 10.1103/PhysRevLett.120.105901} {\bibfield
  {journal} {\bibinfo  {journal} {Phys. Rev. Lett.}\ }\textbf {\bibinfo
  {volume} {120}},\ \bibinfo {pages} {105901} (\bibinfo {year}
  {2018})}\BibitemShut {NoStop}%
\bibitem [{\citenamefont {Shen}\ \emph {et~al.}(2020)\citenamefont {Shen},
  \citenamefont {Xie}, \citenamefont {Zhang}, \citenamefont {Wang},\ and\
  \citenamefont
  {Fang}}]{Shen-Si2GeANewVIITypeClathrateWithUltralowThermalConductivityAndHighThermoelectricProperty-2020}%
  \BibitemOpen
  \bibfield  {author} {\bibinfo {author} {\bibfnamefont {J.}~\bibnamefont
  {Shen}}, \bibinfo {author} {\bibfnamefont {T.}~\bibnamefont {Xie}}, \bibinfo
  {author} {\bibfnamefont {L.}~\bibnamefont {Zhang}}, \bibinfo {author}
  {\bibfnamefont {P.}~\bibnamefont {Wang}}, \ and\ \bibinfo {author}
  {\bibfnamefont {Z.}~\bibnamefont {Fang}},\ }\href@noop {} {\bibfield
  {journal} {\bibinfo  {journal} {Sci. Rep.}\ }\textbf {\bibinfo {volume}
  {10}},\ \bibinfo {pages} {1} (\bibinfo {year} {2020})}\BibitemShut {NoStop}%
\bibitem [{\citenamefont {Calandra}\ \emph {et~al.}(2007)\citenamefont
  {Calandra}, \citenamefont {Lazzeri},\ and\ \citenamefont
  {Mauri}}]{Calandra-AnharmAndNonAdiabaticEffectsInMgB2ImplicForTheIsotopeEffectAndInterpretOfRamanSpectra-2007}%
  \BibitemOpen
  \bibfield  {author} {\bibinfo {author} {\bibfnamefont {M.}~\bibnamefont
  {Calandra}}, \bibinfo {author} {\bibfnamefont {M.}~\bibnamefont {Lazzeri}}, \
  and\ \bibinfo {author} {\bibfnamefont {F.}~\bibnamefont {Mauri}},\
  }\href@noop {} {\bibfield  {journal} {\bibinfo  {journal} {Physica C}\
  }\textbf {\bibinfo {volume} {456}},\ \bibinfo {pages} {38} (\bibinfo {year}
  {2007})}\BibitemShut {NoStop}%
\bibitem [{\citenamefont {Pisana}\ \emph {et~al.}(2007)\citenamefont {Pisana},
  \citenamefont {Lazzeri}, \citenamefont {Casiraghi}, \citenamefont
  {Novoselov}, \citenamefont {Geim}, \citenamefont {Ferrari},\ and\
  \citenamefont
  {Mauri}}]{Pisana-BreakdownOfTheAdiabaticBornOppenhApproximationInGraphene-2007}%
  \BibitemOpen
  \bibfield  {author} {\bibinfo {author} {\bibfnamefont {S.}~\bibnamefont
  {Pisana}}, \bibinfo {author} {\bibfnamefont {M.}~\bibnamefont {Lazzeri}},
  \bibinfo {author} {\bibfnamefont {C.}~\bibnamefont {Casiraghi}}, \bibinfo
  {author} {\bibfnamefont {K.~S.}\ \bibnamefont {Novoselov}}, \bibinfo {author}
  {\bibfnamefont {A.~K.}\ \bibnamefont {Geim}}, \bibinfo {author}
  {\bibfnamefont {A.~C.}\ \bibnamefont {Ferrari}}, \ and\ \bibinfo {author}
  {\bibfnamefont {F.}~\bibnamefont {Mauri}},\ }\href@noop {} {\bibfield
  {journal} {\bibinfo  {journal} {Nat. Mater.}\ }\textbf {\bibinfo {volume}
  {6}},\ \bibinfo {pages} {198} (\bibinfo {year} {2007})}\BibitemShut {NoStop}%
\bibitem [{\citenamefont {Somayazulu}\ \emph {et~al.}(2019)\citenamefont
  {Somayazulu}, \citenamefont {Ahart}, \citenamefont {Mishra}, \citenamefont
  {Geballe}, \citenamefont {Baldini}, \citenamefont {Meng}, \citenamefont
  {Struzhkin},\ and\ \citenamefont
  {Hemley}}]{Somayazulu-EvidForSupercAbove260KInLanthanumSuperhydrAtMegabarPressures-PhysRevLett.122.027001-2019}%
  \BibitemOpen
  \bibfield  {author} {\bibinfo {author} {\bibfnamefont {M.}~\bibnamefont
  {Somayazulu}}, \bibinfo {author} {\bibfnamefont {M.}~\bibnamefont {Ahart}},
  \bibinfo {author} {\bibfnamefont {A.~K.}\ \bibnamefont {Mishra}}, \bibinfo
  {author} {\bibfnamefont {Z.~M.}\ \bibnamefont {Geballe}}, \bibinfo {author}
  {\bibfnamefont {M.}~\bibnamefont {Baldini}}, \bibinfo {author} {\bibfnamefont
  {Y.}~\bibnamefont {Meng}}, \bibinfo {author} {\bibfnamefont {V.~V.}\
  \bibnamefont {Struzhkin}}, \ and\ \bibinfo {author} {\bibfnamefont {R.~J.}\
  \bibnamefont {Hemley}},\ }\href {\doibase 10.1103/PhysRevLett.122.027001}
  {\bibfield  {journal} {\bibinfo  {journal} {Phys. Rev. Lett.}\ }\textbf
  {\bibinfo {volume} {122}},\ \bibinfo {pages} {027001} (\bibinfo {year}
  {2019})}\BibitemShut {NoStop}%
\bibitem [{\citenamefont {Flores-Livas}\ and\ \citenamefont
  {Arita}(2019)}]{Flores-APredictionForHotSuperconductivity-2019}%
  \BibitemOpen
  \bibfield  {author} {\bibinfo {author} {\bibfnamefont {J.~A.}\ \bibnamefont
  {Flores-Livas}}\ and\ \bibinfo {author} {\bibfnamefont {R.}~\bibnamefont
  {Arita}},\ }\href@noop {} {\bibfield  {journal} {\bibinfo  {journal}
  {Physics}\ }\textbf {\bibinfo {volume} {12}},\ \bibinfo {pages} {96}
  (\bibinfo {year} {2019})}\BibitemShut {NoStop}%
\bibitem [{\citenamefont {Errea}\ \emph {et~al.}(2020)\citenamefont {Errea},
  \citenamefont {Belli}, \citenamefont {Monacelli}, \citenamefont {Sanna},
  \citenamefont {Koretsune}, \citenamefont {Tadano}, \citenamefont {Bianco},
  \citenamefont {Calandra}, \citenamefont {Arita}, \citenamefont {Mauri} \emph
  {et~al.}}]{Errea-QuantumCrystalStructureInThe250kelvinSuperconductingLanthanumHydride-2020}%
  \BibitemOpen
  \bibfield  {author} {\bibinfo {author} {\bibfnamefont {I.}~\bibnamefont
  {Errea}}, \bibinfo {author} {\bibfnamefont {F.}~\bibnamefont {Belli}},
  \bibinfo {author} {\bibfnamefont {L.}~\bibnamefont {Monacelli}}, \bibinfo
  {author} {\bibfnamefont {A.}~\bibnamefont {Sanna}}, \bibinfo {author}
  {\bibfnamefont {T.}~\bibnamefont {Koretsune}}, \bibinfo {author}
  {\bibfnamefont {T.}~\bibnamefont {Tadano}}, \bibinfo {author} {\bibfnamefont
  {R.}~\bibnamefont {Bianco}}, \bibinfo {author} {\bibfnamefont
  {M.}~\bibnamefont {Calandra}}, \bibinfo {author} {\bibfnamefont
  {R.}~\bibnamefont {Arita}}, \bibinfo {author} {\bibfnamefont
  {F.}~\bibnamefont {Mauri}},  \emph {et~al.},\ }\href@noop {} {\bibfield
  {journal} {\bibinfo  {journal} {Nature}\ }\textbf {\bibinfo {volume} {578}},\
  \bibinfo {pages} {66} (\bibinfo {year} {2020})}\BibitemShut {NoStop}%
\bibitem [{\citenamefont {Polli}\ \emph {et~al.}(2010)\citenamefont {Polli},
  \citenamefont {Alto{\`e}}, \citenamefont {Weingart}, \citenamefont
  {Spillane}, \citenamefont {Manzoni}, \citenamefont {Brida}, \citenamefont
  {Tomasello}, \citenamefont {Orlandi}, \citenamefont {Kukura}, \citenamefont
  {Mathies} \emph
  {et~al.}}]{Polli-ConicalIntersectionDynamicsOfThePrimaryPhotoisomerizEventInVision-2010}%
  \BibitemOpen
  \bibfield  {author} {\bibinfo {author} {\bibfnamefont {D.}~\bibnamefont
  {Polli}}, \bibinfo {author} {\bibfnamefont {P.}~\bibnamefont {Alto{\`e}}},
  \bibinfo {author} {\bibfnamefont {O.}~\bibnamefont {Weingart}}, \bibinfo
  {author} {\bibfnamefont {K.~M.}\ \bibnamefont {Spillane}}, \bibinfo {author}
  {\bibfnamefont {C.}~\bibnamefont {Manzoni}}, \bibinfo {author} {\bibfnamefont
  {D.}~\bibnamefont {Brida}}, \bibinfo {author} {\bibfnamefont
  {G.}~\bibnamefont {Tomasello}}, \bibinfo {author} {\bibfnamefont
  {G.}~\bibnamefont {Orlandi}}, \bibinfo {author} {\bibfnamefont
  {P.}~\bibnamefont {Kukura}}, \bibinfo {author} {\bibfnamefont {R.~A.}\
  \bibnamefont {Mathies}},  \emph {et~al.},\ }\href@noop {} {\bibfield
  {journal} {\bibinfo  {journal} {Nature}\ }\textbf {\bibinfo {volume} {467}},\
  \bibinfo {pages} {440} (\bibinfo {year} {2010})}\BibitemShut {NoStop}%
\bibitem [{\citenamefont
  {Tully}(2012)}]{Tully-PerspectiveNonadiabaticDynamicsTheory-2012}%
  \BibitemOpen
  \bibfield  {author} {\bibinfo {author} {\bibfnamefont {J.~C.}\ \bibnamefont
  {Tully}},\ }\href@noop {} {\bibfield  {journal} {\bibinfo  {journal} {J.
  Chem. Phys.}\ }\textbf {\bibinfo {volume} {137}},\ \bibinfo {pages} {22A301}
  (\bibinfo {year} {2012})}\BibitemShut {NoStop}%
\bibitem [{\citenamefont {Nelson}\ \emph {et~al.}(2014)\citenamefont {Nelson},
  \citenamefont {Fernandez-Alberti}, \citenamefont {Roitberg},\ and\
  \citenamefont
  {Tretiak}}]{Nelson-NonadiabaticExcitedStateMolecularDynamicsModelingPhotophysicsInOrganicConjugatedMaterials-2014}%
  \BibitemOpen
  \bibfield  {author} {\bibinfo {author} {\bibfnamefont {T.}~\bibnamefont
  {Nelson}}, \bibinfo {author} {\bibfnamefont {S.}~\bibnamefont
  {Fernandez-Alberti}}, \bibinfo {author} {\bibfnamefont {A.~E.}\ \bibnamefont
  {Roitberg}}, \ and\ \bibinfo {author} {\bibfnamefont {S.}~\bibnamefont
  {Tretiak}},\ }\href@noop {} {\bibfield  {journal} {\bibinfo  {journal} {Acc.
  Chem. Res.}\ }\textbf {\bibinfo {volume} {47}},\ \bibinfo {pages} {1155}
  (\bibinfo {year} {2014})}\BibitemShut {NoStop}%
\bibitem [{\citenamefont {Dantus}\ \emph {et~al.}(1987)\citenamefont {Dantus},
  \citenamefont {Rosker},\ and\ \citenamefont
  {Zewail}}]{Dantus-RealTimeFemtosecondProbingOfTransitionStatesInChemicalReactions-1987}%
  \BibitemOpen
  \bibfield  {author} {\bibinfo {author} {\bibfnamefont {M.}~\bibnamefont
  {Dantus}}, \bibinfo {author} {\bibfnamefont {M.~J.}\ \bibnamefont {Rosker}},
  \ and\ \bibinfo {author} {\bibfnamefont {A.~H.}\ \bibnamefont {Zewail}},\
  }\href@noop {} {\bibfield  {journal} {\bibinfo  {journal} {J. Chem. Phys.}\
  }\textbf {\bibinfo {volume} {87}},\ \bibinfo {pages} {2395} (\bibinfo {year}
  {1987})}\BibitemShut {NoStop}%
\bibitem [{\citenamefont {Potter}\ \emph {et~al.}(1992)\citenamefont {Potter},
  \citenamefont {Herek}, \citenamefont {Pedersen}, \citenamefont {Liu},\ and\
  \citenamefont
  {Zewail}}]{Potter-FemtosecondLaserControlOfaChemicalReaction-1992}%
  \BibitemOpen
  \bibfield  {author} {\bibinfo {author} {\bibfnamefont {E.}~\bibnamefont
  {Potter}}, \bibinfo {author} {\bibfnamefont {J.}~\bibnamefont {Herek}},
  \bibinfo {author} {\bibfnamefont {S.}~\bibnamefont {Pedersen}}, \bibinfo
  {author} {\bibfnamefont {Q.}~\bibnamefont {Liu}}, \ and\ \bibinfo {author}
  {\bibfnamefont {A.}~\bibnamefont {Zewail}},\ }\href@noop {} {\bibfield
  {journal} {\bibinfo  {journal} {Nature}\ }\textbf {\bibinfo {volume} {355}},\
  \bibinfo {pages} {66} (\bibinfo {year} {1992})}\BibitemShut {NoStop}%
\bibitem [{\citenamefont
  {Zewail}(2000)}]{Zewail-FemtochemistryAtomicScaleDynamicsOfTheChemicalBond-2000}%
  \BibitemOpen
  \bibfield  {author} {\bibinfo {author} {\bibfnamefont {A.~H.}\ \bibnamefont
  {Zewail}},\ }\href@noop {} {\bibfield  {journal} {\bibinfo  {journal} {J.
  Phys. Chem. A}\ }\textbf {\bibinfo {volume} {104}},\ \bibinfo {pages} {5660}
  (\bibinfo {year} {2000})}\BibitemShut {NoStop}%
\bibitem [{\citenamefont {Barth}\ \emph {et~al.}(2009)\citenamefont {Barth},
  \citenamefont {Hege}, \citenamefont {Ikeda}, \citenamefont {Kenfack},
  \citenamefont {Koppitz}, \citenamefont {Manz}, \citenamefont {Marquardt},\
  and\ \citenamefont
  {Paramonov}}]{Barth-ConcertedQuantumEffectsOfElectronicAndNuclearFluxesInMolecules-2009}%
  \BibitemOpen
  \bibfield  {author} {\bibinfo {author} {\bibfnamefont {I.}~\bibnamefont
  {Barth}}, \bibinfo {author} {\bibfnamefont {H.-C.}\ \bibnamefont {Hege}},
  \bibinfo {author} {\bibfnamefont {H.}~\bibnamefont {Ikeda}}, \bibinfo
  {author} {\bibfnamefont {A.}~\bibnamefont {Kenfack}}, \bibinfo {author}
  {\bibfnamefont {M.}~\bibnamefont {Koppitz}}, \bibinfo {author} {\bibfnamefont
  {J.}~\bibnamefont {Manz}}, \bibinfo {author} {\bibfnamefont {F.}~\bibnamefont
  {Marquardt}}, \ and\ \bibinfo {author} {\bibfnamefont {G.~K.}\ \bibnamefont
  {Paramonov}},\ }\href@noop {} {\bibfield  {journal} {\bibinfo  {journal}
  {Chem. Phys. Lett.}\ }\textbf {\bibinfo {volume} {481}},\ \bibinfo {pages}
  {118} (\bibinfo {year} {2009})}\BibitemShut {NoStop}%
\bibitem [{\citenamefont {Schild}\ \emph {et~al.}(2016)\citenamefont {Schild},
  \citenamefont {Agostini},\ and\ \citenamefont
  {Gross}}]{Schild-ElectronicFluxDensityBeyondTheBornOppenheimerApproximation-2016}%
  \BibitemOpen
  \bibfield  {author} {\bibinfo {author} {\bibfnamefont {A.}~\bibnamefont
  {Schild}}, \bibinfo {author} {\bibfnamefont {F.}~\bibnamefont {Agostini}}, \
  and\ \bibinfo {author} {\bibfnamefont {E.~K.~U.}\ \bibnamefont {Gross}},\
  }\href@noop {} {\bibfield  {journal} {\bibinfo  {journal} {J. Phys. Chem. A}\
  }\textbf {\bibinfo {volume} {120}},\ \bibinfo {pages} {3316} (\bibinfo {year}
  {2016})}\BibitemShut {NoStop}%
\bibitem [{\citenamefont {Scherrer}\ \emph {et~al.}(2017)\citenamefont
  {Scherrer}, \citenamefont {Agostini}, \citenamefont {Sebastiani},
  \citenamefont {Gross},\ and\ \citenamefont
  {Vuilleumier}}]{Scherrer-OnTheMassOfAtomsInMoleculesBeyondTheBornOppenheimerApproximation-PhysRevX.7.031035-2017}%
  \BibitemOpen
  \bibfield  {author} {\bibinfo {author} {\bibfnamefont {A.}~\bibnamefont
  {Scherrer}}, \bibinfo {author} {\bibfnamefont {F.}~\bibnamefont {Agostini}},
  \bibinfo {author} {\bibfnamefont {D.}~\bibnamefont {Sebastiani}}, \bibinfo
  {author} {\bibfnamefont {E.~K.~U.}\ \bibnamefont {Gross}}, \ and\ \bibinfo
  {author} {\bibfnamefont {R.}~\bibnamefont {Vuilleumier}},\ }\href {\doibase
  10.1103/PhysRevX.7.031035} {\bibfield  {journal} {\bibinfo  {journal} {Phys.
  Rev. X}\ }\textbf {\bibinfo {volume} {7}},\ \bibinfo {pages} {031035}
  (\bibinfo {year} {2017})}\BibitemShut {NoStop}%
\bibitem [{\citenamefont {Abedi}\ \emph {et~al.}(2010)\citenamefont {Abedi},
  \citenamefont {Maitra},\ and\ \citenamefont
  {Gross}}]{Abedi-ExactFactorization-PhysRevLett.105.123002-2010}%
  \BibitemOpen
  \bibfield  {author} {\bibinfo {author} {\bibfnamefont {A.}~\bibnamefont
  {Abedi}}, \bibinfo {author} {\bibfnamefont {N.~T.}\ \bibnamefont {Maitra}}, \
  and\ \bibinfo {author} {\bibfnamefont {E.~K.~U.}\ \bibnamefont {Gross}},\
  }\href {\doibase 10.1103/PhysRevLett.105.123002} {\bibfield  {journal}
  {\bibinfo  {journal} {Phys. Rev. Lett.}\ }\textbf {\bibinfo {volume} {105}},\
  \bibinfo {pages} {123002} (\bibinfo {year} {2010})}\BibitemShut {NoStop}%
\bibitem [{\citenamefont {Abedi}\ \emph {et~al.}(2012)\citenamefont {Abedi},
  \citenamefont {Maitra},\ and\ \citenamefont
  {Gross}}]{Abedi-CorrelatedElectronNuclearDynamicsEF-2012}%
  \BibitemOpen
  \bibfield  {author} {\bibinfo {author} {\bibfnamefont {A.}~\bibnamefont
  {Abedi}}, \bibinfo {author} {\bibfnamefont {N.~T.}\ \bibnamefont {Maitra}}, \
  and\ \bibinfo {author} {\bibfnamefont {E.~K.~U.}\ \bibnamefont {Gross}},\
  }\href@noop {} {\bibfield  {journal} {\bibinfo  {journal} {J. Chem. Phys.}\
  }\textbf {\bibinfo {volume} {137}},\ \bibinfo {pages} {22A530} (\bibinfo
  {year} {2012})}\BibitemShut {NoStop}%
\bibitem [{\citenamefont {Gidopoulos}\ and\ \citenamefont
  {Gross}(2014)}]{Gidopoulos-Gross-ElectronicNonAdiabaticStates-2014}%
  \BibitemOpen
  \bibfield  {author} {\bibinfo {author} {\bibfnamefont {N.~I.}\ \bibnamefont
  {Gidopoulos}}\ and\ \bibinfo {author} {\bibfnamefont {E.~K.~U.}\ \bibnamefont
  {Gross}},\ }\href@noop {} {\bibfield  {journal} {\bibinfo  {journal} {Phil.
  Trans. R. Soc. London}\ }\textbf {\bibinfo {volume} {372}},\ \bibinfo {pages}
  {20130059} (\bibinfo {year} {2014})}\BibitemShut {NoStop}%
\bibitem [{\citenamefont {Requist}\ and\ \citenamefont
  {Gross}(2016)}]{Requist-ExactFactorBasedDFTofElectronsAndNuclei-PhysRevLett.117.193001-2016}%
  \BibitemOpen
  \bibfield  {author} {\bibinfo {author} {\bibfnamefont {R.}~\bibnamefont
  {Requist}}\ and\ \bibinfo {author} {\bibfnamefont {E.~K.~U.}\ \bibnamefont
  {Gross}},\ }\href {\doibase 10.1103/PhysRevLett.117.193001} {\bibfield
  {journal} {\bibinfo  {journal} {Phys. Rev. Lett.}\ }\textbf {\bibinfo
  {volume} {117}},\ \bibinfo {pages} {193001} (\bibinfo {year}
  {2016})}\BibitemShut {NoStop}%
\bibitem [{\citenamefont {Li}\ \emph {et~al.}(2018)\citenamefont {Li},
  \citenamefont {Requist},\ and\ \citenamefont
  {Gross}}]{Li-DFTofElectronTransferBeyondTheBOapproximationCaseStudyOfLiF-2018}%
  \BibitemOpen
  \bibfield  {author} {\bibinfo {author} {\bibfnamefont {C.}~\bibnamefont
  {Li}}, \bibinfo {author} {\bibfnamefont {R.}~\bibnamefont {Requist}}, \ and\
  \bibinfo {author} {\bibfnamefont {E.~K.~U.}\ \bibnamefont {Gross}},\
  }\href@noop {} {\bibfield  {journal} {\bibinfo  {journal} {J. Chem. Phys.}\
  }\textbf {\bibinfo {volume} {148}},\ \bibinfo {pages} {084110} (\bibinfo
  {year} {2018})}\BibitemShut {NoStop}%
\bibitem [{\citenamefont {Requist}\ \emph {et~al.}(2019)\citenamefont
  {Requist}, \citenamefont {Proetto},\ and\ \citenamefont
  {Gross}}]{Requist-ExactFactorizationBasedDFTofElectronPhononSystems-PhysRevB.99.165136-2019}%
  \BibitemOpen
  \bibfield  {author} {\bibinfo {author} {\bibfnamefont {R.}~\bibnamefont
  {Requist}}, \bibinfo {author} {\bibfnamefont {C.~R.}\ \bibnamefont
  {Proetto}}, \ and\ \bibinfo {author} {\bibfnamefont {E.~K.~U.}\ \bibnamefont
  {Gross}},\ }\href {\doibase 10.1103/PhysRevB.99.165136} {\bibfield  {journal}
  {\bibinfo  {journal} {Phys. Rev. B}\ }\textbf {\bibinfo {volume} {99}},\
  \bibinfo {pages} {165136} (\bibinfo {year} {2019})}\BibitemShut {NoStop}%
\bibitem [{\citenamefont {Kreibich}\ and\ \citenamefont
  {Gross}(2001)}]{Kreibich-MulticompDFTForElectronsAndNuclei-PhysRevLett.86.2984-2001}%
  \BibitemOpen
  \bibfield  {author} {\bibinfo {author} {\bibfnamefont {T.}~\bibnamefont
  {Kreibich}}\ and\ \bibinfo {author} {\bibfnamefont {E.~K.~U.}\ \bibnamefont
  {Gross}},\ }\href {\doibase 10.1103/PhysRevLett.86.2984} {\bibfield
  {journal} {\bibinfo  {journal} {Phys. Rev. Lett.}\ }\textbf {\bibinfo
  {volume} {86}},\ \bibinfo {pages} {2984} (\bibinfo {year}
  {2001})}\BibitemShut {NoStop}%
\bibitem [{\citenamefont {Kreibich}\ \emph {et~al.}(2008)\citenamefont
  {Kreibich}, \citenamefont {van Leeuwen},\ and\ \citenamefont
  {Gross}}]{Kreibich-MulticompDFTForElectronsAndNuclei-PhysRevA.78.022501-2008}%
  \BibitemOpen
  \bibfield  {author} {\bibinfo {author} {\bibfnamefont {T.}~\bibnamefont
  {Kreibich}}, \bibinfo {author} {\bibfnamefont {R.}~\bibnamefont {van
  Leeuwen}}, \ and\ \bibinfo {author} {\bibfnamefont {E.~K.~U.}\ \bibnamefont
  {Gross}},\ }\href {\doibase 10.1103/PhysRevA.78.022501} {\bibfield  {journal}
  {\bibinfo  {journal} {Phys. Rev. A}\ }\textbf {\bibinfo {volume} {78}},\
  \bibinfo {pages} {022501} (\bibinfo {year} {2008})}\BibitemShut {NoStop}%
\bibitem [{\citenamefont {Butriy}\ \emph {et~al.}(2007)\citenamefont {Butriy},
  \citenamefont {Ebadi}, \citenamefont {de~Boeij}, \citenamefont {van
  Leeuwen},\ and\ \citenamefont
  {Gross}}]{Butriy-MulticomponentDFTforTimeDependentSystems-PhysRevA.76.052514-2007}%
  \BibitemOpen
  \bibfield  {author} {\bibinfo {author} {\bibfnamefont {O.}~\bibnamefont
  {Butriy}}, \bibinfo {author} {\bibfnamefont {H.}~\bibnamefont {Ebadi}},
  \bibinfo {author} {\bibfnamefont {P.~L.}\ \bibnamefont {de~Boeij}}, \bibinfo
  {author} {\bibfnamefont {R.}~\bibnamefont {van Leeuwen}}, \ and\ \bibinfo
  {author} {\bibfnamefont {E.~K.~U.}\ \bibnamefont {Gross}},\ }\href {\doibase
  10.1103/PhysRevA.76.052514} {\bibfield  {journal} {\bibinfo  {journal} {Phys.
  Rev. A}\ }\textbf {\bibinfo {volume} {76}},\ \bibinfo {pages} {052514}
  (\bibinfo {year} {2007})}\BibitemShut {NoStop}%
\bibitem [{\citenamefont {van
  Leeuwen}(2004)}]{vanLeeuwen-FirstPrincElectronPhonon-PhysRevB.69.115110-2004}%
  \BibitemOpen
  \bibfield  {author} {\bibinfo {author} {\bibfnamefont {R.}~\bibnamefont {van
  Leeuwen}},\ }\href {\doibase 10.1103/PhysRevB.69.115110} {\bibfield
  {journal} {\bibinfo  {journal} {Phys. Rev. B}\ }\textbf {\bibinfo {volume}
  {69}},\ \bibinfo {pages} {115110} (\bibinfo {year} {2004})}\BibitemShut
  {NoStop}%
\bibitem [{\citenamefont
  {Sutcliffe}(2000)}]{Sutcliffe-TheDecouplingOfElectronicAndNuclearMotions-2000}%
  \BibitemOpen
  \bibfield  {author} {\bibinfo {author} {\bibfnamefont {B.}~\bibnamefont
  {Sutcliffe}},\ }\href@noop {} {\bibfield  {journal} {\bibinfo  {journal}
  {Adv. Chem. Phys.}\ }\textbf {\bibinfo {volume} {114}},\ \bibinfo {pages} {1}
  (\bibinfo {year} {2000})}\BibitemShut {NoStop}%
\bibitem [{\citenamefont
  {Sutcliffe}(2003)}]{Sutcliffe-CoordinateSystemsAndTransformations-2003}%
  \BibitemOpen
  \bibfield  {author} {\bibinfo {author} {\bibfnamefont {B.~T.}\ \bibnamefont
  {Sutcliffe}},\ }\href@noop {} {\bibfield  {journal} {\bibinfo  {journal}
  {Handbook of Molecular Physics and Quantum Chemistry}\ }\textbf {\bibinfo
  {volume} {1}},\ \bibinfo {pages} {485} (\bibinfo {year} {2003})}\BibitemShut
  {NoStop}%
\bibitem [{\citenamefont {Scivetti}\ \emph {et~al.}(2009)\citenamefont
  {Scivetti}, \citenamefont {Kohanoff},\ and\ \citenamefont
  {Gidopoulos}}]{Scivetti-GeneralLocalAndRectilVibrCoordConsistWithEckartsCond-PhysRevA.79.032516-2009}%
  \BibitemOpen
  \bibfield  {author} {\bibinfo {author} {\bibfnamefont {I.}~\bibnamefont
  {Scivetti}}, \bibinfo {author} {\bibfnamefont {J.}~\bibnamefont {Kohanoff}},
  \ and\ \bibinfo {author} {\bibfnamefont {N.~I.}\ \bibnamefont {Gidopoulos}},\
  }\href {\doibase 10.1103/PhysRevA.79.032516} {\bibfield  {journal} {\bibinfo
  {journal} {Phys. Rev. A}\ }\textbf {\bibinfo {volume} {79}},\ \bibinfo
  {pages} {032516} (\bibinfo {year} {2009})}\BibitemShut {NoStop}%
\bibitem [{\citenamefont
  {Schwinger}(1951)}]{Schwinger-OnTheGreensFunctionsOfQuantizedFieldsI-1951}%
  \BibitemOpen
  \bibfield  {author} {\bibinfo {author} {\bibfnamefont {J.}~\bibnamefont
  {Schwinger}},\ }\href@noop {} {\bibfield  {journal} {\bibinfo  {journal}
  {Proc. Nat. Acad. Sci. U. S. A}\ }\textbf {\bibinfo {volume} {37}},\ \bibinfo
  {pages} {452} (\bibinfo {year} {1951})}\BibitemShut {NoStop}%
\bibitem [{\citenamefont
  {Eckart}(1935)}]{Eckart-SomeStudiesConcerningRotatingAxesAndPolyatomicMolecules-PhysRev.47.552-1935}%
  \BibitemOpen
  \bibfield  {author} {\bibinfo {author} {\bibfnamefont {C.}~\bibnamefont
  {Eckart}},\ }\href {\doibase 10.1103/PhysRev.47.552} {\bibfield  {journal}
  {\bibinfo  {journal} {Phys. Rev.}\ }\textbf {\bibinfo {volume} {47}},\
  \bibinfo {pages} {552} (\bibinfo {year} {1935})}\BibitemShut {NoStop}%
\bibitem [{\citenamefont {Wilson}\ \emph {et~al.}(1955)\citenamefont {Wilson},
  \citenamefont {Decius},\ and\ \citenamefont
  {Cross}}]{Wilson-MolecularVibrationsTheTheoryOfInfraredAndRamanVibrationalSpectra-1955}%
  \BibitemOpen
  \bibfield  {author} {\bibinfo {author} {\bibfnamefont {E.~B.}\ \bibnamefont
  {Wilson}}, \bibinfo {author} {\bibfnamefont {J.~C.}\ \bibnamefont {Decius}},
  \ and\ \bibinfo {author} {\bibfnamefont {P.~C.}\ \bibnamefont {Cross}},\
  }\href@noop {} {\emph {\bibinfo {title} {Molecular vibrations: the theory of
  infrared and Raman vibrational spectra}}}\ (\bibinfo  {publisher}
  {McGraw-Hill Book Company},\ \bibinfo {year} {1955})\BibitemShut {NoStop}%
\bibitem [{\citenamefont {Littlejohn}\ and\ \citenamefont
  {Reinsch}(1997)}]{Littlejohn-GaugeFieldsInTheSeparOfRotatAndIntMotionsIntheNbodyProb-RevModPhys.69.213-1997}%
  \BibitemOpen
  \bibfield  {author} {\bibinfo {author} {\bibfnamefont {R.~G.}\ \bibnamefont
  {Littlejohn}}\ and\ \bibinfo {author} {\bibfnamefont {M.}~\bibnamefont
  {Reinsch}},\ }\href {\doibase 10.1103/RevModPhys.69.213} {\bibfield
  {journal} {\bibinfo  {journal} {Rev. Mod. Phys.}\ }\textbf {\bibinfo {volume}
  {69}},\ \bibinfo {pages} {213} (\bibinfo {year} {1997})}\BibitemShut
  {NoStop}%
\bibitem [{\citenamefont
  {Giustino}(2017)}]{Giustino-ElectronPhononInteractFromFirstPrinc-RevModPhys.89.015003-2017}%
  \BibitemOpen
  \bibfield  {author} {\bibinfo {author} {\bibfnamefont {F.}~\bibnamefont
  {Giustino}},\ }\href {\doibase 10.1103/RevModPhys.89.015003} {\bibfield
  {journal} {\bibinfo  {journal} {Rev. Mod. Phys.}\ }\textbf {\bibinfo {volume}
  {89}},\ \bibinfo {pages} {015003} (\bibinfo {year} {2017})}\BibitemShut
  {NoStop}%
\bibitem [{\citenamefont
  {Sayvetz}(1939)}]{Sayvetz-TheKineticEnergyOfPolyatomicMolecules-1939}%
  \BibitemOpen
  \bibfield  {author} {\bibinfo {author} {\bibfnamefont {A.}~\bibnamefont
  {Sayvetz}},\ }\href@noop {} {\bibfield  {journal} {\bibinfo  {journal} {J.
  Chem. Phys.}\ }\textbf {\bibinfo {volume} {7}},\ \bibinfo {pages} {383}
  (\bibinfo {year} {1939})}\BibitemShut {NoStop}%
\bibitem [{\citenamefont
  {Guzman}(2003)}]{Guzman-DerivativesAndIntegralsOfMultivariableFunctions-2003}%
  \BibitemOpen
  \bibfield  {author} {\bibinfo {author} {\bibfnamefont {A.}~\bibnamefont
  {Guzman}},\ }\href@noop {} {\emph {\bibinfo {title} {Derivatives and
  Integrals of Multivariable Functions}}}\ (\bibinfo  {publisher} {Springer
  Science \& Business Media},\ \bibinfo {year} {2003})\BibitemShut {NoStop}%
\bibitem [{\citenamefont {Krantz}\ and\ \citenamefont
  {Parks}(2012)}]{Krantz-TheImplicitFunctionTheoremHistoryTheoryAndApplications-2013}%
  \BibitemOpen
  \bibfield  {author} {\bibinfo {author} {\bibfnamefont {S.~G.}\ \bibnamefont
  {Krantz}}\ and\ \bibinfo {author} {\bibfnamefont {H.~R.}\ \bibnamefont
  {Parks}},\ }\href@noop {} {\emph {\bibinfo {title} {The Implicit Function
  Theorem: History, Theory, and Applications}}}\ (\bibinfo  {publisher}
  {Springer Science \& Business Media},\ \bibinfo {year} {2012})\BibitemShut
  {NoStop}%
\bibitem [{\citenamefont {Krasnoshchekov}\ \emph {et~al.}(2014)\citenamefont
  {Krasnoshchekov}, \citenamefont {Isayeva},\ and\ \citenamefont
  {Stepanov}}]{Krasnoshchekov-DetermOfTheEckartMolecFixedFrameByUseOfTheApparatusOfQuaternionAlgebra-2014}%
  \BibitemOpen
  \bibfield  {author} {\bibinfo {author} {\bibfnamefont {S.~V.}\ \bibnamefont
  {Krasnoshchekov}}, \bibinfo {author} {\bibfnamefont {E.~V.}\ \bibnamefont
  {Isayeva}}, \ and\ \bibinfo {author} {\bibfnamefont {N.~F.}\ \bibnamefont
  {Stepanov}},\ }\href@noop {} {\bibfield  {journal} {\bibinfo  {journal} {J.
  Chem. Phys.}\ }\textbf {\bibinfo {volume} {140}},\ \bibinfo {pages} {154104}
  (\bibinfo {year} {2014})}\BibitemShut {NoStop}%
\bibitem [{\citenamefont {de~Oliveira}\ \emph {et~al.}(2013)\citenamefont
  {de~Oliveira} \emph
  {et~al.}}]{deOliveira-TheImplicitAndInverseFunctionTheoremsEasyProofs-2013}%
  \BibitemOpen
  \bibfield  {author} {\bibinfo {author} {\bibfnamefont {O.}~\bibnamefont
  {de~Oliveira}} \emph {et~al.},\ }\href@noop {} {\bibfield  {journal}
  {\bibinfo  {journal} {Real Analysis Exch.}\ }\textbf {\bibinfo {volume}
  {39}},\ \bibinfo {pages} {207} (\bibinfo {year} {2013})}\BibitemShut
  {NoStop}%
\bibitem [{\citenamefont {Keldysh}(1965)}]{Keldysh-Contour-1965}%
  \BibitemOpen
  \bibfield  {author} {\bibinfo {author} {\bibfnamefont {L.~V.}\ \bibnamefont
  {Keldysh}},\ }\href@noop {} {\bibfield  {journal} {\bibinfo  {journal} {Sov.
  Phys. JETP}\ }\textbf {\bibinfo {volume} {20}},\ \bibinfo {pages} {1018}
  (\bibinfo {year} {1965})}\BibitemShut {NoStop}%
\bibitem [{\citenamefont {Aryasetiawan}\ and\ \citenamefont
  {Gunnarsson}(1998)}]{Aryasetiawan-TheGWMethod-1998}%
  \BibitemOpen
  \bibfield  {author} {\bibinfo {author} {\bibfnamefont {F.}~\bibnamefont
  {Aryasetiawan}}\ and\ \bibinfo {author} {\bibfnamefont {O.}~\bibnamefont
  {Gunnarsson}},\ }\href@noop {} {\bibfield  {journal} {\bibinfo  {journal}
  {Rep. Prog. Phys.}\ }\textbf {\bibinfo {volume} {61}},\ \bibinfo {pages}
  {237} (\bibinfo {year} {1998})}\BibitemShut {NoStop}%
\bibitem [{\citenamefont {Stan}\ \emph {et~al.}(2009)\citenamefont {Stan},
  \citenamefont {Dahlen},\ and\ \citenamefont
  {Van~Leeuwen}}]{Stan-LevelsOfSelfConsistencyInTheGWApproximation-2009}%
  \BibitemOpen
  \bibfield  {author} {\bibinfo {author} {\bibfnamefont {A.}~\bibnamefont
  {Stan}}, \bibinfo {author} {\bibfnamefont {N.~E.}\ \bibnamefont {Dahlen}}, \
  and\ \bibinfo {author} {\bibfnamefont {R.}~\bibnamefont {Van~Leeuwen}},\
  }\href@noop {} {\bibfield  {journal} {\bibinfo  {journal} {J. Chem. Phys.}\
  }\textbf {\bibinfo {volume} {130}},\ \bibinfo {pages} {114105} (\bibinfo
  {year} {2009})}\BibitemShut {NoStop}%
\bibitem [{\citenamefont {Rostgaard}\ \emph {et~al.}(2010)\citenamefont
  {Rostgaard}, \citenamefont {Jacobsen},\ and\ \citenamefont
  {Thygesen}}]{Rostgaard-FullySelfConsistentGWCalculationsForMolecules-PhysRevB.81.085103-2010}%
  \BibitemOpen
  \bibfield  {author} {\bibinfo {author} {\bibfnamefont {C.}~\bibnamefont
  {Rostgaard}}, \bibinfo {author} {\bibfnamefont {K.~W.}\ \bibnamefont
  {Jacobsen}}, \ and\ \bibinfo {author} {\bibfnamefont {K.~S.}\ \bibnamefont
  {Thygesen}},\ }\href {\doibase 10.1103/PhysRevB.81.085103} {\bibfield
  {journal} {\bibinfo  {journal} {Phys. Rev. B}\ }\textbf {\bibinfo {volume}
  {81}},\ \bibinfo {pages} {085103} (\bibinfo {year} {2010})}\BibitemShut
  {NoStop}%
\bibitem [{\citenamefont {Koval}\ \emph {et~al.}(2014)\citenamefont {Koval},
  \citenamefont {Foerster},\ and\ \citenamefont
  {S\'anchez-Portal}}]{Koval-FullySelfConsistentGWandQuasiparticleSelfConsistentGWforMolecules-PhysRevB.89.155417-2014}%
  \BibitemOpen
  \bibfield  {author} {\bibinfo {author} {\bibfnamefont {P.}~\bibnamefont
  {Koval}}, \bibinfo {author} {\bibfnamefont {D.}~\bibnamefont {Foerster}}, \
  and\ \bibinfo {author} {\bibfnamefont {D.}~\bibnamefont {S\'anchez-Portal}},\
  }\href {\doibase 10.1103/PhysRevB.89.155417} {\bibfield  {journal} {\bibinfo
  {journal} {Phys. Rev. B}\ }\textbf {\bibinfo {volume} {89}},\ \bibinfo
  {pages} {155417} (\bibinfo {year} {2014})}\BibitemShut {NoStop}%
\bibitem [{\citenamefont {Lani}\ \emph {et~al.}(2012)\citenamefont {Lani},
  \citenamefont {Romaniello},\ and\ \citenamefont
  {Reining}}]{Lani-ApproximForManyBodyGreensFunctInsightsFromTheFundamentalEquations-2012}%
  \BibitemOpen
  \bibfield  {author} {\bibinfo {author} {\bibfnamefont {G.}~\bibnamefont
  {Lani}}, \bibinfo {author} {\bibfnamefont {P.}~\bibnamefont {Romaniello}}, \
  and\ \bibinfo {author} {\bibfnamefont {L.}~\bibnamefont {Reining}},\
  }\href@noop {} {\bibfield  {journal} {\bibinfo  {journal} {New J. Phys.}\
  }\textbf {\bibinfo {volume} {14}},\ \bibinfo {pages} {013056} (\bibinfo
  {year} {2012})}\BibitemShut {NoStop}%
\bibitem [{\citenamefont
  {M{\"u}ller}(2006)}]{Muller-InorganicStructuralChemistry-2006}%
  \BibitemOpen
  \bibfield  {author} {\bibinfo {author} {\bibfnamefont {U.}~\bibnamefont
  {M{\"u}ller}},\ }\href@noop {} {\emph {\bibinfo {title} {Inorganic Structural
  Chemistry}}}\ (\bibinfo  {publisher} {John Wiley \& Sons},\ \bibinfo {year}
  {2006})\BibitemShut {NoStop}%
\bibitem [{\citenamefont {Karttunen}\ \emph {et~al.}(2010)\citenamefont
  {Karttunen}, \citenamefont {Fassler}, \citenamefont {Linnolahti},\ and\
  \citenamefont
  {Pakkanen}}]{Karttunen-StructuralPrinciplesOfSemiconductingGroup14ClathrateFrameworks-2010}%
  \BibitemOpen
  \bibfield  {author} {\bibinfo {author} {\bibfnamefont {A.~J.}\ \bibnamefont
  {Karttunen}}, \bibinfo {author} {\bibfnamefont {T.~F.}\ \bibnamefont
  {Fassler}}, \bibinfo {author} {\bibfnamefont {M.}~\bibnamefont {Linnolahti}},
  \ and\ \bibinfo {author} {\bibfnamefont {T.~A.}\ \bibnamefont {Pakkanen}},\
  }\href@noop {} {\bibfield  {journal} {\bibinfo  {journal} {Inorg. Chem.}\
  }\textbf {\bibinfo {volume} {50}},\ \bibinfo {pages} {1733} (\bibinfo {year}
  {2010})}\BibitemShut {NoStop}%
\bibitem [{\citenamefont
  {Hellmann}(1937)}]{Hellmann-EinfuhrungInDieQuantenchemie-1937}%
  \BibitemOpen
  \bibfield  {author} {\bibinfo {author} {\bibfnamefont {H.}~\bibnamefont
  {Hellmann}},\ }\href@noop {} {\emph {\bibinfo {title} {Einf{\"u}hrung in die
  Quantenchemie (Deuticke, Leipzig)}}}\ (\bibinfo {year} {1937})\BibitemShut
  {NoStop}%
\bibitem [{\citenamefont
  {Feynman}(1939)}]{Feynman-ForcesInMolecules-PhysRev.56.340-1939}%
  \BibitemOpen
  \bibfield  {author} {\bibinfo {author} {\bibfnamefont {R.~P.}\ \bibnamefont
  {Feynman}},\ }\href {\doibase 10.1103/PhysRev.56.340} {\bibfield  {journal}
  {\bibinfo  {journal} {Phys. Rev.}\ }\textbf {\bibinfo {volume} {56}},\
  \bibinfo {pages} {340} (\bibinfo {year} {1939})}\BibitemShut {NoStop}%
\bibitem [{\citenamefont
  {Farid}(1999)}]{Farid-GroundAndLowLyingExcitedStatesOfInteractElectronSystemsASurveyAndSomeCriticalAnalyses-1999}%
  \BibitemOpen
  \bibfield  {author} {\bibinfo {author} {\bibfnamefont {B.}~\bibnamefont
  {Farid}},\ }\href@noop {} {\emph {\bibinfo {title} {Electron Correlation in
  the Solid State}}},\ edited by\ \bibinfo {editor} {\bibfnamefont {N.~H.}\
  \bibnamefont {March}}\ (\bibinfo  {publisher} {World Scientific Publishing},\
  \bibinfo {year} {1999})\ pp.\ \bibinfo {pages} {103--261}\BibitemShut
  {NoStop}%
\bibitem [{\citenamefont {Baym}\ and\ \citenamefont
  {Mermin}(1961)}]{Baym-DeterminationOfThermodynGreensFunctions-1961}%
  \BibitemOpen
  \bibfield  {author} {\bibinfo {author} {\bibfnamefont {G.}~\bibnamefont
  {Baym}}\ and\ \bibinfo {author} {\bibfnamefont {N.~D.}\ \bibnamefont
  {Mermin}},\ }\href@noop {} {\bibfield  {journal} {\bibinfo  {journal} {J.
  Math. Phys.}\ }\textbf {\bibinfo {volume} {2}},\ \bibinfo {pages} {232}
  (\bibinfo {year} {1961})}\BibitemShut {NoStop}%
\bibitem [{\citenamefont {Karlsson}\ and\ \citenamefont {van
  Leeuwen}(2016)}]{Karlsson-PartialSelfConsistAndAnalytInManyBodyPertTheorParticNumberConservAndAGenerSumRule-PhysRevB.94.125124-2016}%
  \BibitemOpen
  \bibfield  {author} {\bibinfo {author} {\bibfnamefont {D.}~\bibnamefont
  {Karlsson}}\ and\ \bibinfo {author} {\bibfnamefont {R.}~\bibnamefont {van
  Leeuwen}},\ }\href {\doibase 10.1103/PhysRevB.94.125124} {\bibfield
  {journal} {\bibinfo  {journal} {Phys. Rev. B}\ }\textbf {\bibinfo {volume}
  {94}},\ \bibinfo {pages} {125124} (\bibinfo {year} {2016})}\BibitemShut
  {NoStop}%
\bibitem [{\citenamefont {Marini}\ and\ \citenamefont
  {Pavlyukh}(2018)}]{Marini-FunctApprToTheElectrAndBosonicDynamOfManyBodySystPerturbWithAnArbitStrongElectronBosonInteract-PhysRevB.98.075105-2018}%
  \BibitemOpen
  \bibfield  {author} {\bibinfo {author} {\bibfnamefont {A.}~\bibnamefont
  {Marini}}\ and\ \bibinfo {author} {\bibfnamefont {Y.}~\bibnamefont
  {Pavlyukh}},\ }\href {\doibase 10.1103/PhysRevB.98.075105} {\bibfield
  {journal} {\bibinfo  {journal} {Phys. Rev. B}\ }\textbf {\bibinfo {volume}
  {98}},\ \bibinfo {pages} {075105} (\bibinfo {year} {2018})}\BibitemShut
  {NoStop}%
\bibitem [{\citenamefont {Maradudin}\ \emph {et~al.}(1971)\citenamefont
  {Maradudin}, \citenamefont {Montroll}, \citenamefont {Weiss},\ and\
  \citenamefont {Ipatova}}]{maradudin-harm-appr-1971}%
  \BibitemOpen
  \bibfield  {author} {\bibinfo {author} {\bibfnamefont {A.}~\bibnamefont
  {Maradudin}}, \bibinfo {author} {\bibfnamefont {E.~W.}\ \bibnamefont
  {Montroll}}, \bibinfo {author} {\bibfnamefont {G.~H.}\ \bibnamefont {Weiss}},
  \ and\ \bibinfo {author} {\bibfnamefont {I.~P.}\ \bibnamefont {Ipatova}},\
  }\href@noop {} {\emph {\bibinfo {title} {Theory of The Lattice Dynamics in
  The Harmonic Approximation}}},\ Vol.\ \bibinfo {volume} {Supplement 3}\
  (\bibinfo  {publisher} {Academic Press},\ \bibinfo {year} {1971})\ pp.\
  \bibinfo {pages} {6--81}\BibitemShut {NoStop}%
\bibitem [{\citenamefont
  {Born}(1940)}]{Born-OnTheStabilityOfCrystalLatticesI-1940}%
  \BibitemOpen
  \bibfield  {author} {\bibinfo {author} {\bibfnamefont {M.}~\bibnamefont
  {Born}},\ }\href@noop {} {\bibfield  {journal} {\bibinfo  {journal} {Math.
  Proc. Cambridge Philos. Soc.}\ }\textbf {\bibinfo {volume} {36}},\ \bibinfo
  {pages} {160} (\bibinfo {year} {1940})}\BibitemShut {NoStop}%
\bibitem [{\citenamefont
  {Allen}(1972)}]{Allen-NeutronSpectroscopyOfSuperconductors-PhysRevB.6.2577-1972}%
  \BibitemOpen
  \bibfield  {author} {\bibinfo {author} {\bibfnamefont {P.~B.}\ \bibnamefont
  {Allen}},\ }\href {\doibase 10.1103/PhysRevB.6.2577} {\bibfield  {journal}
  {\bibinfo  {journal} {Phys. Rev. B}\ }\textbf {\bibinfo {volume} {6}},\
  \bibinfo {pages} {2577} (\bibinfo {year} {1972})}\BibitemShut {NoStop}%
\bibitem [{\citenamefont
  {Grimvall}(1981)}]{Grimvall-ElectronPhononInteractionInMetals-Book-1981}%
  \BibitemOpen
  \bibfield  {author} {\bibinfo {author} {\bibfnamefont {G.}~\bibnamefont
  {Grimvall}},\ }\href@noop {} {\emph {\bibinfo {title} {The Electron-Phonon
  Interaction in Metals}}}\ (\bibinfo  {publisher} {North-Holland},\ \bibinfo
  {year} {1981})\BibitemShut {NoStop}%
\bibitem [{\citenamefont
  {Giustino}(2019)}]{Giustino-ErratumElectronPhononInteractFromFirstPrinciples-RevModPhys.91.019901-2019}%
  \BibitemOpen
  \bibfield  {author} {\bibinfo {author} {\bibfnamefont {F.}~\bibnamefont
  {Giustino}},\ }\href {\doibase 10.1103/RevModPhys.91.019901} {\bibfield
  {journal} {\bibinfo  {journal} {Rev. Mod. Phys.}\ }\textbf {\bibinfo {volume}
  {91}},\ \bibinfo {pages} {019901} (\bibinfo {year} {2019})}\BibitemShut
  {NoStop}%
\bibitem [{\citenamefont {Hedin}\ and\ \citenamefont
  {Lundqvist}(1970)}]{Hedin-EffectsOfElectElectrAndElectrPhononInteractOnTheOneElectrStatesOfSolids-1970}%
  \BibitemOpen
  \bibfield  {author} {\bibinfo {author} {\bibfnamefont {L.}~\bibnamefont
  {Hedin}}\ and\ \bibinfo {author} {\bibfnamefont {S.}~\bibnamefont
  {Lundqvist}},\ }\href@noop {} {\bibfield  {journal} {\bibinfo  {journal}
  {Solid State Phys.}\ }\textbf {\bibinfo {volume} {23}},\ \bibinfo {pages} {1}
  (\bibinfo {year} {1970})}\BibitemShut {NoStop}%
\end{thebibliography}%
\end{document}